\documentclass[preprint]{elsarticle}
\usepackage{amsmath, bm}
\usepackage{hyperref}
\usepackage{subfig}
\usepackage{lineno}

\newtheorem{thm}{Theorem}

\long\def\ignore#1{\relax}

\title{Confidence intervals for the Poisson distribution\tnoteref{t1}}
\tnotetext[t1]{\bigskip Work supported in part by the U.S. Department of Energy, Office of Science, Office of High Energy Physics, under Award Number DE-SC0011925.}
\author{Frank C.~Porter\corref{cor1}}
\address{Physics Department 356-48, California Institute of Technology, Pasadena, CA 91125 USA}
\cortext[cor1]{Corresponding author, fcp@caltech.edu}
\ead{fcp@caltech.edu}
\date{\today}

\begin{document}
%\linenumbers

\begin{abstract}
The Poisson probability distribution is frequently encountered in physical science measurements. In spite of the simplicity and familiarity of this distribution,
there is considerable confusion among physicists concerning the description of results obtained via Poisson sampling. 
The goal of this paper is to mitigate this confusion by examining and comparing the properties of both conventional and popular alternative techniques. We concern ourselves in particular with the description of results, as opposed to interpretation. After considering performance with respect to several desirable properties we recommend summarizing the results of Poisson sampling with confidence intervals proposed by Garwood. We note that the $p$-values obtained from these intervals are well-behaved and intuitive, providing for consistent treatment.
We also find that averaging intervals can be problematic if the underlying Poisson distributions are not used. 
\end{abstract}


\maketitle

\tableofcontents

\bigskip

\section{Introduction}
\label{sec:Introduction}

Poisson processes are familiar in physics. In searches for new phenomena, sampling from Poisson distributions with very few expected counts, including background, is common.  We concern ourselves here with sampling\footnote{I consider ``sampling'' from a probability distribution to be synonymous with making an ``observation''.} random variable (RV) $N=0,1,2,\ldots$ from a Poisson distribution (pdf) of the form:
\begin{equation}
f(n;\theta,b) = \frac{\mu^n}{n!}e^{-\mu}=\frac{(\theta+b)^n}{n!}e^{-\theta-b},
\label{eq:Poisson}
\end{equation}
where $n$ is a possible value of $N$, $\mu$ is the mean of the Poisson distribution, equal to the sum of the ``signal strength'' $\theta$, considered to be the unknown quantity of interest, and a background strength, $b\ge0$,
assumed here to be known. In fact it is often a function, $g(\theta)$, of the signal strength that we wish to learn about. We will assume for our discussion that $g(\theta)$ is strictly monotonic. This is the typical situation; other cases may require special treatment. It is also common that the background strength is only imperfectly known, or that there are other uncertainties. However, it will suffice for our discussion to consider the case where $b$ is known. We deliberately restrict to a simple case because there is already much to understand and even
here there is confusion.

We take it as our goal to describe the result of the measurement (we identify the term ``measurement'' as a sampling from a probability distribution). Certainly, quoting the sampled value $n$, or equivalently $n-b$, is an excellent description of the measurement. It is a sufficient statistic for $\theta$ and we have an intuitive grasp of the Poisson RV.
However, when we propagate this into a statement concerning a measurement of the physical quantity of interest, $g(\theta)$, this Poisson intuition tends to be obscured.
This leads us to consider the more general framework of confidence intervals and $p$-values.

Before embarking on this, it is necessary to consider more generally what we would like to accomplish. We have said that the goal is to describe the result of the measurement. This can be contrasted with what we ultimately would like to know: the value of $\theta$, or of $g(\theta)$. In order to avoid confusion, it is useful to distinguish statements about the measurement (sampled value $n$) from statements about the value of the parameter ($\theta$ or $g$), even though both statements may have expressions in terms of the parameter. We consider three aspects that we might be interested in: ({\it i\/}) point estimates; ({\it ii\/}) interval estimates in order to give an idea of how well determined is the point estimate; and ({\it iii\/}) hypothesis tests, for example to address whether a possible signal could be significant.

As mentioned, we are ultimately interested in the value of the parameter. In this case, the point estimate might be regarded as what we believe to be the most probable value of the parameter. The word ``believe'' is important here -- we do not actually know the parameter value, so we can only say what we think (believe) is most likely. Likewise, an interval estimate assigns a probability with which we believe the parameter lies within a stated interval in parameter space. A hypothesis test assigns a probability to a statement about the value of the parameter. For example, we may have a 5\% degree of belief that $\theta$ is greater than zero. It is important to understand that here we are making statements about the actual parameter value. This is the domain of Bayesian statistics.

On the other hand, we here wish to describe the result of the measurement, without making any further inference about the value of the parameter. In this case, an appropriate point estimate could be simply $\hat\theta=n-b$, or $\hat g(\theta) = g(n-b)$. More generally, it could be computed as that value of the parameter for which the sampled value is most probable. A descriptive interval estimate could determine an interval in parameter space such that the interval contains the parameter value with a specified probability over the long run, that is, in the frequency interpretation of repeating the experiment many times. A hypothesis test has a similar interpretation.
Here, we are making no statement about the value of the parameter. For our Poisson sampling, we are only phrasing the fact that we observed $n$ events in different ways. If this appears to be trivial, it is -- no matter how we say it, we are trying to say that we observed $n$ events. The utility of not just saying ``$n$'' is in formulating the statement as a description connected with the physical parameter of interest.
It also provides for a succinct standard framework for describing results independent of the sampling distribution.

Given that what we care about is the actual parameter value, it seems reasonable to suggest that once we have made a measurement, there is no need to make descriptive statements if we can go ahead and quote the result in terms of statements about the parameter value. The problem is that we know of no way to accomplish this without doing so in the framework of degree of belief, and we have no guarantee that two people will agree to believe the same thing. Thus, we find merit in the idea of describing the result of the measurement in some objective way, independent of consumer, which is helpful for each to form their own inferences about the parameter value.

The distinction between ``descriptive'' and ``interpretive'' statistics is nicely made in the context of physics measurements in the paper by James and Roos~\cite{James1991} and earlier by Solmitz in~\cite{Solmitz1964}. As motivated above we here take the point of view that description is a useful thing to do, even though what we really want to know is the actual parameter value. If you disagree, this paper is not for you. We concentrate on descriptive statistics henceforth.

There are many ways to describe something, including something as simple as the result of a measurement in a Poisson process.
We have already mentioned giving the value of $n$, and $g(n-b)$. This is a good idea and should be done. However, we noted that this may not be so simple or complete a description when translated to the parameter of interest. We are typically interested in a point estimate, an estimate of variance, and if the Poisson process is a search for a signal of something new, an estimate of how significant our observation is compared with the no signal hypothesis. We consider these in turn.

Finally, it seems to be necessary to expand on the motivation for this paper, which has two aspects.
First, it may be questioned why we are even interested in the Poisson distribution. Historically, the answer was obvious, but nowadays with the popularity of unbinned likelihood analyses it is possible and even common to do an analysis without reference to the Poisson distribution. 

Consider in particular the familiar situation in which the 
experiment is a measurement of a rare process (e.g., a search for a beyond the standard model process).
In the design phase, such an experiment is typically optimized using ``cut-and-count'' methodology, i.e., Poisson statistics. However, the actual final analysis design is now often approached using an unbinned likelihood methodology. Usually this involves fitting signal and background component distributions to a one or more dimension set of event variables, such as a momentum. Nuisance parameters include background level and signal efficiency, and possibly other things. The nuisance parameters may be determined from auxillary studies or as part of the maximum likelihood fit. Uncertainties due to the nuisance parameters may be included in the fitting methodology, possibly using profile likelihoods, or by incorporating separately as part of the ``systematic uncertainty'' analysis. Model uncertainties are treated as systematic uncertainties. 

While the methodology just summarized is often adopted and has the potential to make effective use of the available data, it is not the only option. Obtaining confidence intervals for the signal strength is complicated by lack of knowledge about the sampling distribution for the likelihood. At high statistics (number of events), we know that this distribution may be simply related to a normal distribution, but we are here trying to make a measurement of a rare process where we must also suppress backgrounds to very low levels. Thus, we are very likely not close to this asymptotic regime. To obtain intervals with frequency validity, we must perform simulations at different points in parameter space to derive the probability distribution of the likelihood.

An alternative is to return to our ``cut-and-count'' analysis, in which we define a signal region in the chosen event variables. Then we know we have a Poisson distribution in the counts and can obtain interval estimates using this fact. Nuisance parameters such as background and efficiency are typically determined to much better precision than that implied by the Poisson fluctuations and are conveniently treated as part of the systematic uncertainty analysis. The likelihood analysis choice is motivated by the notion that an unbinned analysis, which makes use of more information, can provide greater statistical power than the signal bin, though this is generally not demonstrated. I suggest that the possibility that the cut-and-count analysis is essentially as good should be investigated in such rare process analyses, and adopted if found to have comparable statistical power. This simplifies the error analysis. We examine how to do the Poisson analysis in this paper.

The second motivation for this paper is more general, using the Poisson case as a familiar example. We have mentioned above the role of frequency statistics and the distinction between describing a measurement and making a statement about the true sampling distribution. There remains considerable confusion in the physics community about this distinction, even though it is quite crisp. I'll discuss this generally (but with the backdrop of the Poisson distribution) in section~\ref{sec:descriptive}, and the remainder of the paper contains detailed discussion of the Poisson providing illustrations that help with understanding this distinction. I am confident that, in spite of this effort, my point-of-view will be regarded as controversial by some, but I am convinced that it is the clearest and most useful way to view things, and that it is important to put this forward.

The Poisson distribution is discrete, and this presents some interesting features. This has resulted in a number of imaginative approaches to confidence intervals optimizing on various desirable properties. However, the main considerations we discuss are found also in continuous distributions. While the choices are rich for the Poisson, different proposals, including corresponding choices to some of those discussed here, exist as well for continuous distributions, Even the normal distribution is not exempt, and is
controversial given the discussion in section~\ref{sec:descriptive}, and further mention in the conclusions.

\subsection{Point estimates}

It is natural to use $n-b$ or $g(n-b)$ for the point estimator of $\theta$ or $g(\theta)$. This just corresponds to the maximum likelihood estimator, where the likelihood function is
\begin{equation}
\label{eq:likelihood}
L(\theta;n) = \frac{(\theta+b)^n}{n!}e^{-\theta-b},
\end{equation}
treated as a function of $\theta$ for given sampling $n$. Our likelihood function is well-defined for any $\mu\ge0$, that is for any $\theta\ge-b$. If $n=0$ in some sampling, our point estimate is $\hat\theta=-b$. This gives an accurate description of the measurement, i.e., $n=0$, expressed in terms of the signal parameter $\theta$.
It is irrelevant to this description whether we know something (on physical grounds, say) about $\theta$ such as that $\theta\ge a$ (commonly, $a=0$). 

We may contrast this with the point estimator in a Bayesian (degree-of-belief) analysis in a situation where we might know on physical grounds, for example, that $\theta$ cannot be negative. 
In this case we often (depending on the prior distribution) use the value of $\theta$ that maximizes the likelihood as our point estimate, but now we define the likelihood function to be zero for $\theta<0$, since we are certain that the (degree-of-belief) probability is zero for $\theta<0$. It may be supposed that we have incorporated the permitted range of $\theta$ into the prior distribution for $\theta$. 

We regard our choice of point estimator for the Poisson model as obvious.
Some might suggest that it not be permitted to be in a region where $\theta$ cannot lie. This can arise from a misunderstanding of the descriptive goal, perhaps as an attempt to make the estimate also serve as a statement about the true value. 
In particular, if we are aware that we have had a downward fluctuation in the background, it would not be a sensible description to obscure this fact. In general, it is a futile pursuit to try to simultaneously achieve description and interpretation. See section~\ref{sec:descriptive} for further discussion.

\subsection{Interval estimates}

The point of an interval estimate in descriptive statistics is to provide information about how large the variations of the point estimate could be, that is, an idea of how far the point estimator may typically deviate from the expectation of the point estimator (the parameter value, assuming our point estimator is unbiased). A scale for such variation is the standard deviation, which in Poisson statistics is $\sqrt{\mu}$. Given a sampling $n$, an estimate for this is $\sqrt{n}$. This is indeed a fine choice. That is, we could quote an interval for $\theta$ of $[n-b-\sqrt{n}, n-b+\sqrt{n}]$, translating to an interval for $g(\theta)$ of $\left[g(n-b-\sqrt{n}), g(n-b+\sqrt{n})\right]$. The case when $n=0$
is not so much of a problem as one might suppose, since quoting an interval of zero size describes exactly the measurement.

However, the understanding of these intervals is intimately tied to the Poisson distribution and it would be handy if we could devise a methodology for description that 
provides for comparison of results from different sampling models. Quoting estimated variances or standard deviations accomplishes this to some extent, but the associated probabilities depend on the sampling distribution. That is, the probability to be, say, within one standard deviation of the mean, depends on the sampling distribution.
On the other hand, if we can make a statement about probabilities, then we will be providing a distribution-independent description. Thus, we look for such a program.

This problem is solved in the context of the frequency interpretation of statistics~\cite{Neyman1937}: We make statements that are true with stated probability. That is, given any sampling $n$, we devise an algorithm to make a statement depending on $n$, such that the statement is true with the stated probability over (infinitely) repeated instances of the measurement. Sometimes people object that we generally do not actually repeat the measurement procedure, at least not a large number of times, and therefore this is unrealistic. But we do not need to repeat the measurement to do the calculation, and the concept is well-defined in terms of the mathematics. This is the approach that is generally adopted for descriptive statistics and is the one that we shall develop further below. First, however, we will note other possible choices.

Another approach to a descriptive statistic may be derived from Bayesian methodology, even though it may not be thought of as such. We construct a posterior distribution for parameter $\theta$ from the likelihood function corresponding to observation $n$
according to:
\begin{equation}
P(\theta;n) =\frac{L(\theta;n)P(\theta)}{\int_\theta L(\theta;n)P(\theta)\,d\theta},
\end{equation}
where $P(\theta)$ is the prior distribution for $\theta$. In a Bayesian analysis, one uses the prior (to the sampling $n$) degree-of-belief probability in the value of $\theta$ for $P$. Often, in practice one tries to use a prior describing ``complete ignorance'' in the value of $\theta$. How to describe one's complete ignorance in the value of a parameter is a subject of much discussion and ultimately is a matter of personal choice. 

However, people often suggest prescribing such a prior
based only on the sampling distribution. This is often called a reference or objective prior. This is an attempt to make a statement about the parameter without incorporating any prior knowledge or belief.  Such an analysis proceeds as a Bayesian analysis in the sense that it may be interpreted as a statement of belief if one is willing to accept the reference prior as describing one's prior degree of belief. There is no
attempt, or desire, to satisfy a frequency interpretation. Biller and Oser, for example, have advocated this approach~\cite{Biller2015}, based on uniform priors in some choice of parameterization.

On the other hand, as an objective procedure depending on a known (parametrically) sampling distribution, this can also be proposed as a description of the measurement. The interpretation as a descriptive statistic is, however, unclear.  It is possible to investigate the frequency properties, in order to see whether the context of a frequency interpretation makes sense. We shall later discuss two popular reference priors for the Poisson distribution.

We may also consider ``fiducial statistics'' as a possible source of descriptive statistics.
Fiducial statistics were invented to provide a framework for inference without the problem of the subjective prior of the Bayesian methodology~\cite{Fisher1930}. The interpretation, and even the generality, of a ``fiducial interval'' is not clear; further discussion 
may be found, for example, in~\cite{Pedersen1978}, \cite{Wang2000}, and \cite{Kendall1973}. Given certain conditions, a fiducial interval,
$[\ell,u]$, corresponding to a given sampling $n$ of random variable $N$,  is computed according to 
\begin{equation}
\label{eq:fiducial}
\begin{split}
\hbox{Prob}[N\le n | \theta=u(n)] &= \alpha/2\\
\hbox{Prob}[N\ge n | \theta = \ell(n)] &= \alpha/2.
\end{split}
\end{equation}
Here, $1-\alpha$ is the ``fiducial probability'', and we have equalized the probability in the two tails.
We have here expressed this in notation in the context of our discussion of the Poisson distribution.

We mention fiducial intervals here because an important interval we discuss below can be derived as a fiducial interval
and is often referred to as such.

\subsection{Signal significance}

Besides  describing a measurement in terms of a point estimate and an estimate of uncertainty, we are often interested in a description of how significant our observation is compared with the no signal hypothesis or alternatively, how consistent the data is with the no signal hypothesis. As discussed above we're going to adopt the frequency context to provide descriptive  interval estimates with a probabilistic meaning. Thus, we will adopt 
the same context for statements of significance, again with a descriptive probabilistic meaning.

\subsection{Following sections}

It is crucial to understand that our discussion is in the context of frequency statistics. That is, we
are only interested here in providing a description of a measurement. There is no attempt to make a 
statement about the true mean of the Poisson distribution. We suggest that we are providing a description that
could be useful input towards forming such an inference, but we stop short of actually making an inference. We belabor  this point because it is neglected in many discussions, contributing to considerable confusion.

We keep our discussion focused on the basic question, without the complication of unknown nuisance parameters. The hope is that, once some
basic issues are dealt with, then similar methodology may be applied in other situations. A discussion of this is timely, because of 
relatively recent advances in the understanding of some technical matters that we consider important, as will be discussed below.

We will continue to develop the needed concepts. In
section~\ref{sec:construction} we review the construction of frequency statistics. Some controversial points concerning frequency statistics are discussed in section~\ref{sec:descriptive}.
 In section~\ref{sec:desirable} we consider potentially desirable properties. 
In sections~\ref{sec:UL}--\ref{sec:priors} we discuss methods from the statistics community,  followed by two methods invented in, and popular with, the particle physics community in section~\ref{sec:alternative}. Averaging results is discussed in section~\ref{sec:averaging}, with discussion and conclusions in sections~\ref{sec:discussion} and~\ref{sec:conclusions}.

We adopt notational conventions from ref.~\cite{Narsky2014}. Throughout,  $b$ is the known expected ``background'' counts and $\theta$ is the unknown expected signal counts, with $\mu\equiv \theta+b$. Corresponding hypothetical values of $\theta$ and $\mu$ are denoted by $\theta_0$ and $\mu_0$, respectively (and a ``1'' subscript denotes an alternative hypothetical value). In general, we nest our parentheses, $\{[()]\}$. We use $E()$ to denote an expectation value. We will generally denote our confidence intervals as closed intervals, following the convention in~\cite{Shao2003},
though this is not universal (e.g.,~\cite{Thulin2017}), and it does not usually matter when dealing with a continuous parameter space. 
Exceptions will have $($ or $)$ where they occur. Sets are depicted with braces $\{\}$.

The computations have all been implemented on and performed with MATLAB. Where possible, calculations have been validated against published results.

\section{Construction of frequency statistics}
\label{sec:construction}

To prepare for the following discussion, we briefly review the construction of frequency
statistics. We begin with the notion of a confidence interval, or more generally a confidence set, followed by the construction of $p$-values, addressing significance.

\subsection{Confidence sets}

We are interested in confidence sets and $p$-values for a Poisson sampling process,
for random variable (RV) $N=0,1,2,\ldots$ with distribution (pdf):
\begin{equation}
f(n;\theta,b) = \frac{(\theta+b)^n}{n!}e^{-\theta-b},
\label{eq:Poisson2}
\end{equation}
where $n$ is a possible value of $N$, $\theta$ is the unknown quantity (``signal strength'') of interest, and $b\ge0$ is
a background strength, assumed here to be known. Where needed, our point estimator for $\theta$ will be
the maximum likelihood estimator (MLE), designated with a ``hat'', $\hat\theta = N-b$ (possibly restricted to be non-negative).

A defining concept when discussing confidence sets is the ``coverage''. That is, $C_\alpha(N)$ is
a confidence set with confidence level (CL) $1-\alpha$ for parameter $\theta$ corresponding to RV $N$ iff
\begin{equation}
  \hbox{Prob}\left[\theta\in C_\alpha(N)\right] \ge 1-\alpha.
  \label{eq:Calpha}
\end{equation}
The probability statement concerns random variable $C_\alpha(N)$ (independent of $\theta$), and we are working with the frequentist interpretation. This statement must hold for any possible value of $\theta$ (which need not be the same for repeated samplings).
If there is equality in Eq.~\ref{eq:Calpha}, then the coverage of our confidence set is precise.\footnote{This is also referred to as ``exact'' coverage, but we choose the word ``precise'' in order to avoid confusion with the use of ``exact'' introduced  in section~\ref{sec:exact}.} Otherwise, it ``overcovers''.
For discrete distributions such as the Poisson, obtaining precise coverage can be achieved by introducing an additional continuous RV~\cite{Stevens1950, Cousins1994, 1996Porter}, under the notion of a ``randomized confidence set''. However, this does not
add anything of essential value to our description, and we do not advocate this. Thus, for the Poisson, we do not insist on equality.

A common approach to finding confidence sets is as follows (for example, section 2.1.3 in ref.~\cite{Narsky2014}):
Our goal is to construct random sets $C_\alpha(N)$ of values for $\theta$ with the property
in Eq.~\ref{eq:Calpha}. 
Toward finding sets $C_\alpha(N)$, it is helpful to consider constructing sets $S_\alpha(\theta)$ of possible values of RV $N$ for any given $\theta$ with the
property that if a sampled value $n\in S_\alpha(\theta)$, then $\theta$ will be in the confidence set.
That is,
\begin{equation}
\label{eq:Salpha}
S_\alpha(\theta) \equiv \{n | \theta\in C_\alpha(n)\}.
\end{equation}

This can be re-expressed in the language of hypothesis testing. Indeed, a suggested interpretation~\cite{Vos2005} is to regard a confidence interval as a set of parameter values that are not rejected by a hypothesis test. Consider a test for
\begin{equation}
\begin{split}
\hbox{$H_0$:\ } &\theta=\theta_0\\
\hbox{against}\\
\hbox{$H_1$:\ } &\theta\ne \theta_0
\end{split}
\end{equation}
We construct a test at the level $\alpha$ significance by construction of a test statistic $T_{\theta_0}(N)$ taking on values 0 or 1. Hypothesis $H_0$ is accepted whenever $T_{\theta_0}=0$ and rejected when $T_{\theta_0}=1$. The ``acceptance region'', $A_\alpha(\theta_0)$ for the test is defined as the set of values $N=n$ such that $T_{\theta_0}=0$:
\begin{equation}
\label{eq:Aalpha}
A_\alpha(\theta_0) \equiv \{n | T_{\theta_0}(n)=0\}.
\end{equation}
The probability content is that the significance level is the probability of a Type I error, where we reject $H_0$ when $H_0$ is true, i.e.,
\begin{equation}
\alpha = \hbox{Prob}\left[T(N)=1 |\theta=\theta_0\right] = \hbox{Prob}\left[N\notin A_\alpha(\theta_0) |\theta=\theta_0\right].
\end{equation}
We are dealing with a discrete distribution, so instead of equality, we require that the type I error probability be less than or equal to $\alpha$. We may rewrite as
\begin{equation}
\hbox{Prob}\left[N\in A_\alpha(\theta) |\theta\right] \ge 1-\alpha,
\end{equation}
where we have also generalized to any given choice for the value of $\theta$.
We are going to restrict our attention to acceptance regions containing contiguous values of RV $N$, that is, with no gaps in the sequence. Thus, we can rewrite our probability statement as that of obtaining two integers $n_\ell$ and $n_h$, depending on $\theta$ and $\alpha$, such that:
\begin{equation}
\sum_{k=n_\ell}^{n_h} f(k;\theta,b) \ge 1-\alpha,
\end{equation}
and then $A_\alpha(\theta) = [n_\ell,n_h]$.
Finally, looking at Eq.~\ref{eq:Salpha}, we see that $A_\alpha(\theta)$ may be used as set $S_\alpha(\theta)$,
and hence defines a prescription for determining a confidence set at the $1-\alpha$ confidence level,
\begin{equation}
C_\alpha(n) = \{\theta |  n \in A_\alpha(\theta)\}.
\end{equation}

For the Poisson problem there is no real utility in permitting gaps, even though one could potentially get better behavior on some criteria, such as coverage. We are going to immediately require that our confidence set be a continuous set of values of $\theta$, as discussed in section~\ref{sec:desirable}.
Thus, we define random variables $\ell_\alpha(N)$ and $u_\alpha(N)$ with the property that
\begin{equation}
\label{eq:CI}
\hbox{Prob}(\ell_\alpha\le\theta\le u_\alpha) \ge 1-\alpha.
\end{equation}
That is, we construct a ``confidence interval'' at confidence level $1-\alpha$.

\subsection{$p$-values}
\label{sec:pvalues}

In addition to describing the result of a measurement with a confidence interval, it can be useful to quote a $p$-value (\cite{Shao2003, Wasserstein2016}), for example, when we are searching for evidence of something ``new''.
The $p$-value provides a probability statement addressing how consistent a measurement is with a 
given hypothesis, in the context of a set of possible alternative hypotheses. Thus, we consider the question of consistency with a given value of the Poisson mean:
\begin{equation}
\begin{split}
\hbox{$H_0$:\ } &\theta=\theta_0\\
\hbox{against}\\
\hbox{$H_1$:\ } &\theta \ne \theta_0.
\end{split}
\end{equation}

If we have a confidence interval derived from a hypothesis test, we may use the acceptance regions $A_\alpha(\theta_0)$ to obtain a $p$-value as follows: 
Let, for any parameter value $\theta_0$,
\begin{equation}
 A_\alpha(\theta_0)\equiv\{n|\theta_0\in C_\alpha(n)\},
 \end{equation}
 be the acceptance set corresponding to confidence interval $C_\alpha(n)$.
 For a given observation $n_{\rm obs}$, if $n_{\rm obs}<\theta_0$, find the smallest $\alpha$ such that the smallest element of $A_\alpha(\theta_0)$ is $n_{\rm obs}$. If $n_{\rm obs}>\theta_0$, find the smallest $\alpha$ such that the largest element of $A_\alpha(\theta_0)$ is $n_{\rm obs}$. Then $p=\alpha$.
 That is, $A_\alpha(\theta_0)$ is interpreted as the
 acceptance region, at the $\alpha$ significance, for the test $H_0:\theta=\theta_0$.
 The observed value of $n$ is just barely consistent with $H_0$ at significance level $p$~\cite{Shao2003}.

 A small value of $p$ corresponds to a large value of $1-\alpha$, i.e., to a confidence interval with a large confidence level. That is, we must have a large confidence interval in order to contain the observation $n$. The smaller $p$ is, the less probable our observation is under $H_0$. We would like to design our confidence interval such that we obtain $p$-values that can be interpreted as something like the probability of obtaining an observation as far or farther from the value $\theta_0+b$ as the observed value $n$. 

A common case is for a null hypothesis that there is no signal $\hbox{$H_0$:\ }\theta=\theta_0=0$, and the alternative hypothesis of interest is that there is a signal, $\hbox{$H_1$:\ } \theta > 0$. If we have devised a test for this, with acceptance region $A_\alpha(0)$, then again $p$ is computed as the smallest value of
$\alpha$ for which $n\in A_\alpha(0)$. Since this is a one-sided test, there is a unique interpretation that
we would like to have: The most sensible $p$-value is one that gives the probability that $N\ge n$ under the hypothesis   $\hbox{$H_0$:\ }\theta_0=0$, where $n$ is the observed value. That is,
\begin{equation}
p = \hbox{Prob}(N\ge n|\theta=0).
\end{equation}

\section{Descriptive statistics}
\label{sec:descriptive}

The first draft of this paper received a comment from Cousins~\cite{Cousins2025} on two main issues:
\begin{enumerate}
\item I have suggested above that the purpose of frequency statistics is descriptive, that is, it makes statements about the result of the observed sampling. Cousins suggests that it also satisfies an inferential purpose.
\item I have suggested in the discussion below Eq.~\ref{eq:likelihood} that, when describing the result of a sampling, the likelihood function may sometimes be usefully evaluated at parameter values that are known to be incorrect (e.g., ``non-physical''). Cousins suggests that this is forbidden.
\end{enumerate}
I consider these two issues to be related.
I have posted a response~\cite{Porter2025} but it seems  desirable to include an expanded discussion in published form here.

We first address the descriptive vs. inferential property. Kendall, Stuart, Ord, and Arnold define ``statistical inference'' as an ``inductive process from sample to population'' (section 26.1 of~\cite{Kendall1999}). Shao defines ``statistical inference'' in a somewhat self-referential way as  making ``an inference about the unknown population based on the sample $X$'' (section 2.4 of~\cite{Shao2003}). Thus, the notion of inference involves making some statement about the sampling distribution based on an observed sample. Our Poisson case may be regarded as an example. In this case, the specific quantity of the population that we would like to infer is the population parameter $\theta$, or equivalently $\mu =\theta +b$. We note the we are not considering finite populations that might be of interest in other fields.
That is, our inference problem is to make some statement about the Poisson mean given a sampling from the distribution. 

The problem of inference (and induction) has a long history of philosophical thought and controversy.
Karl Popper, for example, grapples with the problem of going from observations to statements about truth and falsity~\cite{Popper1959}. He considers the problems of inductive logic to be ``insurmountable'', and that 
the problems ``inherent in  the doctrine \dots that inductive inference, although not `strictly valid', {\it can attain some degree of `reliability' or of `probability'\/ \/}'' (emphasis his) are likewise insurmountable.\footnote{A recent essay on these matters appears in~\cite{Henderson2024}.} I subscribe to this point of view. That is, in my interpretation, statements about truth always require some level of subjectivity, which can be encoded as ``degree of belief''. In other words, our knowledge results from a process of abductive reasoning.

Physicist E.~T.~Jaynes' substantial work on probability, ``Probability theory: the logic of science''~\cite{Jaynes2003} provides examples demonstrating that frequency statistics (including from Poisson sampling) interpreted as inferences can be nonsensical. As Jaynes was only concerned with inference in his book, and not description, he exhibits little patience for frequency statistics.

Statistics philosopher Deborah Mayo and statistician David Cox published an article entitled ``Frequentist statistics as a theory of inductive inference''~\cite{Mayo2006FrequentistSA}. In spite of the suggestive title, they do not argue that frequentist statistics provides an inference about truth, such as the value of a population parameter. Instead, frequentist statistics provide ``evidence'' that may be useful in forming an inference. This is precisely the distinction that I am making.  A very small $p$-value under some null hypothesis ($H_0$) presents evidence against the null hypothesis, but must be taken in the context of all other available evidence before making the inference that $H_0$ is most likely incorrect. For example, if there is no other relevant evidence or experience, a $p$-value of 0.001 might be construed as strong evidence against $H_0$, but if this same $p$-value is the last in a sequence of 1000 results with an approximately uniform distribution of $p$-values, then the evidence against $H_0$ is weak. 

The classic ``Advanced Theory of Statistics'' text~\cite{Kendall1999}  uses the expression ``frequentist approach (e.g., section 26.4) in their chapter (26) entitled ``Comparative Statistical Inference''. As far as I know, this reference avoids the expression ``frequentist inference'', which I consider to be an astute choice, as I argue that frequentist statistics may be employed in forming an inference, but does not itself provide the inference. Their chapter 26 considers several approaches to estimation, including Bayesian.
They recognize the controversial nature of the subject, and present their personal views in sections 26.58 through 26.78.  I consider that my point of view is consistent with
their discussion. In particular, their section 26.62 states:

\medskip

\begin{minipage}{0.9\textwidth}
``The frequency approach leads then to point and interval estimates and tests of hypotheses that are keyed to an interpretation of performance {\it in the long run} [italics theirs].  The other approaches we have described constitute several attempts to develop an additional, or alternative, notion of probability that enables the investigator to make inferential statements conditionally upon the data recorded in a {\it particular statistical experiment} [italics theirs].''
\end{minipage}

\medskip

In spite of our above definition of inference, we must take some care in reading the statistics literature.
A relatively recent case in point is the work ``Computer Age Statistical Inference: Algorithms, Evidence and Data Science'', by the two prominent statisticians Efron and Hastie~\cite{Efron2016}. Chapter 2 of this work is titled ``Frequentist Inference'', suggesting a possible conflict with the point I have been making.  However, they define this to mean:

\medskip

\begin{minipage}{0.9\textwidth}
``We can now give a first definition of frequentist inference: the accuracy of an observed estimate $\hat\theta=t(x)$ is the probabilistic accuracy of $\hat\Theta=t(X)$ as an estimator of $\theta$. This may seem more a tautology than a definition, but it contains a powerful idea: $\hat\theta$ is just a single number but $\hat\Theta$ takes on a range of values whose spread can define measures of accuracy.''
\end{minipage}

\medskip

\noindent That is, the notion is one of accuracy (in a probabilistic or frequency sense) rather than a statement about the true value of parameter $\theta$. Thus, there is a nuance of language difference in the meaning assigned to inference, but no conflict with our point of view.\footnote{Their ``second'' definition appears to be given on page 14, and involves the conventional frequentist picture of repeating the measurement many times.}

In this context, Efron and Hastie further remark on the distinction between the frequentist inference based on the ensemble and the result of a given observation in footnote 3 on page 37: ``\dots frequentist inferences, which are calibrated in terms of possible observed data sets $X$, may be inappropriate for the actual observation $x$.''

It is interesting to notice also Efron and Hastie's discussion of ``Fisherian inference'', in which a statement about the true value of parameter $\theta$ is made: ``Then everyone  should agree in the absence of prior information that $\hat\theta$ is the best estimate of $\theta$, that $\theta$ has about 95\% chance of lying in the interval $\hat\theta \pm 1.96\hat\sigma/\sqrt{n}$, etc.''~\cite{Efron2016}. As Efron and Hastie remark, Fisher attempted to produce inferences about the true value of the parameter, going beyond frequency statistics, but without embracing Bayesian methods. In spite of Fisher's many important contributions, this program has not been successful (``Fisher's logical system is not in favor these days''~\cite{Efron2016}). I think this is at least partly because the meaning of ``95\% chance'' in the above quote is not clear.\footnote{Lindley~\cite{Lindley1958} proved that, under restricted conditions, Fisher's fiducial method gives the same result as a Bayesian calculation with a uniforma prior.}

In the context of statistical methods for particle physics, Solmitz\footnote{Solmitz is credited with introducing the maximum likelihood method to particle physics~\cite{Pripstein1981}.}~\cite{Solmitz1964} describes both Bayesian and frequentist (``anti-Bayesian'') estimation, making the point in the latter methodology that no statement is actually made concerning the true value of a parameter without making a further step with Bayes' principle. James and Roos~\cite{James1991} discuss reporting of results near an unphysical region, again noting that the frequentist statistics make no statement about the true value of a parameter.

People (including our above mention of Jaynes) have sometimes proven that frequency statistics is ``wrong'', resulting when they treat it as serving an inferential function. For a fairly recent prominent example,
Trafimow and Marks~\cite{Trafimow02012015}, as editors of the journal {\it Basic and Applied Social Psychology} (BASP) famously banned the use of frequentist statistics such as $p$-values and confidence intervals from publication in their journal. They claimed that these statistics are ``invalid''. The essence of the argument is contained in the abstract of~\cite{Trafimow2003}: 

\smallskip

\begin{minipage}{0.9\textwidth}
``Because the probability of obtaining an experimental finding given that the null hypothesis is true [$p(F\backslash H_0)$] is not the same as the probability that the null hypothesis is true given a finding [$p(H_0\backslash F)$], calculating the former probability does not justify conclusions about the latter one. As the standard null-hypothesis significance testing procedure does just that, it is logically invalid''.
\end{minipage}

\smallskip

\noindent Here, by ``standard null-hypothesis significance testing procedure'' they mean frequency statistics such as $p$-values. 
The first sentence is a well-known observation. The second sentence is an assertion that frequency statistics provides an inference, which is incorrect.

It is remarkable that the editors state that no inferential procedures are required in BASP, but that ``BASP will require strong descriptive statistics\dots''. By inferential procedures, they explicitly mean frequentist statistics with a milder stance on Bayesian statistics. Instead of embracing the descriptive frequentist statistics they go against their own requirement and ban them by misunderstanding them as inferential.  In their defense, their ban is the result of finding that too many practitioners appear to also misunderstand the distinction and actually use $p$-values towards making stronger inferential conclusions than justified. This is the argument that people who do understand the distinction sometimes make that frequency statistics are problematic because they are often misunderstood (see, e.g.,~\cite{Aschwanden2016}). I prefer to think that education can mitigate this difficulty, and that we can benefit from use of frequency statistics.

The  BASP editorial banning frequency statistics generated an immediate reaction (and some derision), with some people agreeing and others pointing out the flawed understanding.
The American Statistical Association felt compelled to issue a statement clarifying what $p$-values are and aren't and how they should be used~\cite{Wasserstein2016}.
This statement contains the important point about $p$-values: ``It is
a statement about data in relation to a specified hypothetical explanation, and is not a statement
about the explanation itself.'' That is, it is a description of the data in a relevant context, not an inference about the context.

I have explained the rationale for treating frequency statistics as descriptive rather than inferential. As noted, this does not mean that frequency statistics cannot be used in an inferential process. They just cannot be treated as themselves providing an inference. Instead, as a description of the results of a measurement, they provide evidence upon which conclusions about the parent population may be reached. I suggest that such conclusions in physics can best be expressed in a degree-of-belief paradigm.

Let us turn to the second point of contention, the suggestion that it is forbidden to evaluate the likelihood function in an unphysical region. It is useful here to make the distinction between a descriptive methodology and a degree-of-belief methodology. We will mostly use the example of sampling from the Poisson distribution with mean $\mu= \theta + b$ where $b$ is a known background parameter and $\theta$ is an unknown signal parameter, although the discussion is by no means limited to this example. We suppose that on physical grounds we know that $\theta\ge 0$. It is important to keep in mind that the sampling distribution and the likelihood function are not the same thing -- the former gives probabilities for values of a random variable, given any parameter value(s), and the latter provides the value of the sampling distribution for a given value of the random variable, as a function of parameter value(s).  
%It is possible to have the same likelihood function with samples from different sampling distributions (e.g.,~\cite{1996Porter}).

We may start with degree-of-belief, by which I mean that we wish to make statements about the true value of $\theta$ reflecting our belief (which may be different for different people). As we have complete belief that $\theta\ge 0$, we will never quote a result that includes negative values of $\theta$. In this case, we might as well say that the likelihood function is zero, or not defined, for $\theta<0$. It does not really matter, because we will always impose the $\theta\ge 0$ restriction on our statements of belief. In the context of using Bayes theorem, our prior belief is $P(\theta<0)=0$ and that imposes this constraint independent of what we do with the likelihood function.

The situation is quite different if we are describing our observation (measurement, or sampling from the parent distribution). In this case there is no obvious restriction because we are not trying to make a statement about the value of $\theta$. We are interested in providing a useful description of the result of the measurement. 

It may be debated what constitutes a useful description, and I list several possibly desirable properties, not necessarily all mutually compatible, in section~\ref{sec:desirable}. But at the very least, it seems clear that we should provide a description that gives as much information contained in the measurement as possible, relevant to the question we are trying to address. In our Poisson example, this question is the value of $\theta$. That is, a description should include a sufficient statistic, if doing so is realistic. In our simple Poisson case, this is readily accomplished. Simply quoting the observed counts, $n$, is such a statistic. It is hard to imagine anyone objecting to quoting this statistic as the result of the sampling.
Yet it is precisely the maximum likelihood estimator for $\mu$, if the likelihood is allowed to be evaluated in the non-physical region, $\mu<b$. If this is not to be permitted, the statistic becomes $\max(n,b)$. The corresponding point estimates for $\theta$ are $n-b$ or $\max(n-b,0)$. If the likelihood function cannot be evaluated for $\theta<0$ (or $\mu< b$), the maximum likelihood estimator loses the sufficiency property, i.e., suppresses information. Allowing the likelihood function to be evaluated at negative $\theta$ maintains the sufficiency property of the point estimator, which I consider to be a strong argument for doing so.

Others also suggest that the likelihood function can be evaluated without constraining to physical parameter space. Solmitz~\cite{Solmitz1964} allows the likelihood function to be defined outside of the physical region, referring to the position of the maximum as ``which may be outside of the physical region''. James and Roos~\cite{James1991} argue that when publishing the description of the result of a measurement ``it should even be as much as possible independent of physical constraints. The main reason is that only then can the results be unbiased and therefore easily combined with other results.''\footnote{While this is a commonly expressed reason, I wouldn't necessarily characterize it as the ``main'' reason, e.g., considering my above and below discussion} 

Unbiasedness is generally regarded as a desirable property in a point estimator, and $n-b$ is an unbiased estimator for $\theta$. The property of smallest error, as in the mean squared error (MSE) is also a desirable property and sometimes we wish to use such a statistic (especially in trading off bias and variance in the context of density estimation). For this property, the biased estimator  $\max(n-b,0)$ has a smaller MSE than $n-b$. However, if we have $n-b$ we can also form $\max(n-b, 0)$; we cannot go in the other direction, as we have lost sufficiency.

A familiar situation arises when measuring a signal above a (here assumed to be known) background distribution in some sampled distribution expressed as a histogram with $n$ bins with contents $x=\{x_1,\ldots,x_n\}$. If the statistics of the background is large, the Poisson bin contents are approximately normally distributed. Assume for simplicity here that the shape and location of the signal pdf is known and the only unknown is the signal strength. Suppose that we do a maximum likelihood fit to estimate the signal strength. We wish, for example, to test the goodness of fit of our model to the data. Let $L_{\rm max}$ be the maximum likelihood for our fit to the data according to the model with signal strength varied. Let $L_0$ be given by
\begin{equation}
L_0 = \prod_{i=1}^n \frac{{x_i}^{x_i}}{x_i!}e^{-x_i}.
\end{equation}
Let $\lambda = -2\ln L_{\rm max}/L_0$.
In the asymptotic regime, the goodness of fit of the data to the signal plus background model is obtained according to $\lambda$ being approximately $\chi^2$ distributed with $n-1$ degree of freedom. This statement holds as long as the signal strength is allowed to be negative as well as positive, i.e., $L_{\max}$ is to be computed without constraining the signal to be in the physical region (to satisfy the interior of the maintained hypothesis according to Wilk's theorem). While we can investigate the distribution of $\lambda$ with a non-negative signal constraint, it is certainly convenient to be able to use the $\chi^2$ distribution, as is commonly done. This provides another situation supporting the evaluation of likelihoods in the non-physical region.

In the context of descriptive statistics, we have discussed ways in which it is useful to evaluate the likelihood function in non-physical regions. There is no reason why this should be forbidden in this context, and we will show in this paper that this is indeed desirable in the discussion of descriptive confidence intervals for the Poisson distribution. I repeat that there is no reason to evaluate the likelihood function in a non-physical region in the context of degree-of-belief statistics however, and perhaps this is the source of some resistance to the idea.

\section{Desirable properties of confidence intervals}
\label{sec:desirable}

Over the decades, an extensive discussion on confidence intervals for discrete distributions including Poisson has developed in the statistics literature, with additional
contributions from the physics community.
The discreteness of the sample space has complicated the discussion, although many of the same issues hold as well in the case of continuous distributions. Various properties are
considered desirable by one or another investigator. Unfortunately, these ``desirable'' properties cannot all be optimally satisfied at once and compromises must be made. The situation in physics remains confused, with proponents
of different algorithms and no consensus. In this paper we discuss what properties could be considered desirable and then investigate the properties of several approaches and formulate recommendations.

There are no unknown nuisance parameters in this problem, and obtaining solutions to Eq.~\ref{eq:Calpha} 
is relatively straightforward without having to deal with nuisance parameters.  Indeed, there are many approaches that have been suggested in the statistics literature, plus some in the physics community. Comparative reviews of several methods in the statistics literature appear, for example, in refs.~\cite{Brown2003}, \cite{Hirji2006}, \cite{Patil2012}, and~\cite{Thulin2017}. Here, we discuss matters in the context of physics measurements. In doing so, we aren't suggesting that physics measurement differs fundamentally from other fields, but we are interested in the context of the culture of thought in physical measurement.

\subsection{Exactness}
\label{sec:exact}

 There is no complete consensus on how to compute Poisson confidence intervals, though a major distinction is whether to insist on ``exact'' (defined below)
intervals or not.  

Hence our first choice for physics contexts is  ``exact'' versus
``approximate'' confidence sets. Exact refers to a confidence set determined
using the actual sampling distribution, here, Poisson. In this case, we can ensure that we satisfy the inequality in Eq.~\ref{eq:Calpha}.
On the other hand, it is possible to base sets on approximations to the Poisson distribution, for example, the normal distribution. Such sets
are not ``exact'' and may both overcover and undercover. When the number of counts is large, the normal (or more refined variations) approximation may well be good enough, and is often adopted. However, we are presently mostly concerned with the small statistics situation. Physicists generally prefer to err on the side of ``conservatism'', in order to bias against making false discovery claims. We adopt this view here and in order to ensure this, will think mostly about exact confidence sets. We thus work with the Poisson distribution without approximation and investigate methods that never undercover. The cost will sometimes be substantial overcovering. Over the years a number of solutions have been proposed
for exact Poisson confidence intervals in the statistics community, such as~\cite{Garwood1936}, \cite{Crow1959}, \cite{Casella1989}, \cite{Blaker2000}, \cite{Kabaila2001}, as well proposals for distributions such as the binomial that can readily be adapted to the Poisson, e.g., \cite{Stevens1950}, \cite{Sterne1954}, \cite{Somerville2013}. A fairly recent general resource for exact analysis of discrete sampling distributions is \cite{Hirji2006}.

While we choose to require Eq.~\ref{eq:Calpha}, we emphasize here that approximate methods may well be acceptable in some situations. Our discussion does not preclude adopting these when appropriate. Here, however, we concentrate on sets satisfying  Eq.~\ref{eq:Calpha}.

\subsection{Connectedness}
\label{sec:connectedness}

There is no requirement in Eq.~\ref{eq:Calpha} that a confidence set be connected. For some distributions, disconnected sets provide
useful descriptions. However, the Poisson distribution is unimodal, and our sets will be easiest to understand if we insist on connected confidence sets (e.g.,~\cite{Blaker2000}).
That is, we will insist that the sets $C_\alpha(N)$ contain only single continuous ranges of $\theta$ values. This is not always automatic in practice; we
must enforce it. This requirement will put constraints on our ability to minimize overcovering.  With this requirement, our confidence set can be described as an interval:
\begin{equation} 
C_\alpha(N) = \left[\ell_\alpha(N),u_\alpha(N)\right].
\end{equation}

We henceforth require connected intervals. If we have a confidence set that is not connected, we can always add intermediate parameter values until it is connected. This will be at the cost of performance in terms of overcoverage. For simplicity we will also drop the $\alpha$ subscript on $\ell$ and $u$ from now on as understood. 

\subsection{Contains maximum likelihood estimator}
\label{sec:containMLE}

For any observation, $n$, we require that the confidence interval $(\ell,u)$ for $\mu$ contains $n$: $\ell\le n\le u$. That is, the MLE
for $\mu$ is in the confidence interval, and the confidence interval for $\theta$ is $[\ell-b,u-b]$ is $\ell-b\le n -b \le u-b$, containing the MLE $n-b$ for $\theta$.\footnote{For technical reasons, occurring when a lower endpoint for some observation coincides with an upper endpoint for another observation, Casella and Robert~\cite{Casella1989} argue for half-open confidence intervals, $[\ell(n), u(n))$. In this case our requirement for a $\theta$ confidence interval $[\ell-b,u-b)$ is $\ell-b\le n-b < u-b$.} As in~\cite{Casella1989}, we suspect violation of this will be non-optimal in other respects, and in any event, it is a descriptive property that we would normally expect (it seems bizarre to quote a confidence interval that excludes the point estimate) and hence should probably require. As discussed in section 3, we insist that the maximum likelihood estimator not be restricted to some physical parameter space.

\subsection{Optimal coverage}

While we allow ourselves to err on the side of conservatism, it is wasteful to overdo this. Thus, we prefer intervals for which the coverage is close to the stated confidence level $1-\alpha$.

Over-coverage is in a sense minimized if~\cite{Kabaila2001} 
\begin{equation}
\inf_\theta \hbox{Prob}\left\{\theta\in [\ell(N),u(N)] |\theta\right\} = 1-\alpha.
\label{eq:optimalCoverage}
\end{equation}
That is, for at least some $\theta$, the confidence interval gives exact coverage. This is a reasonable property to strive for,
since if we have an interval that does not achieve this, it seems plausible that we could improve it until it does.

There are in fact many algorithms that satisfy this. Historically, much emphasis has been placed on finding algorithms
that diminish overcoverage, as this implies less wasteful use of data.

\subsection{Length}
\label{sec:length}

Another property that has long been considered desirable is shortness of the interval  (e.g.,~\cite{Kendall1973, Shao2003}). If $d(N)\equiv u(N)-\ell(N)$ is the length of the interval,
short values for $d(N)$ are preferred. 
Along with coverage, interval length has historically been a much-emphasized property.
As $d(N)$ is a random variable, the notion of shortness requires some definition and choice. We may consider the expected length,
\begin{equation}
E(d;\theta,b) = \sum_{n=0}^\infty d(n) f(n;\theta, b).
\end{equation}
Unless we state otherwise, this will be what we mean by ``length'' or ``size''.  We would like to have this length be ``short'' for all values of $\theta$.

Statisticians have defined various notions of shortest length intervals going beyond this. 
One such sense of shortest length is from Kabaila and Byrne~\cite{Kabaila2001}: An interval $[\ell(N),u(N))$ is shortest if we construct an interval $[\ell^\prime(N),u^\prime(N)) \subset [\ell(N),u(N))$ where the subset is proper for at least one $N=n$, then there exists a $\theta$ such that the primed interval undercovers. This is called the ``inability to be shortened'' property.

Casella and Robert~\cite{Casella1989} discuss the idea of comparing sums of lengths as a measure of length optimality, introducing a notion of ``dominance'': A given confidence level interval $C^\prime =[\ell^\prime(n),u^\prime(n))$ is said to dominate another interval, $C=[\ell(n),u(n))$, with the same confidence level if there exists an $N_0$ such that, whenever $N>N_0$ either
\begin{equation}
\sum_{n=0}^N[u^\prime(n) - \ell^\prime(n)] < \sum_{n=0}^N[u(n) - \ell(n)]
\end{equation}
or
\begin{align}
\sum_{n=0}^N [u^\prime(n) - \ell^\prime(n)] &= \sum_{n=0}^N [u(n) - \ell(n)],\quad\hbox{and}\\
\hbox{Prob}(\theta\in C^\prime|\theta) &\ge \hbox{Prob}(\theta\in C|\theta),\quad\hbox{with inequality for some $\theta$.}
\end{align}
These authors show that, if for some observation $n$, confidence procedure $C$ has $\ell(n)\ge \ell(n+1)$ or  $u(n)\ge u(n+1)$, then $C$ can be dominated, i.e., improved in length. Based on this observation, Casella and Robert propose a refinement procedure to improve confidence intervals.

Schilling and Holladay~\cite{Schilling2017} point out a paradox with the proposed refinement methodology, and propose an alternative length criterion:
Interval $C^\prime$ is said to be shorter length than $C$ if $\exists N_0$ such that
\begin{align}
\sum_{n=0}^N[u^\prime(n)-\ell^\prime(n)] &\le \sum_{n=0}^N[u(n)-\ell(n)], \quad\forall N, \hbox{\ and}\\
\sum_{n=0}^N[u^\prime(n)-\ell^\prime(n)] &< \sum_{n=0}^N[u(n)-\ell(n)], \quad\forall N>N_0.
\end{align}

Short lengths are considered desirable to localize the description of the measurement. However, we will see that
it must be done with some care  that some values of $n$ do not lead to absurdly short intervals. That is, we have
intuition that the variance of the Poisson is $\mu$,  and hence intervals should scale like $\sqrt{n}$ for different $n$
values. If we violate this, then our intervals lack the intuitive connection of a good description.

\subsection{Scale and asymptopia}
\label{sec:scale}

It should be easy to make an intuitive connection with the result. For Poisson sampling (Eq.~\ref{eq:Poisson}), the variance of
the distribution is $\theta +b$ and hence $\sqrt{\theta+b}$, or estimator $\sqrt{N}$, sets the scale we expect for the size of a confidence set. Indeed, an acceptable confidence set should most naturally asymptotically
approach $(\hat\theta - f_\alpha\sqrt{\hat\theta+b},\hat\theta - f_\alpha\sqrt{\hat\theta+b})$, where $\hat\theta = N-b$
and $f_\alpha=\sqrt{2}\,\hbox{erf}^{-1}(1-\alpha)$.

\subsection{Ordered} 

A confidence interval is called ``ordered''~\cite{Casella1989, Kabaila2001} if, for any pair of different observations, $n_1<n_2$,
the lower confidence bounds are ordered, $\ell_1\le\ell_2$ and the upper confidence bounds are ordered, $u_1\le u_2$. Reference~\cite{Kabaila2001} actually defines ``ordered'' with strict inequalities, $\ell_1<\ell_2$ and $u_1< u_2$. We will refer to this as ``strict ordering''.
Any reasonable description of a measurement should at least be ordered. It would be counterintuitive to have, e.g., a lower interval bound that decreases when the observation increases, for given confidence level.

\subsection{Symmetry}

Another property that might be desirable for a good description is a notion of symmetry.
Here, following our frequency interpretation, symmetry is taken to mean that the interval is constructed with approximately equal probability content in the high and low tails outside the interval. We will call this ``symmetric in probability''. This is indeed the classic Poisson interval, and we consider it in section~\ref{sec:Garwood}. 

A different notion of symmetry is that of equal size intervals on both sides of the chosen point estimator. We will call this ``symmetry in interval'' to distinguish from the above notion of symmetry. In this case, the result of the measurement can be described as the point estimator ``plus or minus $c$'', where $c$ is one-half the confidence interval size (with confidence level traditionally chosen as 68\%\footnote{Motivated by the normal probability to be within one standard deviation of the mean. When we say 68\% here, we always mean more precisely this normal probability.}). While   perhaps a nice feature, we do not have a good reason to insist on it, although we will use such intervals as convenient in our discussion of averages in section~\ref{sec:averaging}.

We aren't interested in searching explicitly for either one-sided or two-sided intervals, leaving this
property to come naturally out of the algorithm. For the descriptive nature of frequentist statistics, it
is more important to achieve a good description and insisting on two-sided or one-sided intervals
is not helpful. Averaging results from different measurements can be awkward with one-sided intervals.
We discuss limits further in section~\ref{sec:UL}. 
On the other hand, one-sided intervals, especially upper limits, are often very useful in interpretive statements, such
as ``I am 95\% sure that the true value is less than five.'' This is a degree-of-belief statement, forming the basis for
making a decision, and is the domain of Bayesian analysis. While we will investigate the frequency properties of two
choices of uninformative prior in section~\ref{sec:priors}, Bayesian analysis is outside the scope of our present interest.

\subsection{Nested}
\label{sec:nested}

A criterion that is discussed in the statistics literature is that of ``nesting'', or more strongly, ``strictly nesting''. A confidence interval is said to be
``nested'' if the interval for a larger confidence level includes all intervals with smaller confidence levels~\cite{Blaker2000, Thulin2017} for a given observation.
 Strict nesting means that the interval with the smaller confidence level is a proper subset of the interval with the larger
 confidence level~\cite{Thulin2017}. 
The nesting property can be expressed in terms of test acceptance regions: Confidence intervals
derived from a hypothesis test are nested iff 
\begin{equation}
A_\alpha(\theta) \supseteq A_{\alpha^\prime}(\theta),\quad\forall\theta\hbox{ and }\forall\alpha^\prime>\alpha.
\end{equation}

Crow~\cite{Crow1956}, in discussing the binomial distribution, considers lack of nesting ``interesting and theoretically permissible, but practically undesirable'', and takes steps to ensure nesting. 
Blaker~\cite{Blaker2000}, while critical of the steps taken by Crow, argues that nesting is sufficiently important to justify sacrificing optimal lengths. We will discuss nesting further, including the more recent work of Thulin and Zwanzig~\cite{Thulin2017}, in section~\ref{sec:discussion}.

\subsection{Continuity and monotonicity}
\label{sec:Continuity}

Another concern that has been raised~\cite{Thulin2017} is that with some methodologies, the interval may be a discontinuous
function of the choice of confidence level, or the $p$-value may be a discontinuous function of the null hypothesis~\cite{Vos2008}. A slight change in confidence level could lead to a large change in the interval, or a slight change in null hypothesis could lead to a large change in $p$-value. Since, for any given observation, the pdf is a continuous function of $\theta$, the confidence interval might be expected to be continuous in the confidence level. Two models which are essentially identical could result in large differences in $p$-value if continuity is violated. Thus, continuity of the interval with respect to confidence level, or of the $p$-value as a function of null hypothesis, is desirable.

We also consider it important for the upper and lower interval bounds to be monotonic functions of the confidence level, $1-\alpha$. Otherwise we could have the counterintuitive situation in which the upper bound decreases or the lower bound increases when the CL increases. It also seems desirable for the interval length to monotonically increase as $n$ increases, but we do not consider this to be as important, at least if exceptions are not too glaring.

\subsection{Gives sensible $p$-values}
\label{sec:sensiblep}

A good confidence interval will also yield sensible $p$-values. For example, the $p$-value should be a monotonic function of the observation for observations on either side of $H_0$. That  is, it should be a bimonotonic function of $n$, for any $H_0$ not at the boundary of parameter space.
This property is related to other properties we have listed.

\subsection{Discussion}

It has been shown~\cite{Blaker2000} that, for discrete distributions such as Poisson and binomial,  it is not generally possible to simultaneously 
achieve properties of continuous intervals, optimal lengths, and nesting.\footnote{Violation of nesting from inverting a hypothesis test can also happen with continuous distributions~\cite{Chernoff1951}.} We must
decide on what is most important. 
Since we usually discuss intervals in the context of a given choice of confidence level, it might be argued that nesting and continuity
do not appear as a practical matter. However, we sometimes do look at different confidence level intervals for a given observation, and then we could encounter issues. The issue also becomes crisper when we look at $p$-values. We return to this in section~\ref{sec:discussion}.

So far, we have only considered properties that do not depend on the presence of background. That is, so far, 
 we can do all the calculation as if $b=0$, obtaining a confidence interval for
$\theta+b$, and then subtract $b$ in the end to obtain a confidence interval for $\theta$. The desired probability
statement, Eq.~\ref{eq:Calpha}, holds.
Thus, if we wish, we can pre-compute a table of confidence intervals for all $N=n$, independent of $b$, 
and use it to obtain the desired confidence interval for any $b$. 

Often, we know on physical grounds that $\theta$ cannot
be negative (or cannot be less than, or perhaps not be greater than some value) even though the mathematically allowed parameter space in Eq.~\ref{eq:Poisson} extends to $\theta=-b$.
This need not be a concern until one makes statements about the true value of $\theta$, which we are not doing; we are
only describing a measurement. Nevertheless, sometimes people are uncomfortable with the idea that the description could
include ``unphysical'' values in the parameter space. Of course, if this happens for some $n$, we can just cut the interval
off to exclude the unphysical values. This is mathematically  fair, because it does not invalidate the probability statement in Eq.~\ref{eq:Calpha}.
However, we recommend against doing this truncation, because it actually makes the descriptive statement more difficult to understand, that is, it obscures the intuitive connection with the sampled $n$.
Nonetheless, physicists have invented special techniques incorporating the background in a manner that avoids dealing with
unphysical regions, and we examine two popular such approaches in sections~\ref{sec:CLs} and~\ref{sec:FC}.

\section{One-sided intervals}
\label{sec:UL}

A one-sided confidence interval is a confidence set that includes all parameter values less than some value (``upper limit'') or greater than some value (``lower limit''). Being one-sided, such intervals are generally less useful descriptions of a measurement than an interval that constrains both sides. Nevertheless, the one-sided interval provides a useful starting point for our discussion. 

The discussion of upper and lower limits is very similar; we will consider the concrete case of an upper limit here.
If we want to construct an upper limit, as our confidence interval, there really is not much ambiguity for a frequentist
limit.
For a Poisson distribution with mean $\mu=\theta + b$ we look for sets of the form 
\begin{equation}
S_\alpha(\mu) =\{n| n=n_{\rm min}, n_{\rm min}+1,\ldots\},
\end{equation}
where $n_{\rm min}$ is determined by
\begin{equation}
\hbox{Prob}(N\in S_\alpha(\mu) |\mu) \ge 1-\alpha.
\end{equation}
That is, for the Poisson distribution, we find $n_{\rm min}$ as the largest number such that
\begin{equation}
\hbox{Prob}(N\ge n_{\rm min} |\mu) = \sum_{n=n_{\rm min}}^\infty f(n;\mu)\ge 1-\alpha.
\end{equation}
Note that we have constructed the smallest such set of values of $N$ containing all values greater than some value.

Then the upper confidence limit corresponding to a given observation $N=n$ is determined by
\begin{equation}
C_\alpha(n) = \{\mu|n\in S_\alpha(\mu)\} = [0,\mu=\mu_{\rm max}],
\end{equation}
where $\mu_{\rm max}$ is determined as the largest $\mu$ such that
\begin{equation}
n \in S_\alpha(\mu).
\end{equation}
We will refer to this method as the ``standard'' upper limit. We note that $C_\alpha(n)$ has been constructed as the smallest possible one-sided set of values of $\mu$ with the desired probability statement.  

Besides being smallest, we can check whether our standard upper limit satisfies the other desirable properties we have listed as well. It is by construction an exact interval, satisfying Eq.~\ref{eq:Calpha}. It is also connected. It is not guaranteed to contain the maximum likelihood estimator. For example, if we observe a large $n$, the maximum likelihood estimator is $\hat\mu =n$, but if $1-\alpha$ is small the upper limit may be less than $n$. However, as long as $1-\alpha \ge 1/2$, the upper limit is guaranteed to include $n$, the maximum likelihood estimator, as may readily be demonstrated numerically. Since we usually concern ourselves with  $1-\alpha \ge 1/2$, this feature may be considered to be acceptable. We may demonstrate optimal coverage, i.e., Eq.~\ref{eq:optimalCoverage}. Among possible upper limits, as we have noted in the construction, it gives intervals of shortest length.
The upper limit would satisfy the asymptotic requirement if we rephrased the requirement for a one-sided interval.
Our upper limit is ordered in the sense that, if $n_2>n_1$, then the corresponding upper limits have the property $u_2>u_1$. The property of ``symmetry'' is meaningless for a one-sided interval.
Our upper limits are strictly nested, that is, if $1-\alpha_2>1-\alpha_1$ then $u_2>u_1$ for any given observation. The upper limit is a continuous function of $\alpha$. 

We find that our prescription for an upper limit does well in satisfying the applicable ``desirable'' properties, with the exception that it may not include the maximum likelihood estimator.

In doing the calculation, it is convenient to use the
connection between the Poisson distribution and the $\chi^2$ distribution. To see this connection, consider the probability statements for RVs $X\sim\chi^2_{2n}$ and $N\sim\hbox{Poisson}(\mu)$:
\begin{equation}
\begin{split}
\hbox{Prob}(X<2\mu) &=\int_0^{2\mu} \frac{x^{n-1}e^{-x/2}}{\Gamma(n)2^n}dx\\
\hbox{Prob}(N\ge n) &= 1 - \sum_{k=0}^{n-1} \frac{\mu^k e^{-\mu}}{k!}.
\end{split}
\end{equation}
Differentiating both equations with respect to $\mu$ yields the same result, and we conclude
that
\begin{equation}
\hbox{Prob}(X<2\mu) = \hbox{Prob}(N\ge n).
\end{equation}
Thus, given an observation $N=n$, an upper limit on $\mu$ at the $1-\alpha$ confidence level is given by
\begin{equation}
u(n;1-\alpha) = F^{-1}\left[1-\alpha;2(n+1)\right]/2,
\end{equation}
where $F(u;\nu)$ is the $\chi^2$ cdf for $\nu$ degrees of freedom. The corresponding result for a lower limit on $\mu$ at the $1-\alpha$ confidence level is given by
\begin{equation}
\ell(n;1-\alpha) = F^{-1}(\alpha;2n)/2.
\end{equation}

\begin{figure}
\centering
\includegraphics[width=0.8\columnwidth]{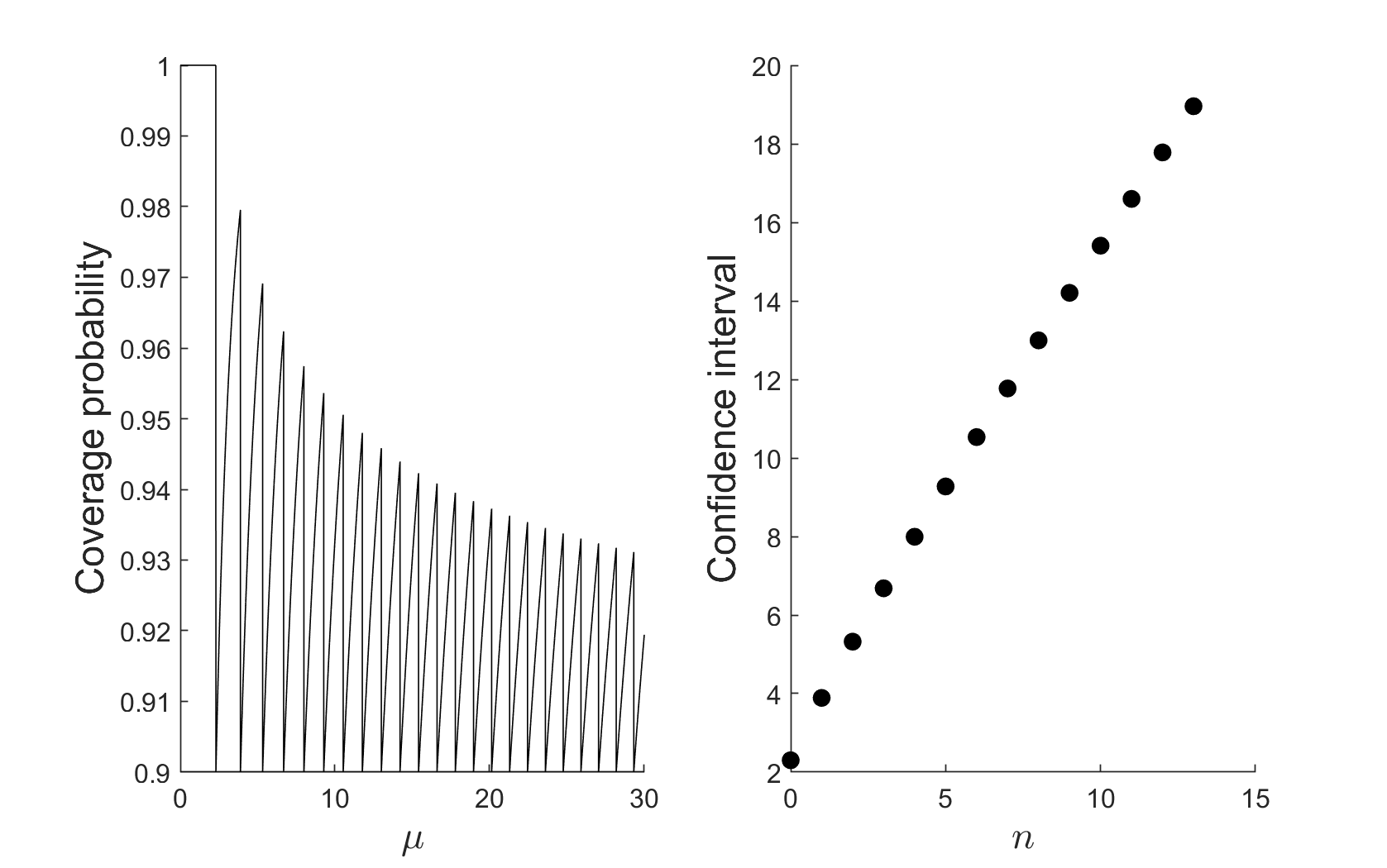}
\caption{Performance of 90\% confidence standard upper limits. Left: coverage probability as a function of $\mu$. The light horizontal line indicates $1-\alpha$. Right: upper limit as a function of observed counts.\label{fig:UL}}
\end{figure}

The upper limit and its coverage probability as a function of $\mu$ are shown in Fig.~\ref{fig:UL}, for 90\% confidence level. If $b$ is known and an upper limit on $\theta$ is desired, we simply subtract $b$ from the upper limit on $\mu$.

We note that our upper limits on $\theta$ (when background is present) could end up being negative (or less than some physical minimum for $\theta$), even if it is known that $\theta\ge 0$ (or some other value) on physical grounds. This is not an issue, the limits are performing as designed (that is, as frequency intervals). If you nonetheless find this concerning, I urge you to re-examine your goal in computing a limit. The fact that you care about consistency with physical values of $\theta$, that is with ``truth'', strongly suggests a goal of interpretation, not description. As discussed in the Introduction and section 3, ``truth'' is not the domain of frequency statistics.
If the goal is to make a statement concerning how large the true value of $\theta$ might be, this is interpretation.
For example, when designing a new experiment to search for a rare process, one starts with some idea for how large the true value might be. This informs the design of the new effort, including decisions around whether it is worth pursuing.  Resort to Bayesian methods.

In fact, frequency upper limits are not typically especially interesting as descriptions of a measurement. Our two-sided intervals are much more informative.   James and Roos~\cite{James1991} also point this out, including the observation that if ``only an upper limit is published, the result cannot be combined with those of other experiments in a straightforward way\dots'' However, the upper limit we have described provides the often useful one-sided $p$-value discussed at the end of section~\ref{sec:pvalues}.

\section{Conventional methods}
\label{sec:conventional}

We summarize and contrast here several algorithms for Poisson confidence intervals that have been developed 
in the statistics community. These are all ``exact'', in the sense that they never undercover.

For all of the methods discussed in this section, the intervals in $\mu$ and $\theta$ are simply related. If $[\ell,u]$ is a confidence interval for $\mu$ then $[\ell-b,u-b]$ is the interval for $\theta$ at the same confidence level. Thus, we need only discuss one parameter, which we will take to be $\mu$. Again, $b$ is a known quantity -- we are not thinking, for example, about a difference of two Poisson RVs.

\subsection{The ``equal-tailed'' or ``fiducial'' interval}
\label{sec:Garwood}

Putting together two one-sided intervals (upper and lower), we may obtain two-sided intervals, and this is the earliest approach to defining a two-sided confidence interval for the Poisson distribution.
The classic Poisson confidence interval is attributed to Garwood~\cite{Garwood1936}\footnote{Sometimes rediscovered in specific fields, e.g.,~\cite{Ulm1990}.}, who derived it as a fiducial interval. It is readily checked that it satisfies equation~\ref{eq:Calpha} (e.g., \cite{1996Porter}). 
This commonly used approach to confidence intervals emphasizes symmetry (in probability) about the observed counts. The idea is to combine a lower
limit with CL $1-\alpha/2$ with an upper limit with CL  $1-\alpha/2$ to obtain a central interval with
CL $1-\alpha$.

As a fiducial interval, we start with Eq.~\ref{eq:fiducial} and find, for given $N=n$, the confidence interval $[\ell(n),u(n)]$, where
\begin{equation}
\begin{split}
\hbox{Prob}[N\ge n | \mu = \ell(n)]  &=  \sum_{k=n}^\infty \frac{\ell^k}{k!}e^{-\ell} = \alpha/2\\
 \hbox{Prob}[N\le n | \mu = u(n)]  &= \sum_{k=0}^n \frac{u^k}{k!}e^{-u} = \alpha/2.
 \end{split}
 \end{equation}
If $n=0$, we set $\ell=0$.
Following the discussion in Sec.~\ref{sec:UL}, this may be computed according to:
\begin{equation}
\left[\ell(n),u(n)\right] =\left[F^{-1}(\alpha/2;2n)/2, F^{-1}(1-\alpha/2;2(n+1))/2\right],
\end{equation}
where $F(x;\nu)$ is the $\chi^2$ cdf for $\nu$ degrees of freedom. 

\begin{figure}
\centering
\includegraphics[width=0.9\columnwidth]{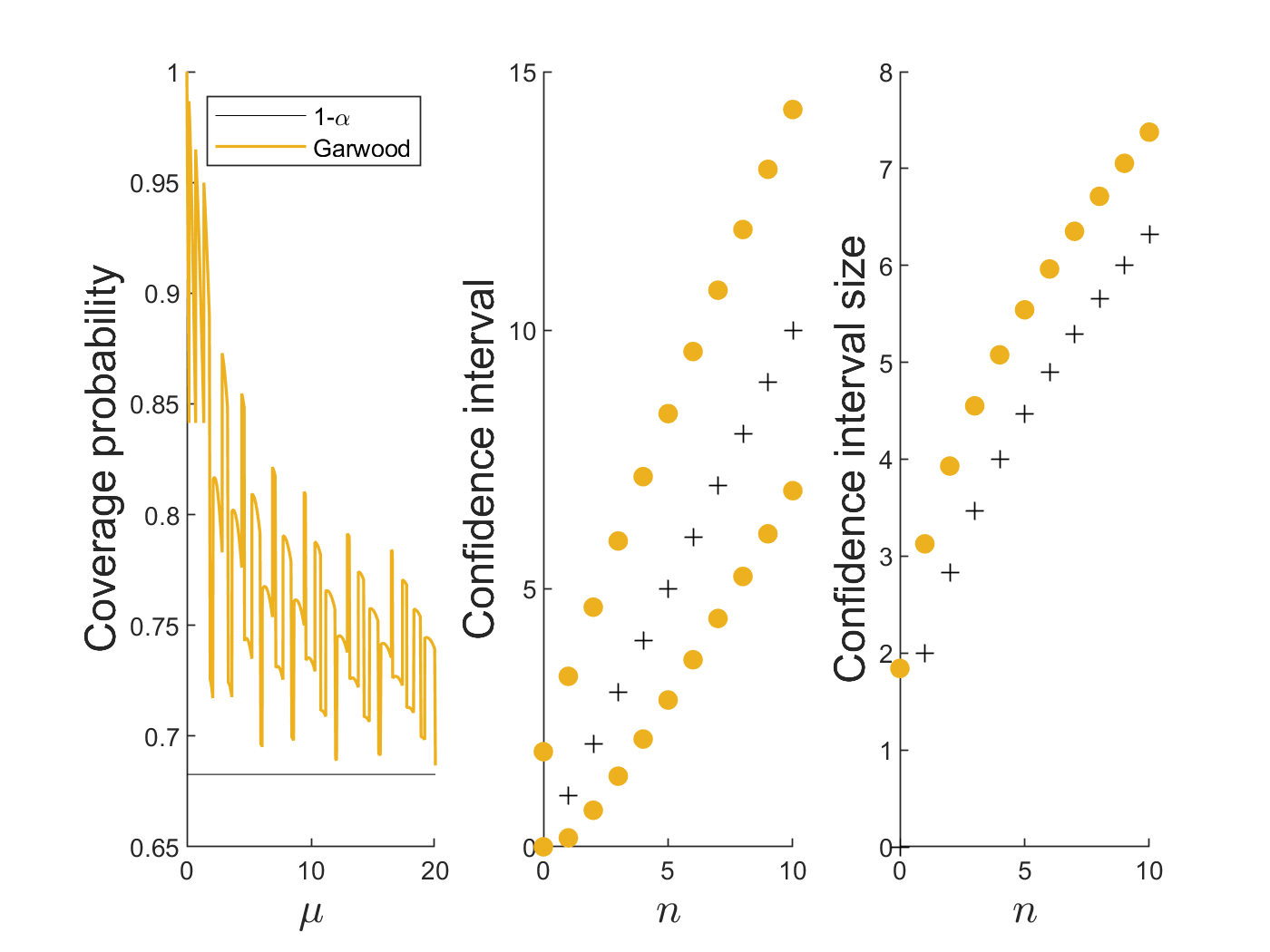}
\caption{Performance of Garwood intervals. Left: coverage probability as a function of $\mu$; Middle: 68\% confidence interval as a function of observed counts. The plus symbols show the maximum likelihood estimator for the mean;  Right: size of the 68\% confidence interval as a function of observed counts. The plus symbols show the values of $2\sqrt{n}$.\label{fig:Garwood}}
\end{figure}

The Garwood intervals are the intervals returned by MATLAB for Poisson sampling.\footnote{Determined by comparing with the output of the MATLAB version R2019b {\tt poissfit} function, which calls {\tt statpoisci}.}
The {\tt tsmethod="central"} option in $R$'s {\tt poisson.exact} function also returns the Garwood intervals~\cite{Rpoissonexact2020}. In particle physics, the Garwood interval is presented in the Particle Data Group~\cite{PDG2024} review.

 The Garwood interval is exact, connected, strictly nested, ordered, and continuous in the confidence level.  
 It is symmetric in probability by construction. It also has the desired asymptotic property and leads to bimonotonic $p$-values.
Thus, it meets many of our desired criteria.

However, a common criticism of the symmetric interval approach is that it yields intervals that considerably
overcover. This is largely due to the fact that it corresponds to the inversion of two one-sided tests. Statisticians have devoted a significant effort towards improving coverage, and the related property of interval length. 
A modification to the fiducial interval procedure known as the ``Mid-P'' method is sometimes advocated~\cite{Kulkarni1998,Hirji2006}. However, this method allows the intervals to undercover for some parameter values, and we do not pursue it here.

It is possible to reduce the overcoverage by relaxing the requirement that each tail have probability $\ge 1-\alpha/2$. That is, we can directly invert a two-sided test, referred to in~\cite{Thulin2017} as ``strictly two-sided intervals''.
Thus, several two-sided intervals (discussed below) with better coverage properties have been devised, abandoning the symmetry property. On the other hand, Thulin and Zwanzig~\cite{Thulin2017} remark that the Garwood interval 
is optimal (i.e., shortest) among strictly nested intervals.

\subsection{Smallest size $S_\alpha(\mu)$ by ordering on probability }
\label{sec:SterneCrowGardner}

Often, we try to make $S_\alpha(\mu)$ as small as possible, where ``small'' may need definition. 
For example, ``small'' could be taken in the sense of the size of the acceptance region for a given $\mu$.\footnote{This is not the same as minimizing overcoverage.}. This is accomplished by letting $S_\alpha(\mu)$ be the smallest set such that
\begin{equation}
\label{eq:probSA}
\hbox{Prob}(N\in S_\alpha(\mu)) = \sum_{n\in S_\alpha(\mu)}  f(n;\theta,b) \ge 1-\alpha,
\end{equation}
where $\mu=\theta +b$ and
\begin{equation}
\label{eq:condSA}
n\in S_\alpha(\mu)\ \forall n \quad\hbox{s.t.}\quad  f(n;\theta,b) \ge \min_{n\in S_\alpha(\mu)}  f(n;\theta,b).
\end{equation}
It may happen for special values of $\mu$ that 
there are two values of $n$ satisfying equality in Eq.~\ref{eq:condSA}, where only one is needed to
satisfy the probability statement in Eq.~\ref{eq:probSA}. If this occurs, we here accept the lower value of $n$.

The idea of Eqs.~\ref{eq:probSA} and \ref{eq:condSA} is that, for given $\mu$, we find those values of $n$ (there are either one or two such values) for which the pdf is maximal. These values are added to $S_\alpha(\mu)$. Then we look for the next highest values of $f$ and add the corresponding values of $n$ to $S_\alpha(\mu)$. This is 
continued until we get a set with probability at least $1-\alpha$.

The sets $C_\alpha(N)$ are then constructed as the sets of values of $\mu$ such that, for any given $N=n$, $n\in S_\alpha(\mu)$. With this construction, Eq.~\ref{eq:Calpha} is satisfied. 

We now explicitly construct sets $S_\alpha(\mu)$. Because $f$ is monotonically decreasing on either side of the peak, $S_\alpha(\mu)$ is a set of adjacent values of $n$. Thus, we look for the lowest 
element of the set, $n_\ell$ and the highest element, $n_h$ to obtain
\begin{equation}
S_\alpha(\mu) = \{n_\ell(\mu),n_\ell(\mu)+1,\ldots,n_h(\mu)\}.
\end{equation}

Given a sampled value, $n$, we look through sets $S_\alpha(\mu)$ for those which 
contain $n$. Because of the monotonicity of $n_\ell$ and $n_h$ with respect to $\mu$,
the sets that contain $n$ will include a continuous range of $\mu$. Thus, our confidence set
is an interval, $C_\alpha(n) = [\ell,u]$, where $n_\ell = \max_n S_\alpha(\ell)$ and
$n_h = \min_n S_\alpha(u)$. The analog of this method was proposed by Sterne~\cite{Sterne1954}
for the binomial distribution, and the method has been discussed for the Poisson distribution by Crow and Gardner~\cite{Crow1959}. We refer to this interval as ``Sterne''.

\begin{figure}
\centering
\includegraphics[width=0.9\columnwidth]{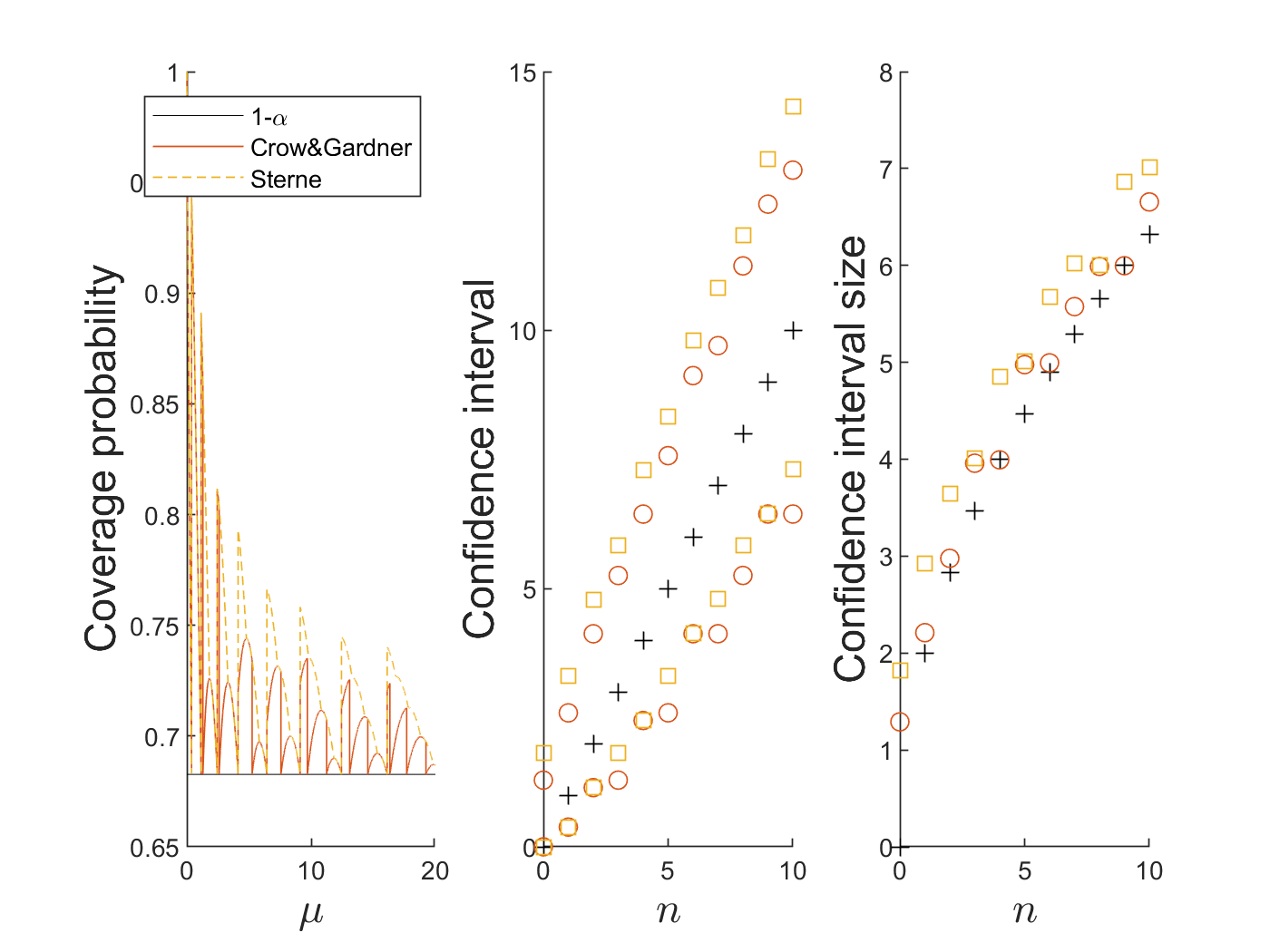}
\caption{Performance of Sterne and Crow\&Gardner 68\% confidence intervals. Left: coverage probability as a function of $\mu$; Middle: confidence interval boundaries as a function of observed counts. Circles are for Crow\&Gardner, squares for Sterne. The plus symbols show the maximum likelihood estimator for the mean; Right: size of the confidence interval as a function of observed counts. Circles are for Crow\&Gardner, squares for Sterne.  The plus symbols show the values of $2\sqrt{n}$.\label{fig:SterneCrow}}
\end{figure}

\begin{figure}
\centering
\includegraphics[width=0.9\columnwidth]{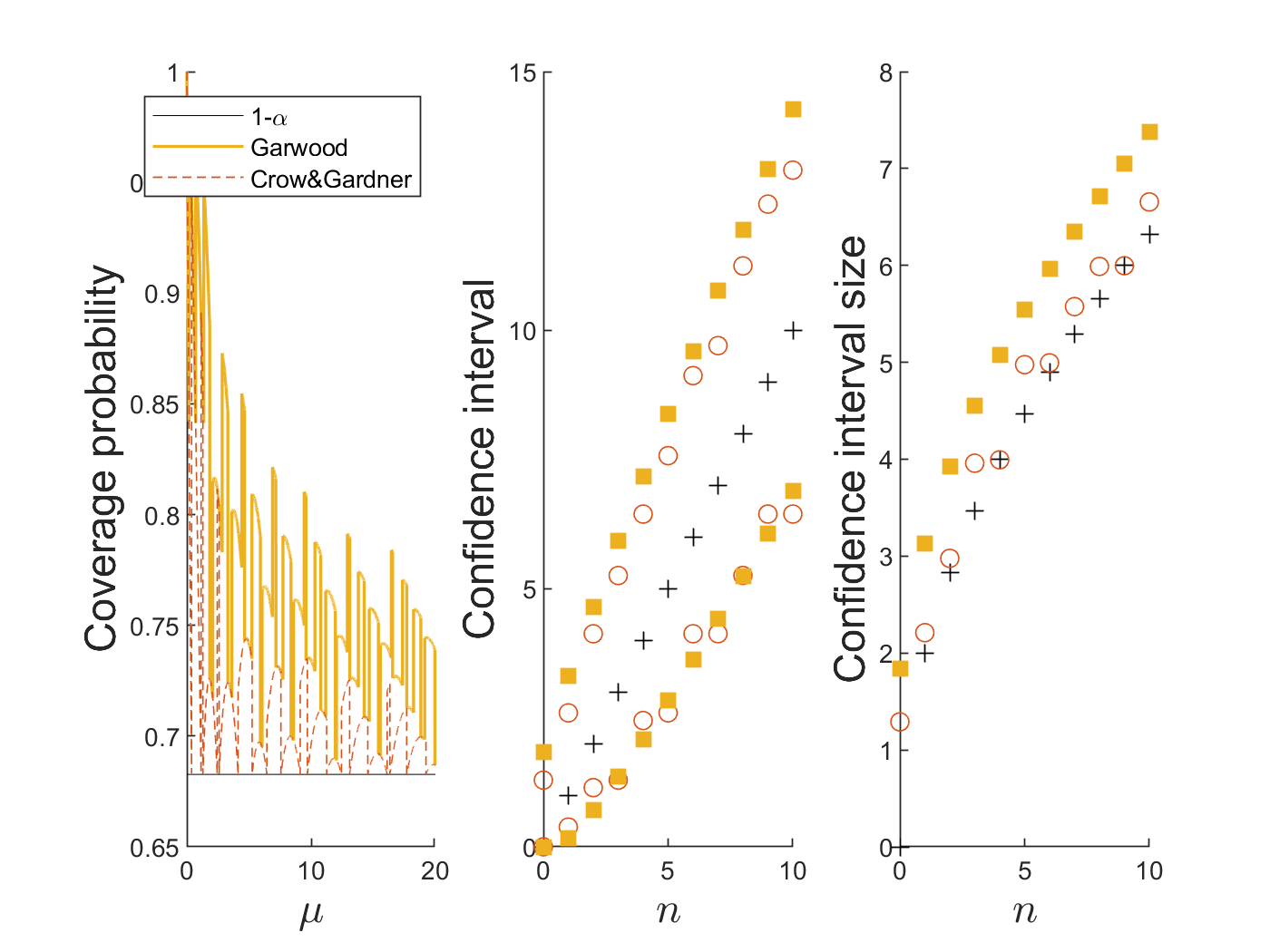}
\caption{Performance of Garwood and Crow\&Gardner 68\% confidence intervals. Left: coverage probability as a function of $\mu$; Middle: confidence interval boundaries as a function of observed counts. Circles are for Crow\&Gardner, filled squares for Garwood. The plus symbols show the maximum likelihood estimator for the mean; Right: size of the confidence interval as a function of observed counts. Circles are for Crow\&Gardner, filled squares for Garwood. The plus symbols show the values of $2\sqrt{n}$.\label{fig:GarwoodCrow}}
\end{figure}

Crow and Gardner~\cite{Crow1959} proposed a variation on this idea that preserves the minimum size of $S_\alpha(\mu)$ but also minimizes the upper bound on the confidence interval among all such smallest sets instead of strictly ordering on probability. We refer to this interval as ``Crow\&Gardner''. 
It may be noted that this modification destroys the correspondence of the Crow\&Gardner intervals with a test-inversion. Instead, effectively a different test is defined for each $\alpha$. This suggests that deriving $p$-values from these intervals may be ill-defined. We will discuss this further in section~\ref{sec:discussion}.

The properties for the Sterne and Crow\&Gardner intervals are compared in Fig.~\ref{fig:SterneCrow}.
The Crow\&Gardner intervals overcover less than the Sterne intervals. This is because we have allowed ourselves to include less probability in our construction of $S_\alpha(\theta)$ by relaxing strict ordering on probability. They are also more symmetric in interval than the Sterne intervals. The Crow\&Gardner intervals also are shorter (and more similar to $\sqrt{n}$) than the Sterne intervals.
On all of these properties, the Crow\&Gardner intervals are preferred. However, with the modification introduced in Crow\&Gardner, the intervals do not necessarily include $n$. This exclusion of $n$ occurs at small confidence levels.

\begin{figure}
\centering
\includegraphics[width=0.8\columnwidth]{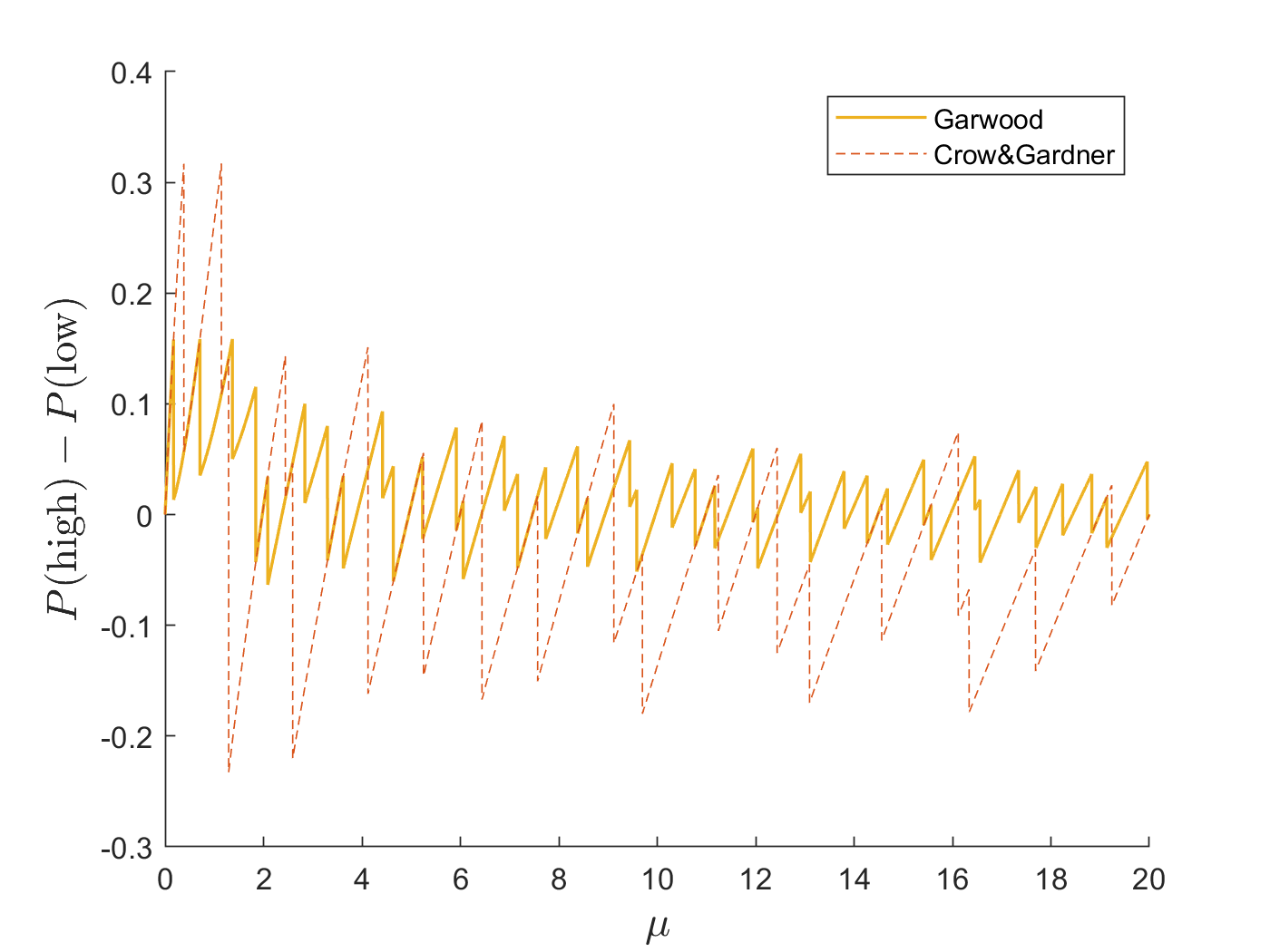}
\caption{Comparison of the symmetry of the Garwood and Crow\&Gardner 68\% confidence intervals. The difference between the probability of an observation in the high side tail and the probability of an observation in the low side tail is plotted as a function of $\mu$.\label{fig:GarwoodCrowSymmetry}}
\end{figure}

In fact, the same observations again hold in a comparison of the Crow\&Gardner intervals with the Garwood intervals, that is, Crow\&Gardner intervals are preferred based on size and coverage, Fig.~\ref{fig:GarwoodCrow}. However, the Garwood intervals are constructed to achieve symmetry in probability, and we may look at this as well. Figure~\ref{fig:GarwoodCrowSymmetry} shows a comparison between the Garwood and Crow\&Gardner intervals. For a given value of $\mu$ we plot the probability to obtain an observation in the tail on the high side of the acceptance region, minus the probability to obtain an observation in the tail on the low side of the  acceptance region. This measure of symmetry is centered more closely about zero for the Garwood than for the Crow\&Gardner intervals.

Schilling and Holladay~\cite{Schilling2017} find, with their alternative to the Casella and Robert length criterion, (section~\ref{sec:length}) that the Crow\&Gardner intervals~\cite{Crow1959} are optimal on this criterion, but also have endpoints that are not strictly increasing in $n$. They argue for a modified procedure, designated ``Modified Crow-Gardner (MCG)'', to obtain strictly increasing endpoints while keeping small the resulting violations of length optimality: 
Start with limits $\ell_{\rm MCG}(n)=\ell_{\rm CG}(n)$ for all $n$, where ``CG'' indicates the Crow-Gardner intervals. Starting with $n=0$, whenever $\ell_{\rm MCG}(n+1) < ml(n) \equiv \ell_{\rm MCG}(n) + \min\left[0.01\ell_{\rm MCG}(n), 0.1)\right]$, make the increase $\ell_{\rm MCG}(n+1) = ml(n)$. The corresponding coincidental endpoint, in which $u_{\rm CG}(m) = \ell_{\rm CG}(n+1)$ for some $m$, is simultaneously increased.

\subsection{Blaker}

Blaker~\cite{Blaker2000} (along with Thulin and Zwanzig~\cite{Thulin2017}) criticizes the Crow\&Gardner intervals as not
guaranteed to be nested, and proposes a different approach, based on an ``acceptability function'', defined in the theorem below. We state the theorem providing the algorithm here. The theorem quoted by Blaker is more generally
for an arbitrary statistic $T(N)$, but we will use for the Poisson case (as does Blaker) the sufficient choice $T=N$, which we restrict to here:

\begin{thm} (Blaker)
Define function $\gamma(\mu,n)\equiv \min\left[\hbox{Prob}(N\ge n|\mu), \hbox{Prob}(N\le n|\mu)\right]$ and ``acceptability function'' $\alpha(\mu; n)\equiv \hbox{Prob}\left[\gamma(\mu,N)\le \gamma(\mu,n)|\mu\right]$.
Then
\begin{enumerate}
\item Set $C_\alpha(n) \equiv \{\theta: \alpha(\mu;n)>\alpha\}$ is a $1-\alpha$ confidence set for $\mu$.
\item $\alpha(\mu; n)$ may be computed according to whichever of the following three cases applies:
\begin{enumerate}
\item If $\hbox{Prob}(N\ge n|\mu) < \hbox{Prob}(N\le n|\mu)$, then
\begin{equation}
\alpha(\mu; n) = \hbox{Prob}(N\ge n|\mu) + \hbox{Prob}(N\le n^*|\mu),
\end{equation}
where $n^*$ is the largest $m$ such that $\hbox{Prob}(N\le m|\mu) \le \hbox{Prob}(N\ge n|\mu)$
\item If $\hbox{Prob}(N\ge n|\mu) = \hbox{Prob}(N\le n|\mu)$, then
\begin{equation}
\alpha(\mu; n) = 1,
\end{equation}
\item If $\hbox{Prob}(N\ge n|\mu) > \hbox{Prob}(N\le n|\mu)$, then
\begin{equation}
\alpha(\mu; n) = \hbox{Prob}(N\le n|\mu) + \hbox{Prob}(N\ge n^{**}|\mu),
\end{equation}
where $n^{**}$ is the smallest $k$ such that $\hbox{Prob}(N\ge k|\mu) \le \hbox{Prob}(N\le n|\mu)$.
\end{enumerate}
\end{enumerate}
\end{thm}
We note that then we also have $S_\alpha(\mu) = \{n: \alpha(\mu,n)>\alpha\}$.

\begin{figure}
\centering
\includegraphics[width=0.85\columnwidth]{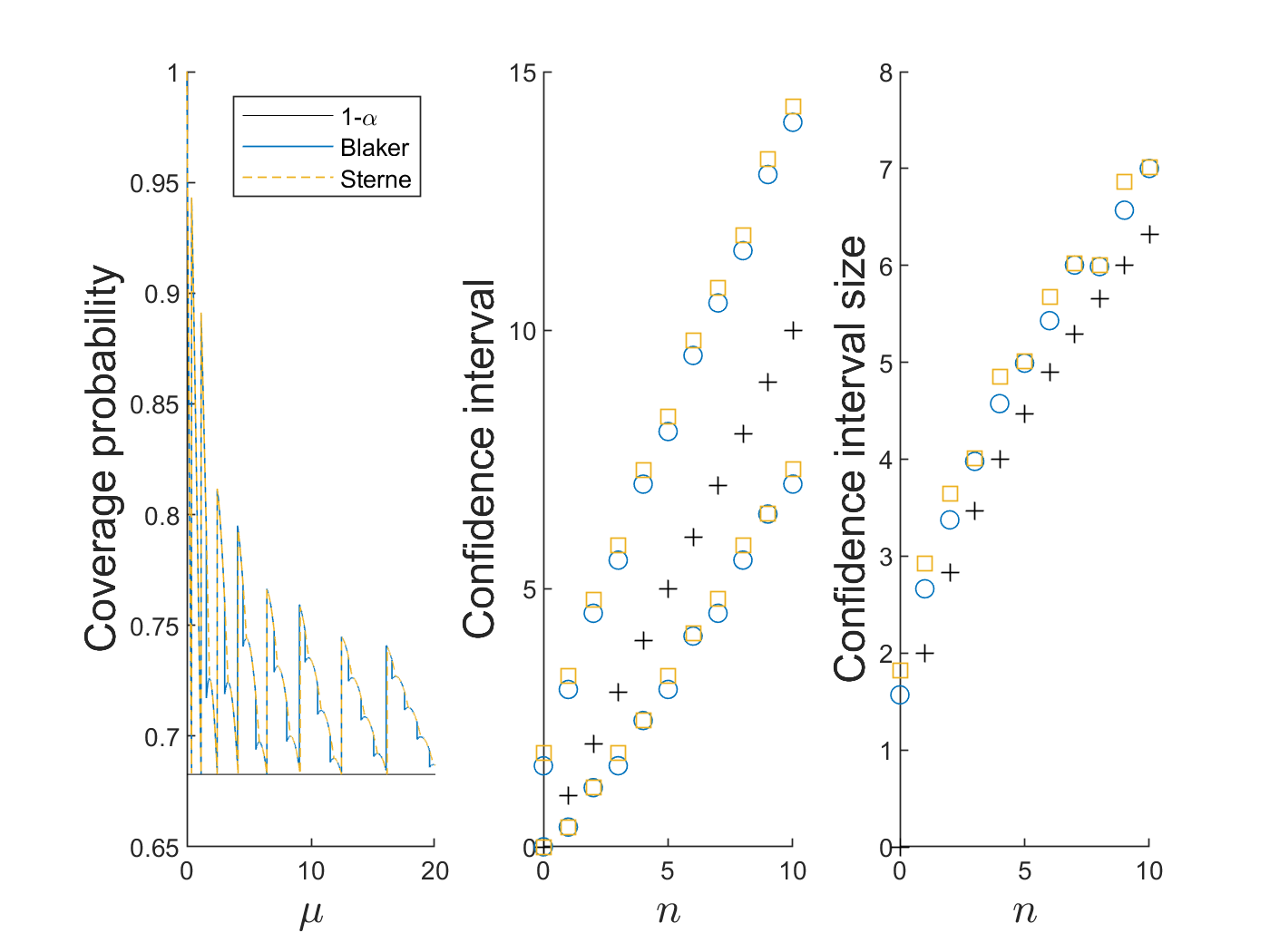}
\caption{Performance of Blaker and Sterne 68\% confidence intervals. Left: coverage probability as a function of $\theta$; Middle: confidence interval boundaries as a function of observed counts. Circles are for Blaker, squares for Sterne. The plus symbols show the maximum likelihood estimator for the mean; Right: size of the confidence interval as a function of observed counts. Circles are for Blaker, squares for Sterne. The plus symbols show the values of $2\sqrt{n}$.\label{fig:SternBlaker}}
\end{figure}

\begin{figure}
\centering
\includegraphics[width=0.8\columnwidth]{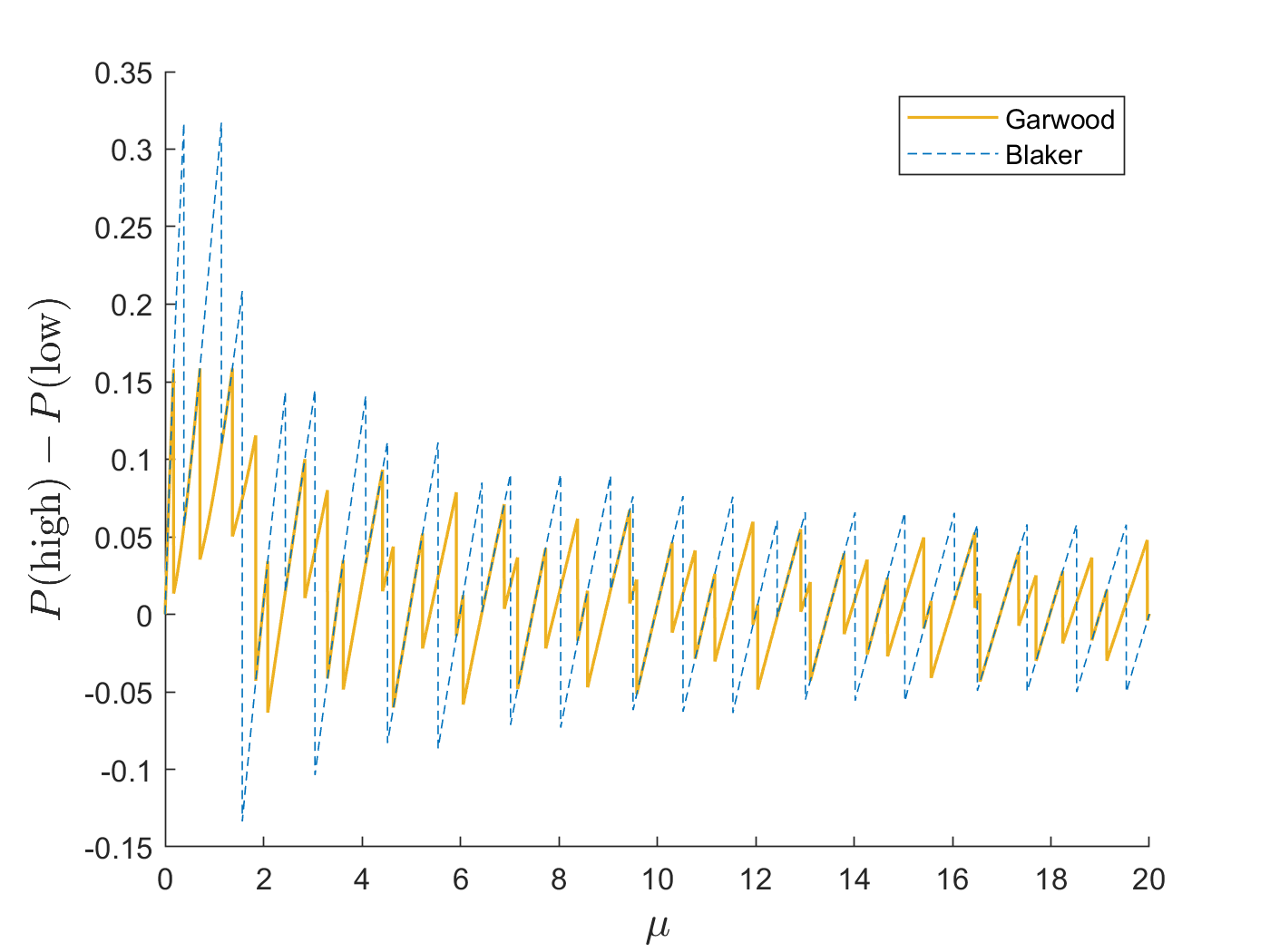}
\caption{Comparison of the symmetry of the Garwood and Blaker 68\% confidence intervals. The difference between the probability of an observation in the high side tail and the probability of an observation in the low side tail is plotted as a function of $\mu$.\label{fig:GarwoodBlakerSymmetry}}
\end{figure}

The Blaker algorithm results in intervals that are similar to the Sterne intervals, Fig.~\ref{fig:SternBlaker}. The Blaker intervals have slightly better coverage and are slightly more symmetric in interval. However, on these properties, the Crow\&Gardner intervals are even better (see also Fig.~\ref{fig:SterneCrow}). On the other hand, the Blaker intervals are much closer to the Garwood intervals in symmetry (compare Fig.~\ref{fig:GarwoodBlakerSymmetry} with Fig.~\ref{fig:GarwoodCrowSymmetry}).

\begin{figure}
\centering
\subfloat[\centering Blaker\label{fig:BlakerNesting}]{{\includegraphics[width=0.5\columnwidth]{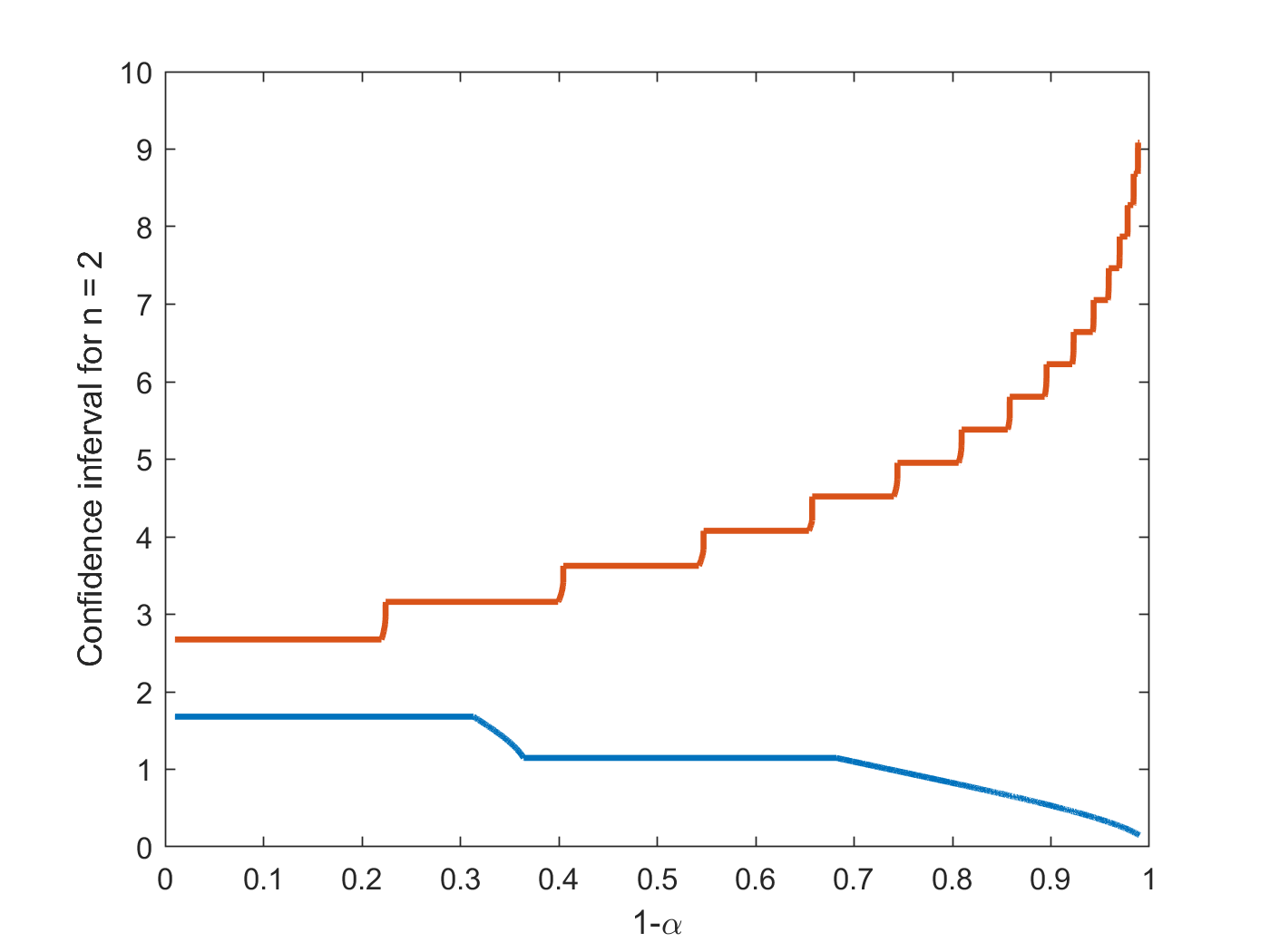}}}
\subfloat[\centering Crow\&Gardner\label{fig:CGNesting}]{{\includegraphics[width=0.5\columnwidth]{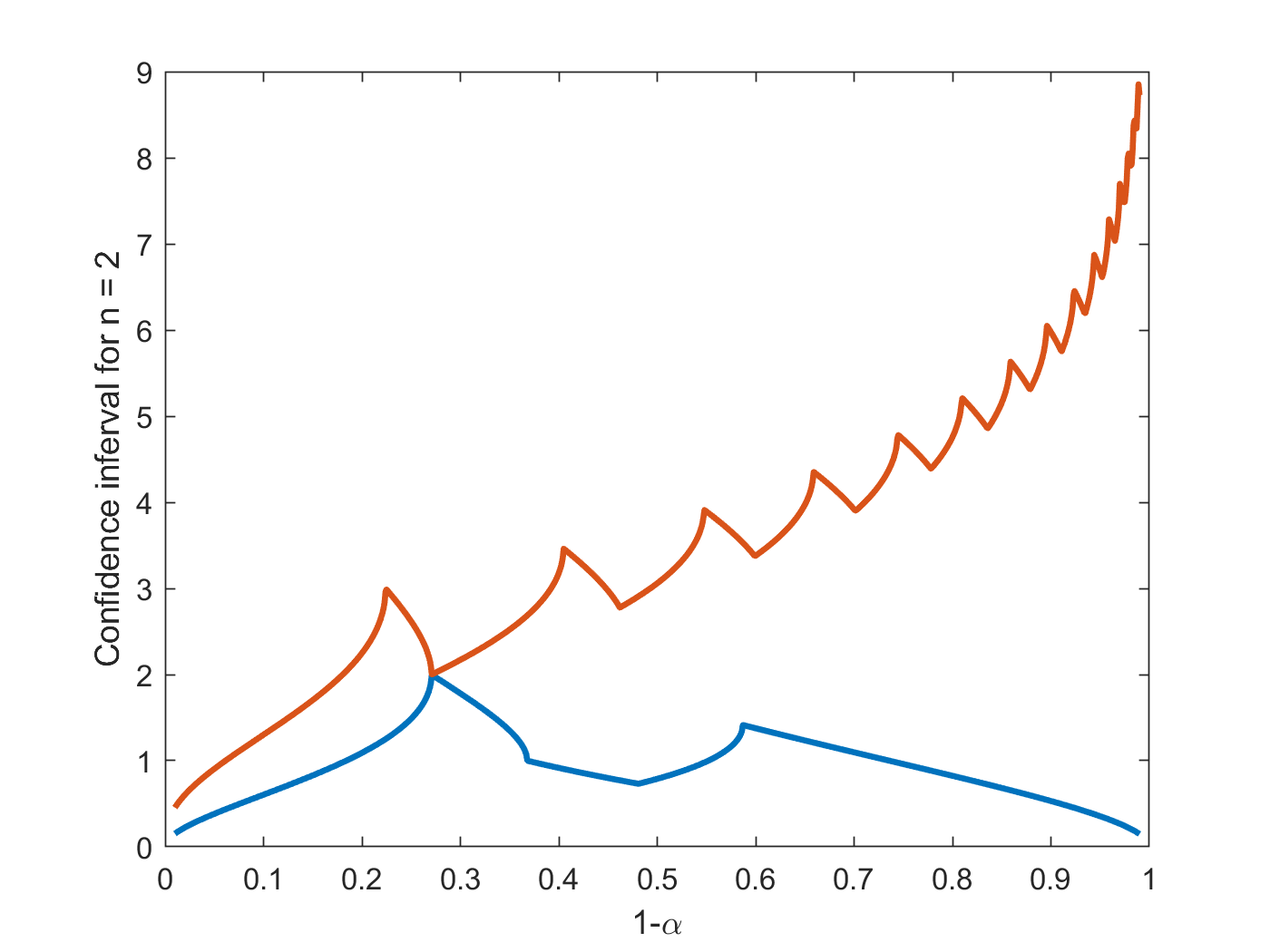}}}
\caption{Confidence intervals as a function of confidence level $1-\alpha$, for observation $n=2$. The curves show the upper and lower bounds in the intervals.}
\label{fig:Nesting}
\end{figure}

When we look at continuity and nestedness, we find that the Blaker intervals are nested, though not strictly, as a function of confidence level, but exhibit discontinuous behavior, Fig.~\ref{fig:BlakerNesting}. The Crow\&Gardner intervals are continuous, but neither nested nor monotonic with CL, Fig.~\ref{fig:CGNesting}. Comparing these figures, we may decide that sacrificing nesting to obtain continuity is not a good trade-off if it comes at the expense of monotonicity.

\subsection{Variations}

We could use a different definition of ``small'' for our sets $S_\alpha(\mu)$, now in the sense of the 
overcoverage probability. However, this approach can lead to very asymmetric results without some additional
constraints. We could combine the idea with the minimal number of observations of section~\ref{sec:SterneCrowGardner}. We look for the smallest number of contiguous observations satisfying the probability content,
selecting the end points that minimize the overcoverage. That is, we relax the strict ordering on probability of Sterne
but also do not require the minimal upper bound on the confidence interval property of Crow\&Gardner. There are many
other variations of these general ideas we could try. Indeed, it is possible to improve coverage, for example by working farther into the tails of the distribution. However, the intervals then tend to be large, and less symmetric in interval, to the point where such results do not seem attractive.

\subsection{Kabaila-Byrne intervals }
\label{sec:KabailaByrne}

\begin{figure}
\centering
\includegraphics[width=1.1\columnwidth]{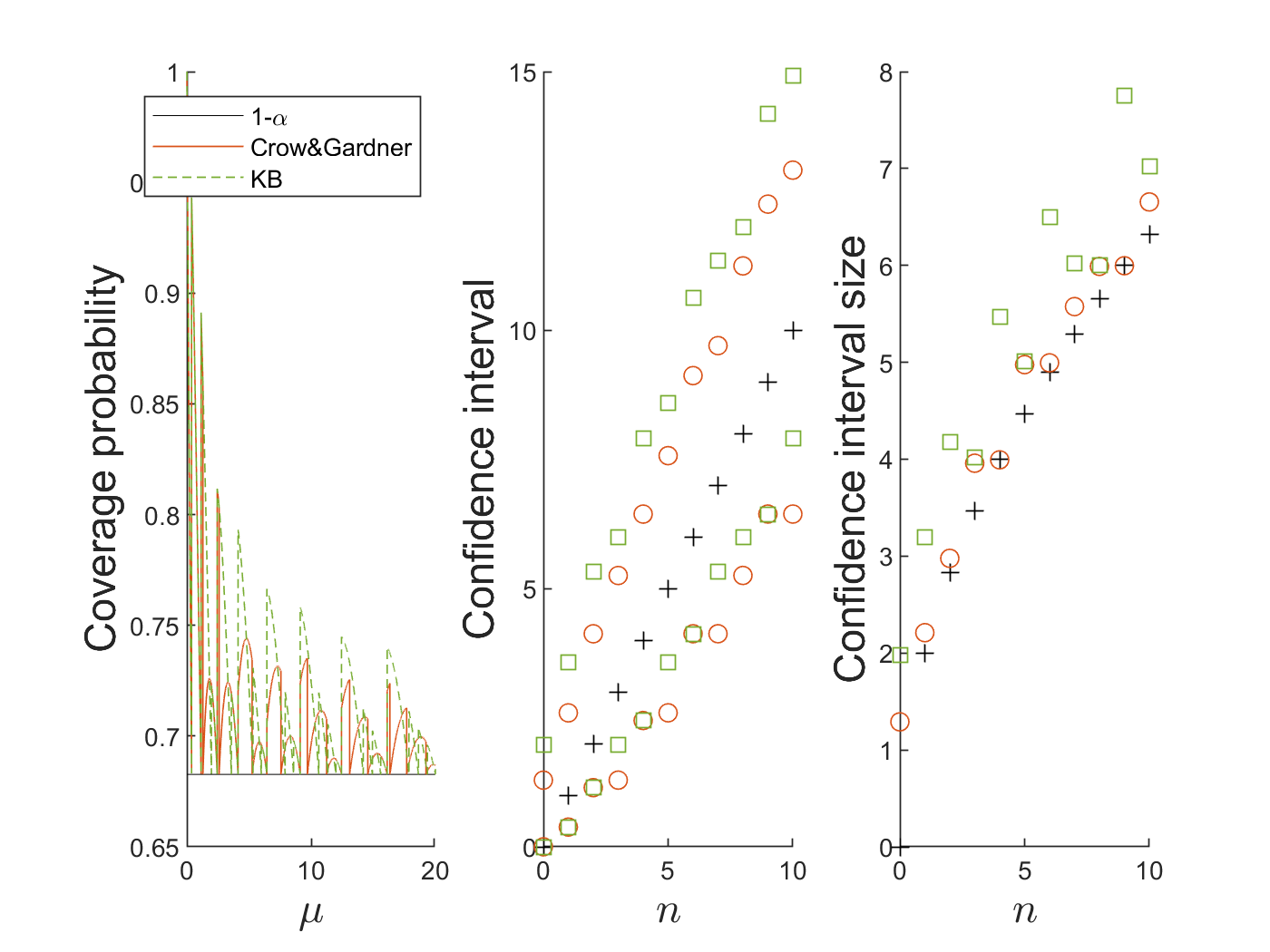}
\caption{Performance of Kabaila\&Byrne and Crow\&Gardner 68\% confidence intervals. Left: coverage probability as a function of $\mu$; Middle: confidence interval boundaries as a function of observed counts. Circles are for Crow\&Gardner, squares for Kabaila\&Byrne. The plus symbols show the maximum likelihood estimator for the mean; Right: size of the confidence interval as a function of observed counts. Circles are for Crow\&Gardner, squares for Kabaila\&Byrne.  The plus symbols show the values of $2\sqrt{n}$.\label{fig:KBCrow}}
\end{figure}

Kabaila and Byrne~\cite{Kabaila2001} propose a confidence interval that is ``shortest'' in the sense that any further shortening would cause the interval to undercover for some values of $\mu$. More completely, their interval is designed to satisfy the criteria of optimal coverage, shortest length, and strict ordering. The reader is referred to~\cite{Kabaila2001} for the algorithm, which we have implemented in MATLAB. Fig.~\ref{fig:KBCrow} shows a comparison of Kabaila\&Byrne and Crow\&Gardner confidence intervals. They generally track, but the Kabaila\&Byrne intervals tend to be shifted higher and are longer intervals than Crow\&Gardner. However, the  Kabaila\&Byrne intervals are constructed requiring strict ordering, which is not satisfied by 
Crow\&Gardner intervals~\cite{Kabaila2001,Schilling2017}.

The Kabaila\&Byrne interval, similarly to what we mentioned earlier for Crow\&Gardner, is not derivable from inverting a test statistic. Instead, it corresponds to a family of different test statistics for each value of $\alpha$~\cite{Thulin2017}.

\subsection{Inverting a likelihood ratio test}
\label{sec:LR}

The likelihood ratio is a ubiquitous concept in analysis of data in the physical sciences. It is known that a hypothesis test for a simple test
based on the likelihood ratio is uniformly most powerful. We often employ the likelihood ratio statistic in other situations with the hope that
it will continue to perform well.
Thus, consider the hypothesis test:
\begin{equation}
\begin{split}
\hbox{$H_0$:\ } &\mu=\mu_0\\
\hbox{against}\\
\hbox{$H_1$:\ } &\mu\ne \mu_0.
\end{split}
\label{eq:LRhypotheses}
\end{equation}
We are going to take the likelihood ratio between the two hypotheses as our test statistic.
Because the alternative is not completely specified, we use the maximum likelihood estimator
for $\mu$ in $H_1$.

Let $L(\mu;n) = f(n;\mu)$ be the likelihood function for a given sample $N=n$, and possible parameter value $\mu$.
As a function of $\mu$, this likelihood is maximal at $\mu=\mu_{\rm ML} = n$.
We form the likelihood ratio according to
\begin{equation}
\lambda(\mu_0;n)\equiv\frac{L(\mu_0;n)}{L(\mu_{\rm ML};n)}.
\label{eq:Lratio}
\end{equation}
With this definition, $0\le\lambda(\mu_0;n)\le1$. 
For the Poisson distribution in particular, the likelihood ratio statistic corresponding to observation $N=n$ is
\begin{equation}
\lambda(\mu_0;n) = \left(\frac{\mu_0}{n}\right)^ne^{n-\mu_0}.
\end{equation}

Now we build the acceptance region for the test. If the likelihood ratio is large, the evidence favors
$H_0$. Thus, we will include in the acceptance region those values of $N=n$ for which the likelihood
ratio is large. We will stop adding values of $n$ when we have reached the desired probability.
That is, we have acceptance region
\begin{equation}
A_\alpha(\mu_0) = \{n|\lambda(\mu_0;n)\ge\lambda_{\alpha;\rm crit}(\mu_0)\},
\end{equation}
where $\lambda_{\alpha;\rm crit}(\mu_0)$ is defined as the largest ratio such that
\begin{equation}
\label{eq:LRacceptanceProbability}
\hbox{Prob}(N\in A_\alpha(\mu_0) |\mu_0) \ge 1-\alpha.
\end{equation}
We can perform this procedure for any desired value of $\mu=\mu_0$.
We could check whether this procedure results in sets of contiguous values of $n$.
However, we will adopt an algorithm that guarantees it, at the cost of possibly not
strictly ordering on the ratio for some values of $\mu$.

To implement this procedure, we build acceptance regions as a function of $\mu$, that is,
different choices for $\mu_0$. For each $\mu$ we start by adding that value of $n$ to 
$A_\alpha(\mu)$
for which the likelihood ratio is maximal. Then we work our way out in $n$ values to find the
next largest likelihood ratio, adding the corresponding $n$ value, until we reach that value of $n$
for which the probability statement in Eq.~\ref{eq:LRacceptanceProbability} is first satisfied.
Our algorithm will search only contiguous values of $n$, giving a set with no gaps in $n$ values.

\begin{figure}
\centering
\includegraphics[width=0.95\columnwidth]{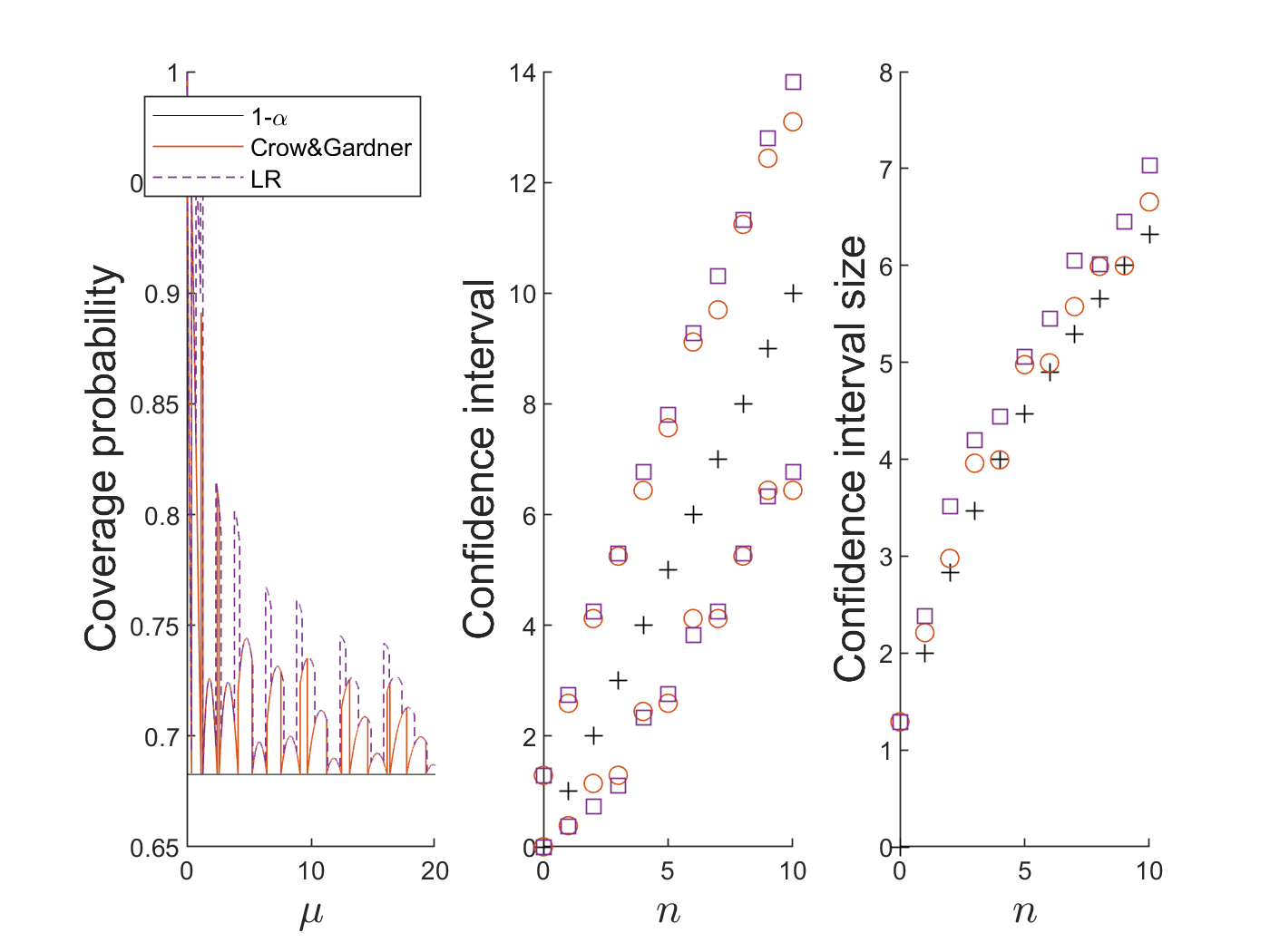}
\caption{Performance of LR and Crow\&Gardner 68\% confidence intervals. Left: coverage probability as a function of $\mu$; Middle: confidence interval boundaries as a function of observed counts. Circles are for Crow\&Gardner, squares for LR. The plus symbols show the maximum likelihood estimator for the mean; Right: size of the confidence interval as a function of observed counts. Circles are for Crow\&Gardner, squares for LR.  The plus symbols show the values of $2\sqrt{n}$.\label{fig:LRCrow}}
\end{figure}

Because this method is based on the likelihood ratio, we denote it as the ``likelihood ratio'' (LR) method. Fig.~\ref{fig:LRCrow} compares the properties of the LR method with Crow\&Gardner. We see that the coverage for LR intervals is often somewhat larger than for Crow\&Gardner, and the intervals are also somewhat longer.

\subsection{Inverting a score test}
\label{sec:score}

The score function is another important concept in statistical analysis. We can devise a test based on the score, again for the hypothesis:
\begin{equation}
\begin{split}
\hbox{$H_0$:\ } &\mu=\mu_0\\
\hbox{against}\\
\hbox{$H_1$:\ } &\mu\ne \mu_0.
\end{split}
\label{eq:scoreHypotheses}
\end{equation}

We define our test statistic as follows: In general, the score function for a parameter $\theta$ and observation $x$ is
\begin{equation}
U(\theta)\equiv \frac{\partial \log L(\theta|x)}{\partial\theta},
\end{equation}
providing a measure for how rapidly the likelihood changes with the parameter for a given observation.
The (Fisher) information number for a doubly-differentiable distribution (pdf) $f(X|\theta)$ may be written~\cite{Shao2003}:
\begin{equation}
I(\theta)\equiv -E\left[\partial_\theta^2\log f(X|\theta) \right].
\end{equation}
The statistic for the score test is defined as (e.g.,~\cite{Narsky2014}):
\begin{equation}
T(\theta_0;X) =\frac{\left[U(\theta_0,X)\right]^2}{I(\theta_0)}.
\end{equation}

For the Poisson distribution with mean $\mu$ in particular, the score statistic corresponding to observation $N=n$ is
\begin{equation}
T(\mu_0;n) = \mu_0\left(\frac{n}{\mu_0} - 1\right)^2.
\end{equation}
Note the squared-deviation from the null hypothesis nature of this statistic.
We proceed as before to define an acceptance region for $\mu_0$ of possible observations:
\begin{equation}
A_\alpha(\theta_0) = \{n|T(\mu_0;n)\le T_{\alpha;\rm crit}(\mu_0)\},
\end{equation}
where $T_{\alpha;\rm crit}(\mu_0)$ is defined as the smallest $T$ such that
\begin{equation}
\label{eq:scoreAcceptanceProbability}
\hbox{Prob}(N\in A_\alpha(\mu_0) |\mu_0) \ge 1-\alpha.
\end{equation}

\begin{figure}
\centering
\includegraphics[width=0.9\columnwidth]{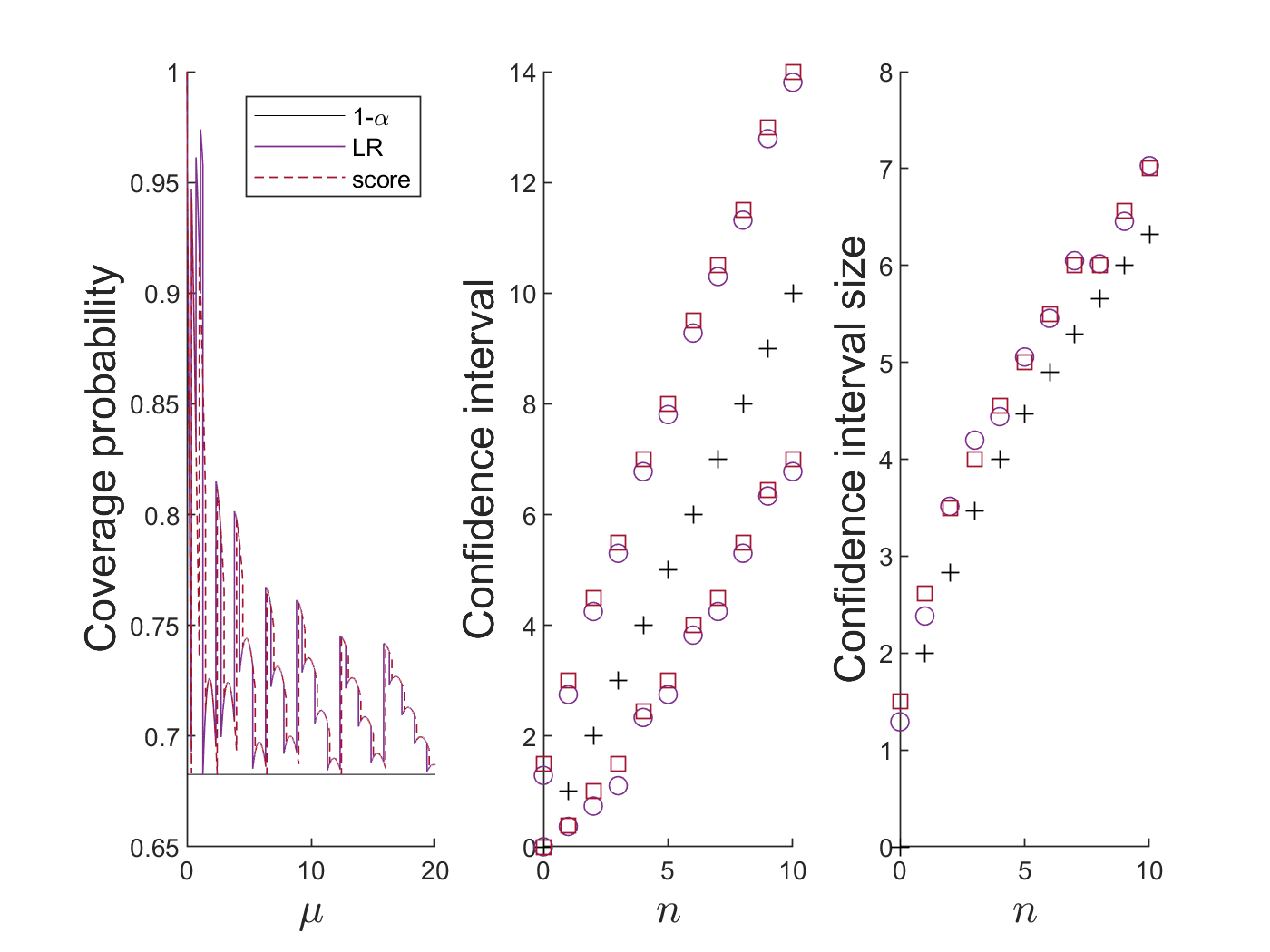}
\caption{Performance of LR and score 68\% confidence intervals. Left: coverage probability as a function of $\mu$; Middle: confidence interval boundaries as a function of observed counts. Circles are for LR, squares for score. The plus symbols show the maximum likelihood estimator for the mean; Right: size of the confidence interval as a function of observed counts. Circles are for LR, squares for score.  The plus symbols show the values of $2\sqrt{n}$.\label{fig:scoreLR}}
\end{figure}

The score intervals turn out to be similar to the likelihood ratio intervals for our Poisson problem, see Fig.~\ref{fig:scoreLR}. The coverage is 
sometimes better with one, sometimes better with the other, but generally has a similar
dependence on $\mu$. The intervals tend to be shifted slightly higher for the score test,
with similar interval sizes. 

\section{Square root of $N$}
\label{sec:sqrtn}

The standard deviation of a Poisson distribution with mean $\mu$ is $\sqrt{\mu}$. Hence, if we use
statistic $N$ as an estimator for $\mu$, we often also use $\sqrt{N}$ as an estimator for the standard deviation. We know that asymptotically the interval $N\pm\sqrt{N}$ provides a 68\% confidence interval for $\mu$. The $\sqrt{N}$ provides a common choice for displaying an error bar on a Poisson-sampled
quantity, and we have an intuitive calibration on the interpretation of such displays. It is natural then to ask what the frequency properties of such intervals are in the context of our present discussion. This is an inexact method. However, since this is such a commonly used statistic, we include it for comparison.

We implement these intervals by defining our sets $C_\alpha(N=n)$, with nominally $\alpha=0.68$:
\begin{equation}
C_{0.68}(n) = \{n-\sqrt{n}\le\mu\le n+\sqrt{n}\}.
\end{equation}
Then we obtain sets $S_\alpha(\mu)$ as
\begin{align}
S_{0.68}(\mu) &=\{n|\mu\in C_{0.68}(n)\}\\
 &= \left\{n|n-\sqrt{n}\le\mu\le n+\sqrt{n}\right\}\\
 &= \left\{n\Big| \frac{1}{2}\left(1+2\mu-\sqrt{1+4\mu}\right)\le n \le  \frac{1}{2}\left(1+2\mu+\sqrt{1+4\mu}\right)\right\}.
 \end{align}
 We could generalize this approach to other confidence levels, still using the asymptotic normal approximation. 
 
\begin{figure}
\centering
\includegraphics[width=0.65\columnwidth]{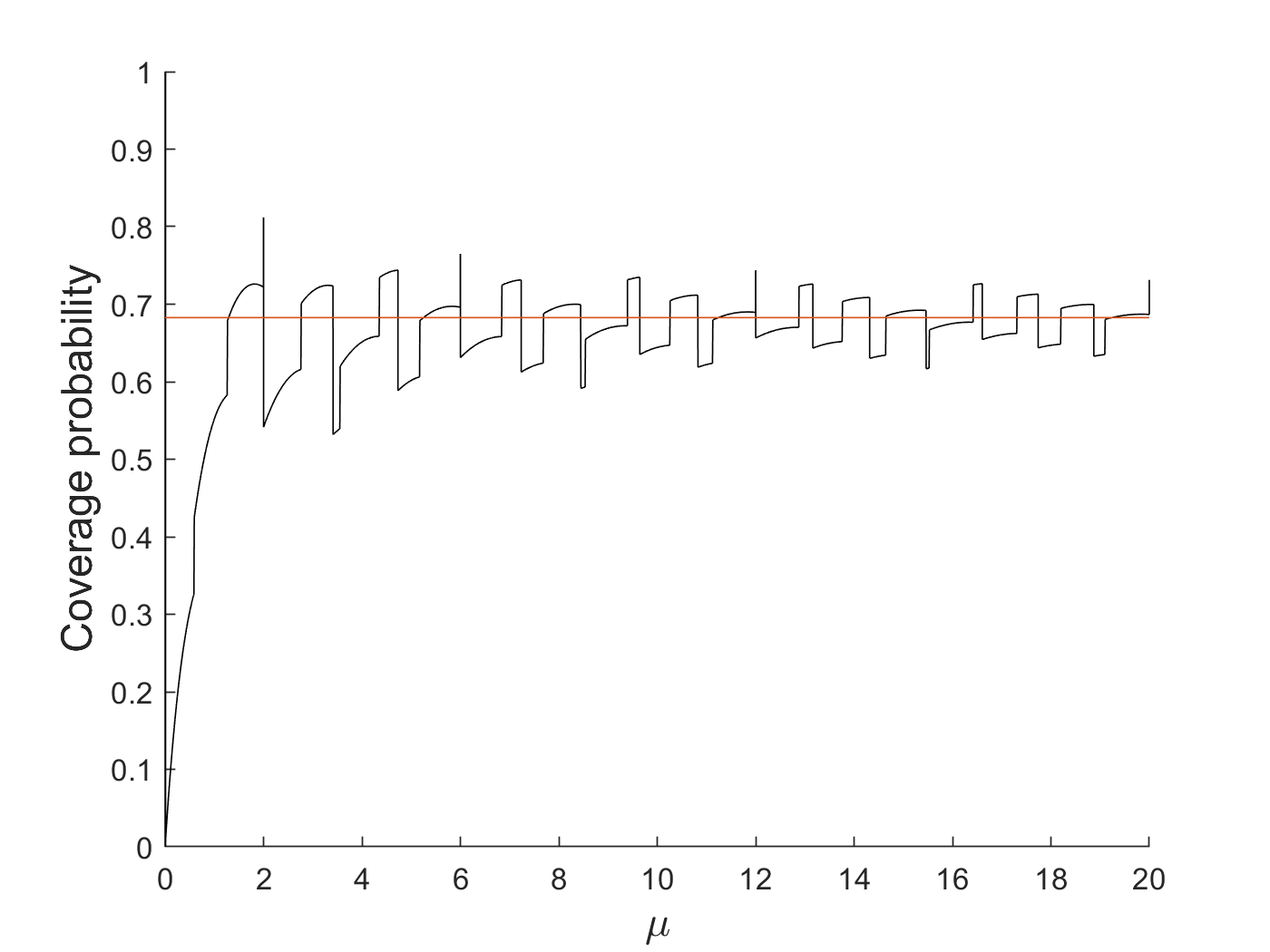}
\caption{Coverage probability of $\sqrt{n}$ confidence intervals as a function of $\mu$. The horizontal line shows $1-\alpha$.\label{fig:sqrtn}}
\end{figure}

The coverage probability of the $\sqrt{n}$ intervals is shown as a function of $\mu$ in Fig.~\ref{fig:sqrtn}.
As expected, the intervals both undercover and overcover, depending on the value of $\mu$, hence do not meet
our coverage requirement. However, at least for large enough $\mu$ the approximation might be good enough for certain purposes, such as graphical representation.

\section{Frequency properties of Bayes' intervals with uninformative priors}
\label{sec:priors}

As mentioned in the introduction, we may examine the frequency properties of objective Bayes methods.
For illustration and comparison, we consider two popular choices commonly referred to as  ``uninformative'' reference priors, the uniform prior and the Jeffreys prior. Here, we need to explicitly include the background parameter. We'll assume that $\theta\ge0$.

The general procedure, for an observation $N=n$, is to form the posterior distribution according to Bayes' theorem:
\begin{equation}
P(\theta;n,b) = \frac{L(\theta;n,b) P(\theta;b)}{\int_0^\infty L(\theta;n,b) P(\theta;b) d\theta},
\end{equation}
where $P(\theta;b)$ is the prior distribution. The Bayes interval for a given probability level $1-\alpha$ is then determined by
including all values of $\theta\in(\theta_{\rm min},\theta_{\rm max})$ such that
\begin{equation}
\int_{\ell}^{u} P(\theta;n,b)\,d\theta = 1-\alpha,
\end{equation}
where  $P(\theta;n,b)$ is larger for any $\theta\in(\theta_{\rm min},\theta_{\rm max})$ than for any $\theta$ outside this interval. That is, we add values of $\theta$ to our confidence set based on ordering of the posterior distribution. We are assuming here that such a solution exists, as it does for posteriors considered here. Otherwise, we would consider potentially disconnected sets in our probability ordering. In some cases, for large enough $n$, we will have $P(\ell;n,b) = P(u;n,b)$, but this is not guaranteed.

\subsection{Uniform prior}

With an assumption on physical grounds that $\theta\ge 0$, the uniform prior is proportional to
\begin{equation}
P_U(\theta) \propto
  \begin{cases}
  0 &\theta < 0 \\
  1 &\theta \ge 0.
  \end{cases}
\end{equation}  
The posterior distribution for $\theta$ assuming a sampled value $n$ from a Poisson sampling with mean $\theta+b$ is
\begin{equation}
P_U(\theta;n,b) = \frac{(\theta+b)^n}{n!}e^{-(\theta+b)}\Big/I_n(b),\quad\theta\ge 0,
\end{equation}
where
\begin{equation}
 I_n(x) \equiv \int_x^\infty \frac{y^n}{n!}e^{-y}dy =e^{-x} \sum_{j=0}^n\frac{x^j}{j!} = \Gamma(n+1,x)/n!,
 \end{equation}
 and $\Gamma(n+1,x)$ is the upper incomplete gamma function.
 The cumulative posterior distribution is
\begin{equation}
\int_0^\theta P_U(\theta^\prime;n,b)d\theta^\prime = 
  1 - \frac{\Gamma(n+1,\theta+b)}{\Gamma(n+1,b)}.
\end{equation}

\subsection{Jeffreys prior}

The Jeffreys prior for sampling distribution $f$ is proportional to the square root of Fisher's information number and is given by~\cite{Jeffreys1946}
\begin{equation}
P_J(\theta) \propto \sqrt{E\left\{\left[\frac{d}{d\theta} \log f(n;\theta)\right]^2\right\}}.
\end{equation}
For our Poisson distribution with background $b$, this works out to:
\begin{equation}
P_J(\theta) \propto
\begin{cases}
  0 &\theta < 0\\
 \frac{1}{\sqrt{\theta+b}} &\theta\ge0,
\end{cases}
\end{equation}
again assuming on physical grounds that $\theta\ge 0$.
We note that the prior depends on the background parameter, suggesting that it is not really capturing a degree-of-belief in $\theta$.\footnote{One could also consider a prior proportional to $1/\sqrt{\theta}$, but this is not the Jeffreys prior unless $b=0$.} In any event, we are looking here at descriptive properties of the method.

The posterior distribution is thus
\begin{equation}
P_J(\theta;n,b) = A\frac{(\theta+b)^{n-1/2}}{n!}e^{-(\theta+b)},\quad\theta\ge 0,
\end{equation}
where the normalization factor $A$ is given by
\begin{equation}
A = n!/\Gamma(n+1/2,b).
\end{equation}
The cumulative posterior distribution is
\begin{equation}
\int_0^\theta P_J(\theta^\prime;n,b)d\theta^\prime = 
  1 - \frac{\Gamma(n+1/2,\theta+b)}{\Gamma(n+1/2,b)}.
\end{equation}

\begin{figure}
\centering
\includegraphics[width=0.9\columnwidth]{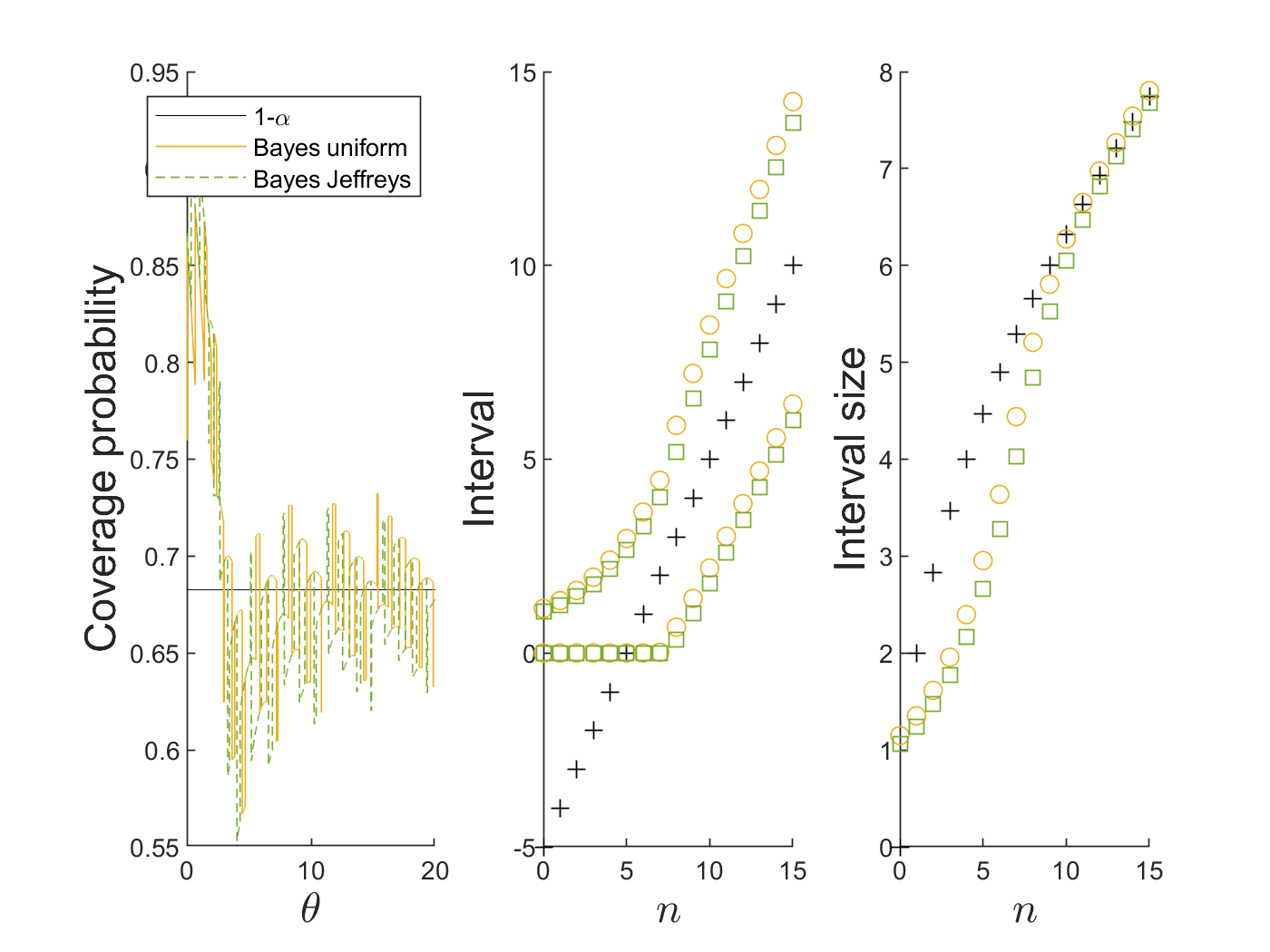}
\caption{Bayes 68\% probability credible intervals for $b=5$, comparing uniform and Jeffreys priors. Left: coverage probability as a function of $\theta$; Middle: confidence interval boundaries as a function of observed counts. Circles are for uniform prior, squares for Jeffreys prior. The plus symbols show the maximum likelihood estimator for $\theta$; Right: size of the confidence interval as a function of observed counts. Circles are for uniform prior, squares for Jeffreys prior. The plus symbols show the values of $2\sqrt{n}$.\label{fig:Bayesb5}}
\end{figure}

Figure~\ref{fig:Bayesb5} shows a comparison of the Bayes 68\% confidence credible intervals for the uniform and Jeffreys priors, for a background
parameter $b=5$.
They are similar, but the preference for $\theta=0$ of the Jeffreys prior is seen in the lower values of the interval boundaries and the smaller interval sizes.
The coverage is observed to both under and over cover, depending on $\theta$. This is unsurprising, as there is no requirement on coverage for Bayes intervals. The Bayes intervals also tend to be shorter in length than $2\sqrt{n}$ for small $n$, unlike the behavior of the conventional frequency intervals of section~\ref{sec:conventional}. 

\begin{figure}
\centering
\includegraphics[width=0.9\columnwidth]{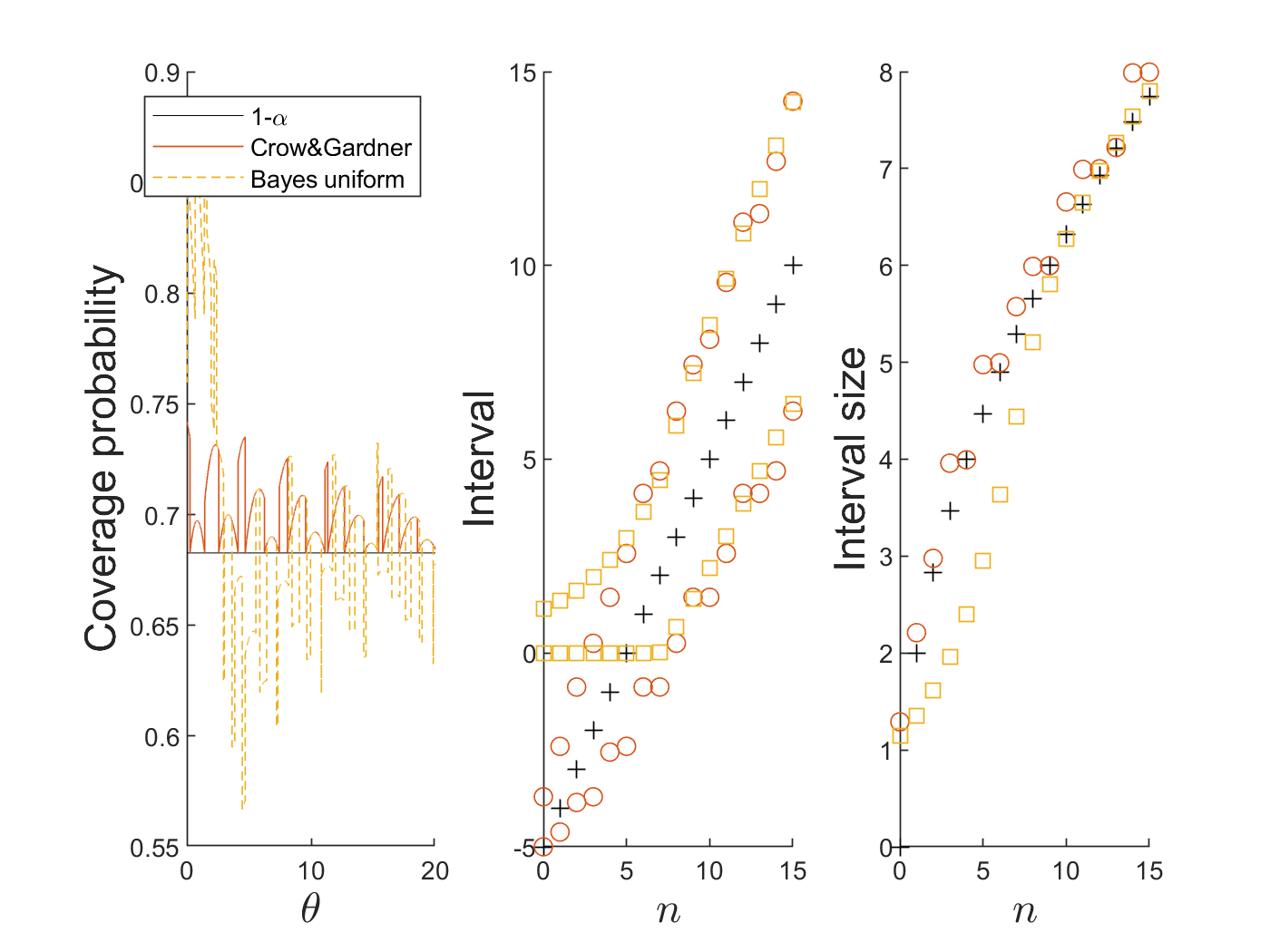}
\caption{Comparison of Crow\&Gardner 68\% confidence intervals and Bayes 68\% probability credible intervals (with a uniform prior) for $b=5$. Left: coverage probability as a function of $\theta$; Middle: confidence interval boundaries as a function of observed counts. Circles are for Crow\&Gardner, squares for uniform prior. The plus symbols show the maximum likelihood estimator for $\theta$; Right: size of the confidence interval as a function of observed counts.  Circles are for Crow\&Gardner, squares for uniform prior. The plus symbols show the values of $2\sqrt{n}$.\label{fig:CrowBayesUb5}}
\end{figure}

The Bayes interval with uniform prior is also compared to the Crow\&Gardner interval in Fig.~\ref{fig:CrowBayesUb5}. For small $\theta$, the coverage of the Bayes interval is substantially higher than Crow\&Gardner, becoming smaller at intermediate $\theta$. The Bayes interval is never negative,
by design as an interpretive statement, while the frequency interval can be negative as a descriptive statement. 

\section{Methods from particle physics}
\label{sec:alternative}

\subsection{The $CL_s$ method}
\label{sec:CLs}

The $CL_s$ method~\cite{Read2000, Read2002, Zech1989} (a discussion in conventional statistics language appears in~\cite{Qian2016}) 
has a similar appearance to the likelihood ratio method, but is quite different. It has been motivated as a means to 
compute exclusion regions on the parameter $\theta$, in comparison with the no signal null hypothesis.

Consider the hypothesis test
\begin{equation}
\begin{split}
\hbox{$H_0$:\ } &\theta=\theta_0\\
\hbox{against}\\
\hbox{$H_1$:\ } &\theta= \theta_1.
\end{split}
\end{equation}
Note that this is not the same test as we considered in Eq.~\ref{eq:LRhypotheses},
since $\theta$ is specified under the alternative. The likelihood ratio is uniformly most powerful for a simple test, so it is
natural to use here.
We thus take as our test statistic, the likelihood ratio between the two hypotheses,\footnote{We define this here to be consistent
with the convention for the $CL_s$ method and attach a prime to flag the difference in convention with Eq.~\ref{eq:Lratio}.}
\begin{equation}
\lambda^\prime(N)\equiv\frac{L(\theta_1;N)}{L(\theta_0;N)}.
\end{equation}
This ratio is denoted as $Q$ in papers on the $CL_s$ method. Here we have possible values $0<\lambda^\prime<\infty$.
We generally work with $R\equiv -2\log \lambda^\prime$, with range $(-\infty,\infty)$, where negative values favor $H_1$ and
positive values favor $H_0$.

The $CL_s$ statistic is defined for observation $N=n$ by
\begin{equation}
CL_s(n) \equiv \frac{1-p_1(n)}{1-p_0(n)},
\end{equation}
where\footnote{Reference~\cite{Qian2016} is confusing here, saying ``larger'' instead of ``larger or equal''.}
\begin{equation}
\begin{split}
1-p_0(n) &\equiv \hbox{Prob}\left[R(N)\ge R(n)|H_0\right]\\
1-p_1(n) &\equiv \hbox{Prob}\left[R(N)\ge R(n)|H_1\right].
\end{split}
\end{equation}
We may note that the quantity $1-p_1$ is simply the $p$-value for hypothesis $H_1$. The quantity
\begin{equation}
p_0(n) = \hbox{Prob}\left[R(N)< R(n)|H_0\right]
\end{equation}
is the $p$-value for $H_0$ in our hypothesis test, except 
for the non-zero measure at $N=n$ according to the discreteness of the distribution.

The $CL_s$ statistic is used to build exclusion regions for $\theta$. The $H_0$ hypothesis is here the hypothesis that the background-only
model ($\theta_0=0$) is correct. We can then scan over the possible signal hypotheses, $\theta_1\ne\theta_0$, and exclude any such hypothesis for which
$CL_s(n;\theta_1)<0.05$, say.

The idea of dividing by $1-p_0$ protects against excluding $\theta_1$ in the following situation: Suppose $\theta_0$ and $\theta_1$ are not
so different, that is the possible signal is ``close'' to zero. Suppose the $p$-value for $\theta_1$ is very small, that is, it provides a poor fit to the data.
In this case, $H_0$ and $H_1$ are about the same, so $1-p_0$ and $1-p_1$ will be nearly the same. Hence $CL_s\approx1$, and we do
not exclude the possibility of a small signal. That is, we are protected against the possibility of excluding the presence of a small signal even
if the fit is poor. We see that the $CL_s$ method is designed to provide ``limits'', never excluding the possibility that there is no signal. This method, motivated by the desire to avoid incorrect statements about the parameter, clearly goes beyond simply description. If a fit to the data is poor, that is an important descriptive statement.

We see that $CL_s$ is generally greater than $1-p_1$, the $p$-value for the signal hypothesis. Thus, an interval based on
$CL_s(n;\theta_1)<\alpha$ is conservative -- the exclusion region corresponds to probability smaller than $\alpha$.

We now apply the $CL_s$ method to our Poisson problem. We have $\theta_0=0$ in this case. Then
\begin{equation}
\lambda^\prime(N) =\left(\frac{\theta_1+b}{b}\right)^Ne^{-\theta_1},
\end{equation}
or
\begin{equation}
R(N) = -2N\log\left(1+\frac{\theta_1}{b}\right) + 2\theta_1.
\end{equation}
Then 
\begin{equation}
\begin{split}
1-p_0 &= \hbox{Prob}(N\le n|H_0)\\
1-p_1 &= \hbox{Prob}(N\le n|H_1).
\end{split}
\end{equation}
The confidence interval (complement of exclusion set) for $\theta$ at the $1-\alpha$ confidence level is then the set of all $\theta$ such that
\begin{equation}
CL_s \equiv \frac{1-p_1}{1-p_0} \ge \alpha.
\end{equation}
As $\theta\to 0$, $CL_s \to 1$, so we always obtain an upper limit by this method. This remains the case if we consider negative $-b\le \theta_1 < 0$,
the interval just extends down to $-b$ in this case.

\begin{figure}
\centering
\includegraphics[width=0.7\columnwidth]{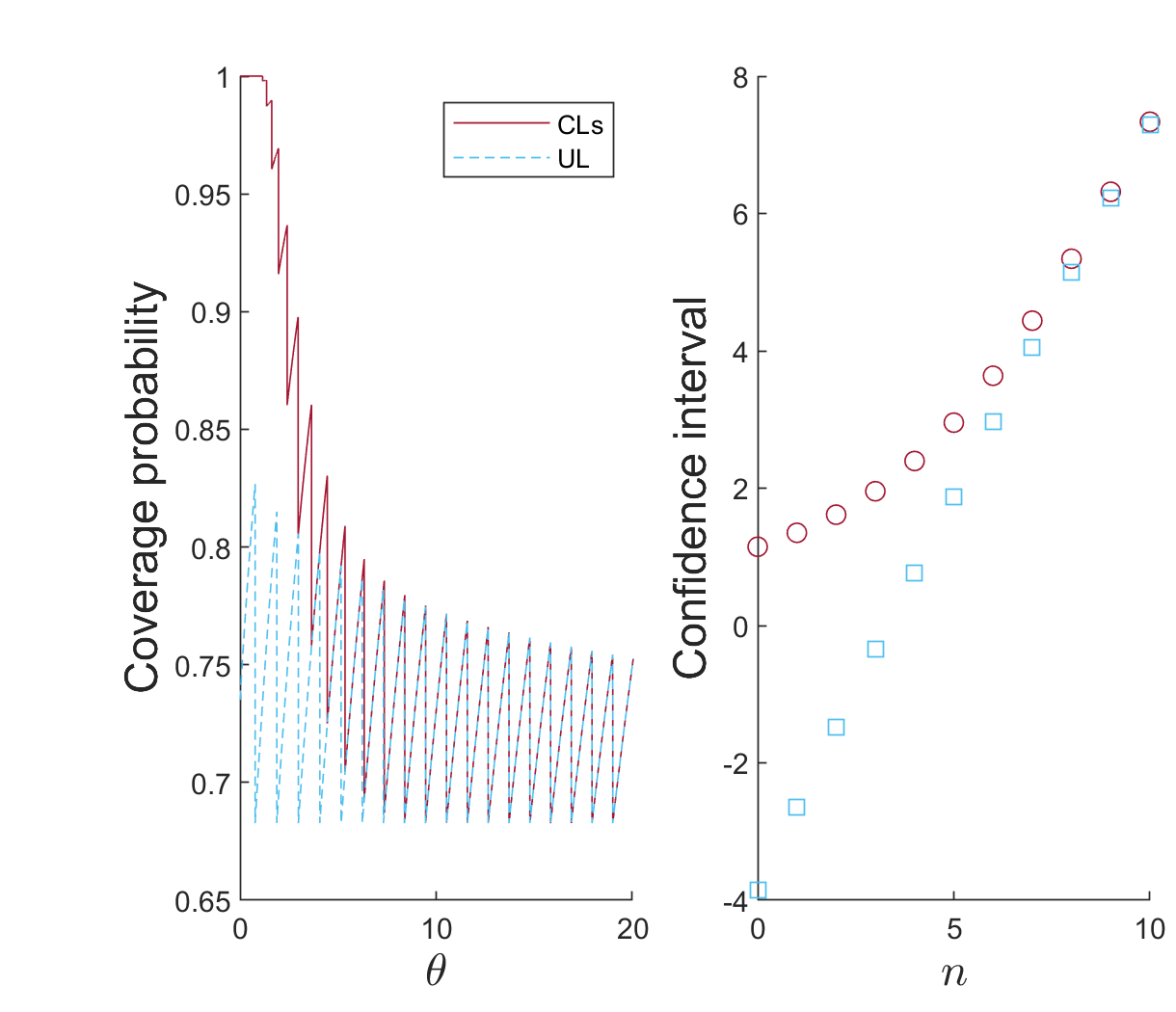}
\caption{Performance of 90\% confidence upper limits, comparing $CL_s$ with the standard upper limit (section~\ref{sec:UL}) for $b=5$. Left: coverage probability as a function of $\theta$; Right: upper limit as a function of observed counts. Circles are for $CL_s$, squares for standard upper limit. \label{fig:CLsUL}}
\end{figure}

The $CL_s$ limit converges on the standard upper limit of section~\ref{sec:UL} in the limit of no background.
For $b>0$, however, the two limits are not the same. The largest difference, for a given background is that the
$CL_s$ limit tends to overcover substantially more than the standard limit for small values of $\theta$, as seen for
$b=5$ in Fig.~\ref{fig:CLsUL}. This is consistent with our observation above that the division by $1-p_0$ leads to a
conservative upper limit. As a descriptive statement, the limit of section~\ref{sec:UL} is preferred.

\subsection{Feldman-Cousins}
\label{sec:FC}

The Feldman-Cousins (FC) method~\cite{Feldman1997} is the same as the likelihood ratio test inversion in section~\ref{sec:LR}, except with an additional condition designed to prevent intervals overlapping a ``non-physical'' region. Here, it is assumed that $\theta<0$ is ``unphysical'', as we interpret it here as a physical rate for some signal process, which cannot be negative. In the case that $b=0$, the FC intervals and the LR intervals are identical, because in this case the ``physical'' region for $\theta$ is the same as the  parameter space of the Poisson model. The statistic is

\begin{equation}
\lambda_{\rm FC}(\theta;n)\equiv\frac{L(\theta;n)}{L(\theta_{{\rm ML}\ge0};n)},
\label{eq:FClambda}
\end{equation}
where 
$\theta_{{\rm ML}\ge0}$ indicates the maximum likelihood estimator for $\theta$, restricted to the  non-negative region of possible ``physical''
values of $\theta$. Only non-negative values of $\theta$ are considered in constructing the acceptance regions, hence $\lambda_{\rm FC}$ is never
larger than one.

In our implementation of the FC method, we do not force the upper endpoints of the intervals to decrease with increasing $b$ for a given observation $n$. 
Hence,
our intervals are sometimes smaller than in~\cite{Feldman1997} and our intervals do not over-cover as much. As everywhere in this article, we do always quote continuous
intervals, consistent with ref.~\cite{Feldman1997}.

\begin{figure}[h]
\centering
\includegraphics[width=0.8\columnwidth]{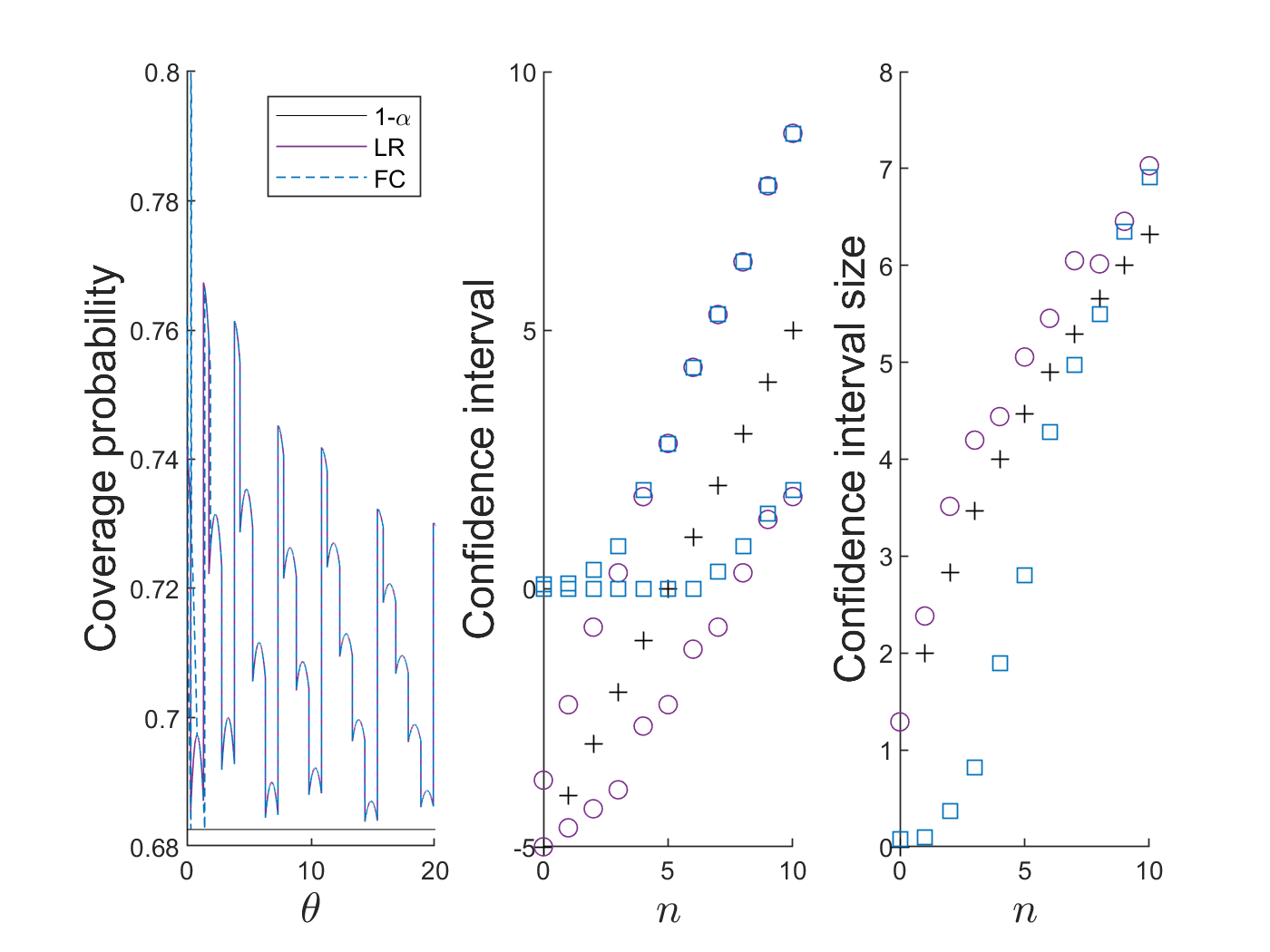}
\caption{Performance of 68\% confidence intervals for $\theta$, comparing FC with the likelihood ratio intervals for $b=5$. Left: coverage probability as a function of $\theta$; Middle: confidence interval as a function of observed counts. Circles are for LR, squares for FC. The plus symbols show the maximum likelihood estimator for $\theta$ (allowed to be negative); Right: interval size as a function of observed counts. Circles are for LR, squares for FC. The plus symbols show the values of $2\sqrt{n}$.\label{fig:FCvsLR}}
\end{figure}

As mentioned above, if there is no background the FC and LR intervals are the same, because Eqn.~\ref{eq:Lratio} and Eqn.~\ref{eq:FClambda}
are identical for zero background. For a background of $b=5$, the comparison between the properties of the FC and LR intervals is shown in Fig.~\ref{fig:FCvsLR}. With background, there are substantial differences between the two intervals.
The coverage is very similar between the two, except for small values of $\theta$, where the FC coverage is greater than the LR, due to the interval being forced positive for small $n$. The intervals are very different for small observations.
The FC intervals, by design, do not include negative values, while the LR intervals can include negative values, reflecting occurrences of negative fluctuations in the background. For small values of observation $n$, the FC intervals are considerably shorter than the LR intervals.

\begin{figure}[h]
\centering
\includegraphics[width=0.8\columnwidth]{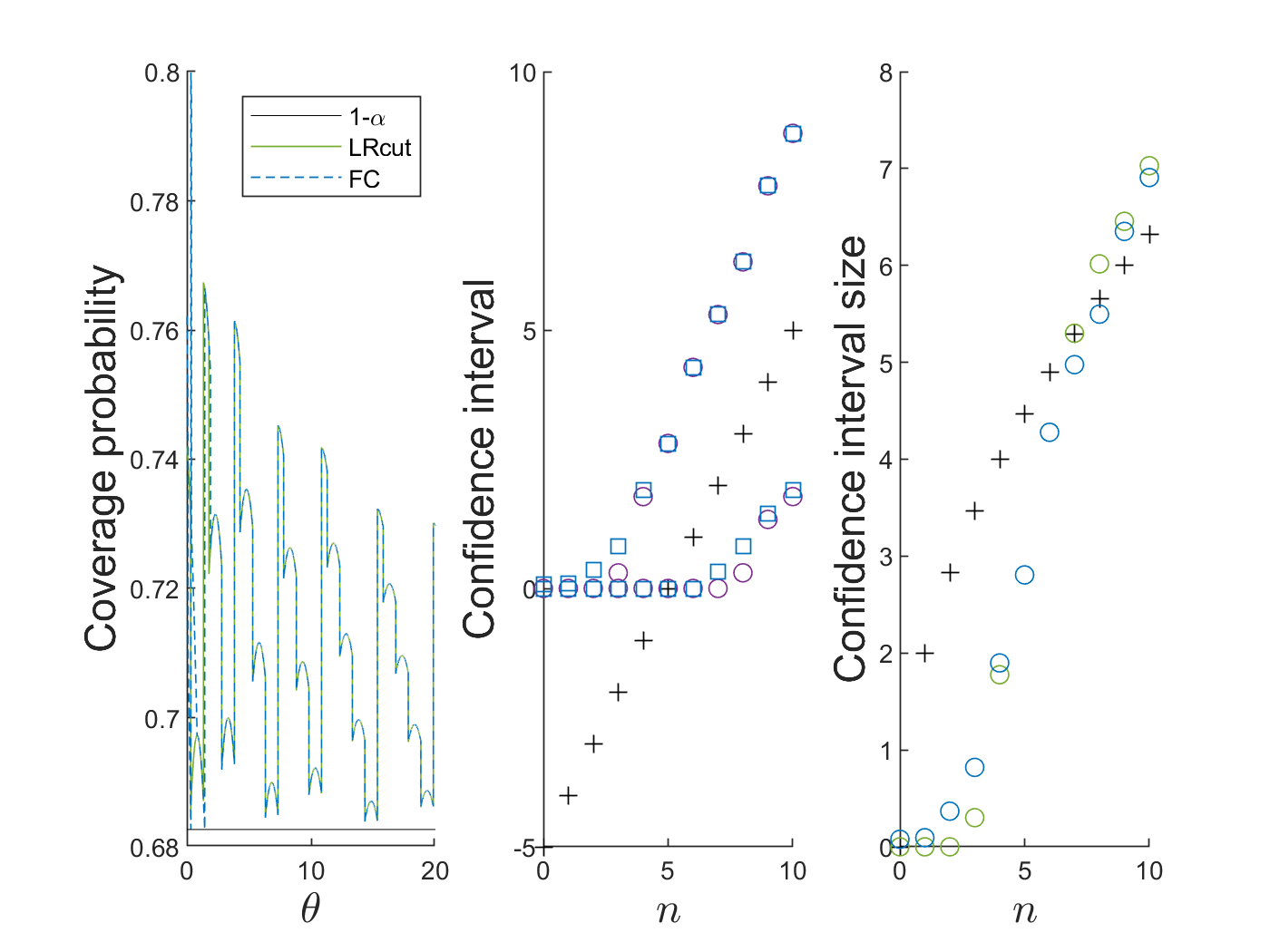}
\caption{Performance of 68\% confidence intervals for $\theta$, comparing FC with the cut likelihood ratio intervals for $b=5$. The likelihood ratio intervals have been cut so that only the positive portion of the interval is retained. Left: coverage probability as a function of $\theta$; Middle: confidence interval as a function of observed counts. Circles are for cut LR, squares for FC. The plus symbols show the maximum likelihood estimator for $\theta$ (allowed to be negative); Right: interval size as a function of observed counts. Circles are for cut LR, squares for FC. The plus symbols show the values of $2\sqrt{n}$.\label{fig:FCvsLRcut}}
\end{figure}

However, we may observe that the two intervals are not so different in the following sense: The coverage (and hence the validity of the interval as a frequency confidence interval) is not affected if we simply cut the LR interval at $\theta=0$, since we are assuming that the true value of $\theta$ must be non-negative. We will refer to these intervals as cut LR intervals. Note that this procedure can yield intervals of zero size, when a sufficiently large negative fluctuation has occurred. For the graphical indication of this, we will plot
this as upper and lower bounds equal to zero, with the understanding that it is a null interval. The comparison with FC, again for $b=5$,
is shown in Fig.~\ref{fig:FCvsLRcut}.
As expected, the coverage is identical with Fig.~\ref{fig:FCvsLR}. However, the intervals and their sizes are
now very similar between the FC and cut LR methods, on the visual scale of the graphs. When the cut LR intervals are very small, or null (occurring when a large negative fluctuation has occurred), the FC interval is also very small, though by design never null.

What do we make of this? The (uncut) LR interval has similar behavior to the other conventional methods of Section~\ref{sec:conventional}, differing in relatively small ways in performance. In particular, all of these intervals 
will become negative in the presence of large downward background fluctuations. As we have discussed, this is quite natural and provides an easy-to-understand description of the measurement. When we cut it off, 
we do not lose frequency validity, but the value as a description of the measurement is diminished: Null intervals may be obtained for multiple possible outcomes, and how much background (plus signal) fluctuation has occurred has been lost. Even when the interval is not completely null, the precision of the measurement is obscured.

The FC interval was designed, in large part at least, to overcome the tendency of physicists to change their analysis depending on the outcome of the measurement, which leads to biased results (in the frequency sense). In particular, the issue addressed is the tendency to decide whether to quote a central interval or a one-sided interval depending on the observation. The FC interval ``makes this decision'' for you and preserves the strict frequency
interpretation. Of course, this is not the only way to accomplish this.\footnote{But far better is to design the analysis ahead of time so that such bias is avoided.} Any of the conventional methods of Section~\ref{sec:conventional} will also do this, if we simply exclude any negative values from the interval.
It could be remarked that there is a distinction, because the FC intervals are never empty, while simply cutting out
negative values in the other methods leads to null intervals for some observations. However, these can also readily be 
made to be non-empty if that is desired. For example, one way to do this is, whenever an empty interval is encountered, to simply modify it to an interval $[0,\epsilon]$, where $\epsilon$ is a small positive number. This could be done in a way that maintains sufficiency, e.g., $\epsilon=1/[\hbox{large number}\times(1+n)]$.  This is of course contrived, but that is really the point, attempting to restrict description to non-negative intervals leads to cryptic results. 

To be clear, I am not advocating use of the cut LR intervals even if modified to ensure no null intervals. The small intervals when the observation is small suggest greater precision than justified. The point is that the FC interval also suffers from this issue. FC~\cite{Feldman1997} suggest to quote the ``sensitivity'' of the experiment to alert the consumer when this occurs.

\begin{figure}[h]
\centering
\includegraphics[width=0.85\columnwidth]{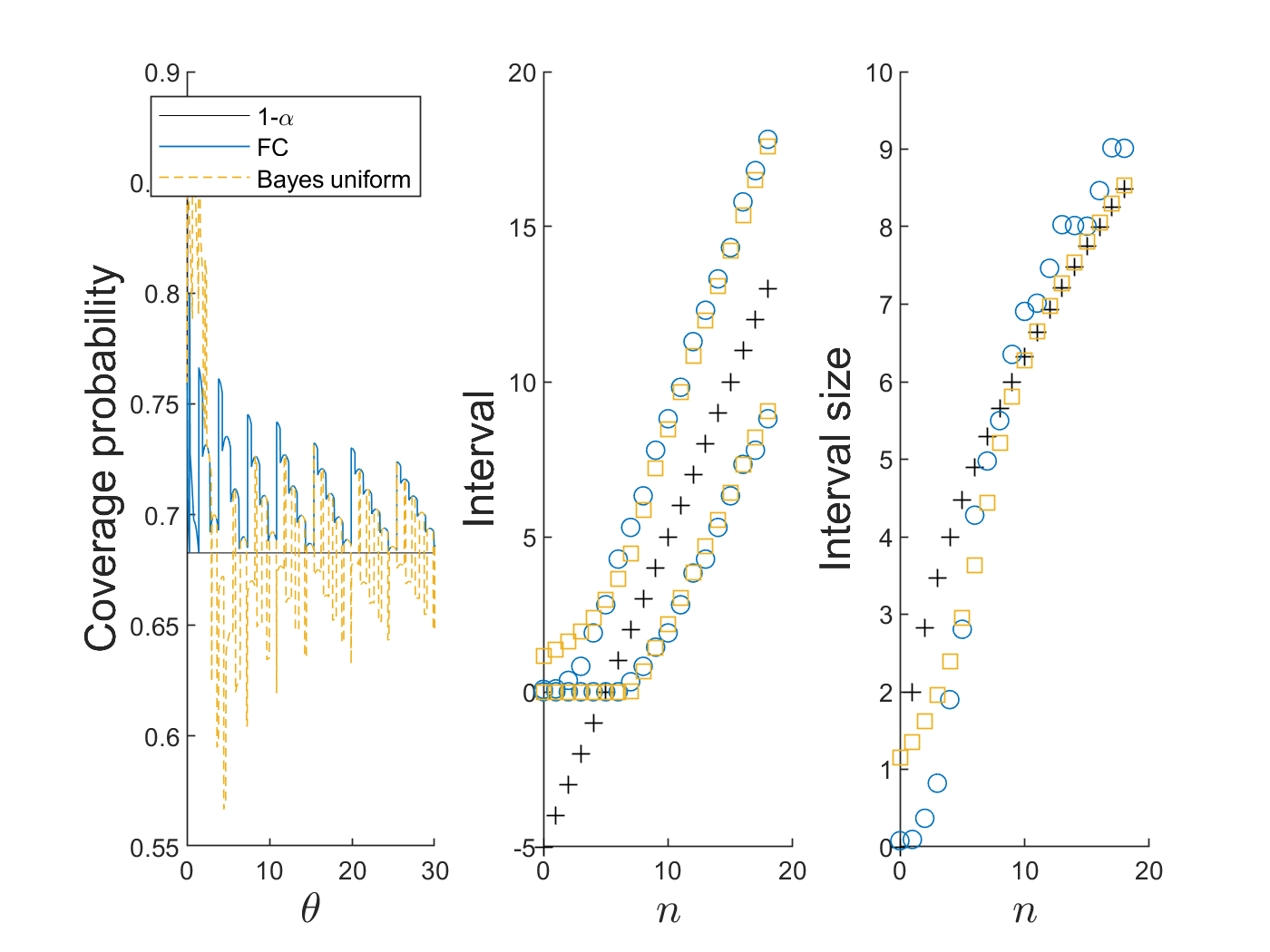}
\caption{Performance of 68\% intervals for $\theta$, comparing FC confidence intervals with Bayes (uniform prior) credible intervals for $b=5$. Left: coverage probability as a function of $\theta$; Middle: confidence interval as a function of observed counts. Circles are for FC, squares for Bayes. The plus symbols show the maximum likelihood estimator for $\theta$ (allowed to be negative); Right: interval size as a function of observed counts. Circles are for FC, squares for Bayes. The plus symbols show the values of $2\sqrt{n}$.\label{fig:FCvsBayesU}}
\end{figure}

We also compare the FC intervals with Bayes intervals with uniform prior in Fig.~\ref{fig:FCvsBayesU}, for background $b=5$.
For small observations $n$, the FC intervals are much shorter than the Bayes intervals, becoming somewhat larger at higher $n$. The comparison with Bayes intervals is consistent with the above observation about apparent precision.

\section{Averaging observations}
\label{sec:averaging}

Suppose we have a set of observations, $n = n_1,n_2,\ldots, n_K$ drawn from Poisson distributions with means $\theta_1,\theta_2,\ldots,\theta_K$,
where $\theta_k=\Gamma T_k$. That is, the physical parameter of interest is here considered to be a ``rate'' $\Gamma$,  and $T_k$ is just the amount
of ``time'' we devoted to observation $k$. It will suffice here to consider the case of no background. The maximum likelihood estimator for $\Gamma$ from each measurement is $\hat\Gamma_k=n_k/T_k$.
Because the Poisson distribution is reproductive, $\sum_{k=0}^K n_k$ is a sampling from a Poisson distribution with mean $\theta\equiv\sum_{k=0}^K \theta_k$.
The overall maximum likelihood estimator for $\Gamma$ is thus $\hat\Gamma =\sum_{k=0}^K n_k/T$, where $T\equiv\sum_{k=0}^K T_k$. This is equivalent to a weighted average, $\hat\Gamma =\sum_{k=0}^K w_k\hat\Gamma_k$, with weights $w_k=T_k/T$. The confidence intervals on the
combined result are then derived according to the combined Poisson. The variance of a single measurement of $\Gamma$ is $\hbox{var}\left(\hat\Gamma_k\right)=\theta_k/T_k^2 = \Gamma/T_k$, and
of the combined result $\theta/T^2=\Gamma/T$. This can be rewritten 
\begin{equation}
\hbox{var}\left(\hat\Gamma\right)=\left[\sum_{k=0}^K \frac{1}{\hbox{var}\left(\hat\Gamma_k\right)}\right]^{-1},
\end{equation}
and our above weighted average is equivalent to a weighted average where the weights are the inverses of the variances.

However, results for a parameter of interest are often described with confidence intervals and it may not always be clear how to go back
to the underlying Poisson distributions. In this case, we would like to base averages of observations on the quoted intervals, and to obtain
confidence intervals for such averages that
continue to respect the claimed coverage. That is, we hope to combine two 68\% confidence intervals and obtain another 
68\% confidence interval, while making efficient use of the combined information. At the least, we would like our procedure to asymptotically converge on correct coverage.

Unfortunately, this is not straightforward. For example, we could try to use our confidence intervals to provide estimates of variance and use inverse square weighting
to produce averaged results.  Let's continue to work with the MLE as the point estimator in the context of the above discussion, with $\Gamma$ our physical parameter of interest. Suppose we have $k$ experiments measuring $\Gamma$, with maximum likelihood estimators
$\hat\Gamma_1,\ldots,\hat \Gamma_K$, and 68\% confidence intervals $(\ell_1,u_1),\ldots,(\ell_K,u_K)$. As above, we compute a weighted average of
our results according to
\begin{equation}
\label{eq:wtdAvg}
\bar\Gamma=\frac{\sum_{k=0}^K w_k\hat\Gamma_k}{\sum_{k=0}^K w_k}.
\end{equation}

For weights, we will use inverse square weighting in a measure of the interval size. The individual confidence intervals may not be symmetric about the MLEs, and there
are several ways we could address this. 
For the sake of the example, we will use the (conservative) procedure of symmetrizing the interval, taking the maximum deviation from the MLE as our error bar. That is, we let
\begin{equation}
\begin{split}
s_k &= \max(u_k-\hat\Gamma_k,\hat\Gamma_k-\ell_k)\\
w_k &= 1/s_k^2.
\end{split}
\end{equation}
In this procedure, we enlarge our confidence interval for a single measurement to the symmetric interval $\hat\Gamma_k\pm s_k$.

With these weights, we compute the point estimator according to the weighted average in Eq.~\ref{eq:wtdAvg}. Following typical practice, we attempt
to compute a 68\% confidence interval for our averaged result as $\bar\Gamma \pm \bar\delta$, where
\begin{equation}
\bar\delta=\left(\frac{1}{\sum_{k=0}^K w_k}\right)^{1/2}.
\end{equation}
It remains to be determined whether this in fact yields correct coverage. 

\begin{figure}[h]
\centering
\includegraphics[width=0.7\columnwidth]{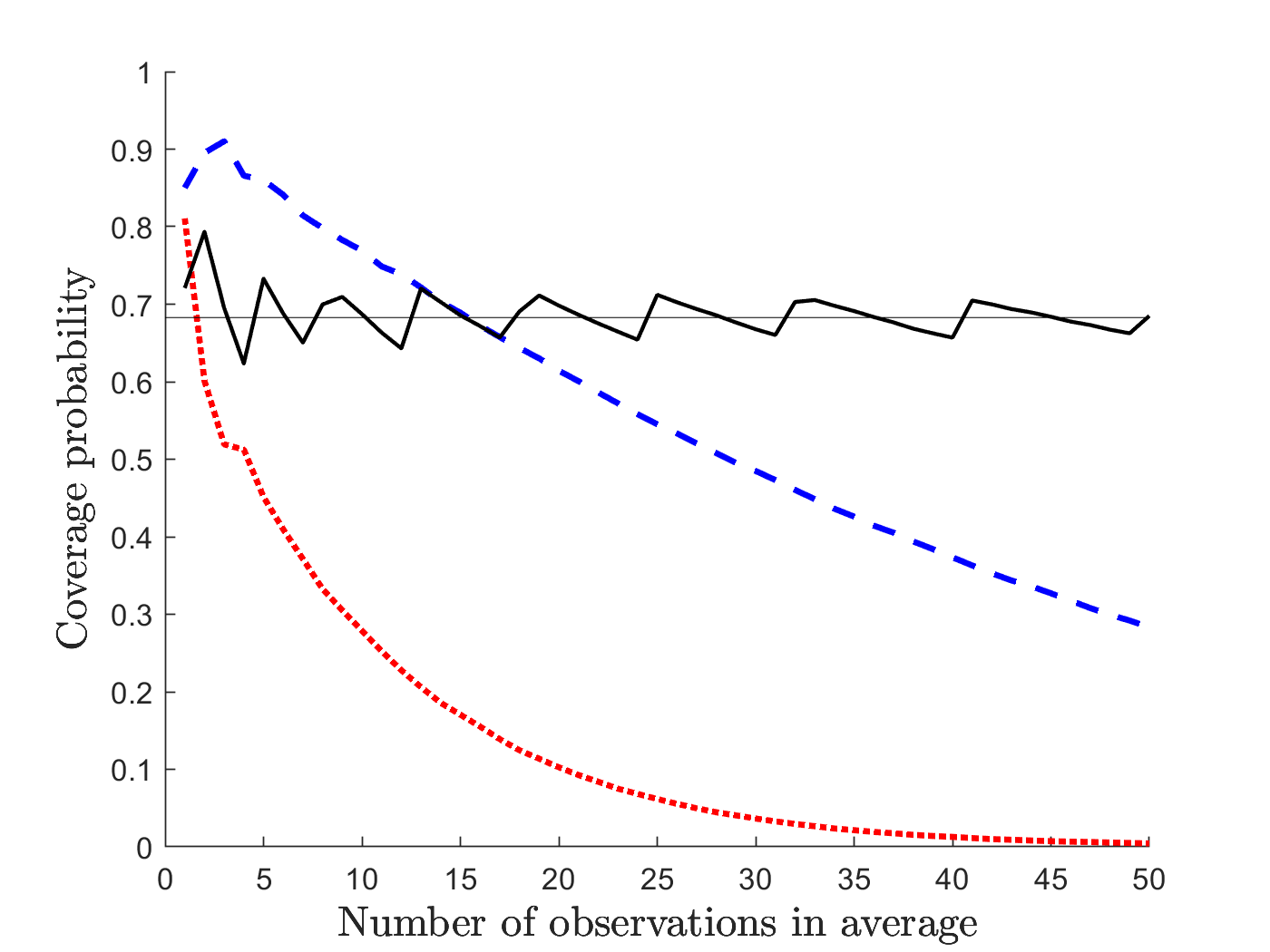}
\caption{Coverage of averaging (symmetrized) Garwood intervals (blue dashed curve) according to the procedure described in the text. For comparison, the solid black curve shows the coverage when the true variance is used as weights, and the red dotted curve shows the coverage when the variances used in the weights is estimated with $\max(1,n)$. The faint horizontal line shows the target coverage.\label{fig:avgGTS}}
\end{figure}

To study the coverage of our averaged result, we simulate a simple situation. As above, we suppose we make several, $K$, observations
of $\hat\Gamma_k=n_k/T_k, k=1,\ldots,K$ with corresponding 68\% confidence intervals $(\ell_k,u_k)$. We apply our weighted averaging procedure and ask how often the interval  $\bar\Gamma \pm \bar\delta$ includes $\Gamma$. It will be sufficient for our demonstration to choose $\Gamma=1$
and $T_k=2, \forall k$. We will use the Garwood 68\% confidence intervals for $(\ell_k,u_k)$ and consider averages of up to 50 observations. Figure~\ref{fig:avgGTS} shows the result for coverage as a function of the number of observations being averaged. We show the  target (0.68) coverage as a horizontal line; values above this correspond to overcovering and values below to undercovering.

Even though the single measurement confidence intervals overcover (coverage probability 0.85 for our symmetrized intervals; note that this is quite a bit higher than the coverage of 0.72 for the Garwood interval for $\mu=2$) for this example, the coverage decreases as more results are averaged, and eventually intervals that severely undercover are obtained. In spite of the conservatism of our
Poisson interval, the averaging process fails to maintain frequency coverage. This may seem counter to expectations based on the central limit theorem.
However, we cannot expect the central limit theorem to apply simply because we make a large number of observations. We also need to know the
variances, which we do not. It may be noted that doing the same simulation with normal samplings with known variance gives the expected 68\% coverage for all the averages.

We show two other calculations on Fig.~\ref{fig:avgGTS}. First, we compute the weighted averages using the true variances ($\theta$) as the weights. In this case, the coverage gradually converges on the normal limit as we increase the number of observations in the average. Of course, the variance is the unknown parameter so we cannot actually do this in practice.
Second, we use observation $n_k$ (except that we use one if $n_k=0$) to estimate the variance in obtaining the weight. In this case, we find that as we average more results the coverage rapidly decreases, similarly to the Garwood interval estimates. We are giving more weight to downward fluctuations, and this is biasing our average. 

Often, we only average a handful of observations, and often subsequent measurements are substantial improvements over previous measurements.
In this situation, coverage of the averages obtained with our procedure may be satisfactory. However, we have demonstrated
that attempting to weight results based on the observed confidence intervals can be dangerous. If the information is available,
it is much better to weight the observations according to the livetime (other things being equal, and accounting for backgrounds) corresponding to each observation.  In particular, it is desirable to go back to the original Poisson observations and perform the averaging based on the joint pdf of those observations. It is strongly recommended that this be done if at all possible. This also has the benefit that it keeps the overcoverage under better control than our ``conservative'' averaging process.
  
%\section{Graphical display of Poisson results}

\section{Discussion}
\label{sec:discussion}

We have investigated several of the most commonly used methods for quoting confidence intervals for the mean of a Poisson distribution, possibly including a background contribution. Even demanding ``exactness'' there is still a large amount of choice in algorithms. We wish to reach some conclusion among the trade-offs and arrive at a recommendation. We embark on a trip through our ``desirable properties'' of section~\ref{sec:desirable} to sort this out, noting again that they may not all be achieved simultaneously.

\subsection{Exactness}
\label{sec:Dexact}
We decided at the outset to only consider exact confidence intervals, in order to guarantee we satisfy the coverage requirement in Eq~\ref{eq:Calpha}. We will insist on this. Thus, $\sqrt(n)$ and Bayesian intervals, as well as a normal approximation, are ruled out in the present context.

\subsection{Connectedness}
\label{sec:Dconnectedness}

We have already argued that for the Poisson distribution, connected intervals are the sensible choice.
If we start with a disconnected interval we can always fill in the values in the gap(s). All of our methods allow
for this property and satisfy it as implemented here. We can improve coverage by relaxing this requirement, but this would be at the cost of unintuitive complexity and we do not consider doing this. 

\subsection{Contains maximum likelihood estimator}

We argued (section~\ref{sec:containMLE}, as does~\cite{Casella1989}) that an acceptable descriptive interval should contain the point estimator, i.e., the maximum likelihood estimator.
The Crow\&Gardner intervals are not guaranteed to include the MLE, e.g., Fig.~\ref{fig:CGNesting}.  While we generally regard this as a serious flaw, in this case it appears only at low confidence levels, below those usually used in quoting confidence intervals. Hence, it may not  be an issue in practice.
The FC intervals also may not include the MLE, in our usage of the term.
However, it is a design feature of FC intervals that $\theta<0$ is excluded and the MLE is defined to never be quoted below zero~\cite{Cousins2025}. Hence, in this sense, the FC intervals do include the MLE.
We have argued in section~\ref{sec:descriptive} that such a restriction does not well serve the goal of description. 

\subsection{Optimal coverage}

The original Garwood interval has been much criticized for substantial overcoverage. Thus, a large amount of the statistics literature has been directed at inventing algorithms that minimize the coverage without
undercovering. We have investigated several of the resulting algorithms. All of them have regions of parameter space with significant overcoverage, while doing better than the Garwood intervals.

From Fig.~\ref{fig:GarwoodCrow} (for 68\% confidence level) we find that the Crow\&Gardner intervals have substantially better coverage than the Garwood intervals.  It also has better coverage in general than the Sterne intervals (Fig.~\ref{fig:SterneCrow}) and the likelihood ratio intervals (Fig.~\ref{fig:LRCrow}). As the score and likelihood ratio intervals
have similar coverage (Fig.~\ref{fig:scoreLR}), the Crow\&Gardner intervals also improve coverage over the score intervals.
The comparison of coverage for  Crow\&Gardner and Kabaila\&Byrne intervals (Fig.~\ref{fig:KBCrow})  is somewhat more mixed, with one or the other performing better depending on $\mu$. However, the general impression favors  Crow\&Gardner once again.
The Blaker intervals perform somewhat better than the Sterne intervals (Fig.~\ref{fig:SternBlaker}), but not so much as to make them better than  Crow\&Gardner.
Thus,   Crow\&Gardner  intervals appear to be among the best in terms of coverage. We note that the Garwood interval only satisfies Eq.~\ref{eq:optimalCoverage} asymptotically (besides perhaps some fortuitous solutions).

However, as will be discussed in sections~\ref{sec:orderedD}--\ref{sec:Dsensiblep} and in  the Conclusions, section~\ref{sec:conclusions}, we must consider the incompatibility of the methods with smallest coverage with other desirable properties, and will decide not to insist on optimal coverage.

\subsection{Length}

\begin{figure}[h]
\centering
\includegraphics[width=0.82\columnwidth]{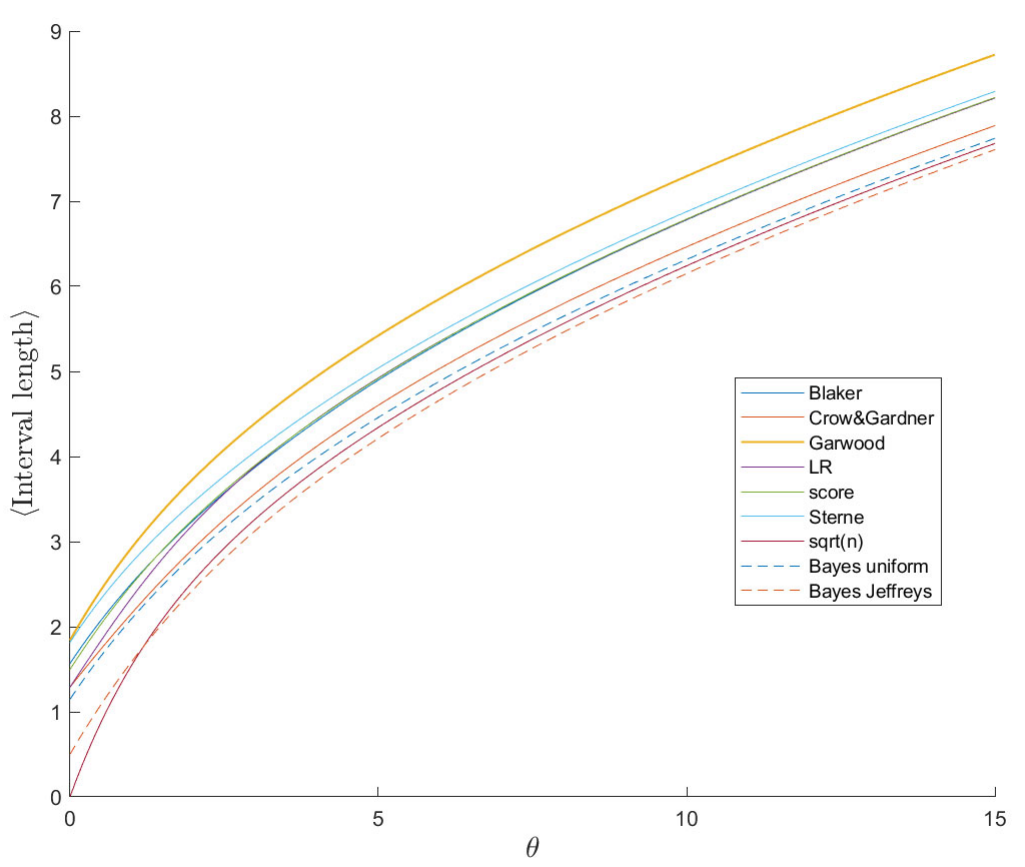}
\caption{Expected length of 68\% CL intervals, as a function of $\theta$ (assuming $b=0$)..\label{fig:avgLength}}
\end{figure}

\begin{figure}[h]
\centering
\includegraphics[width=0.82\columnwidth]{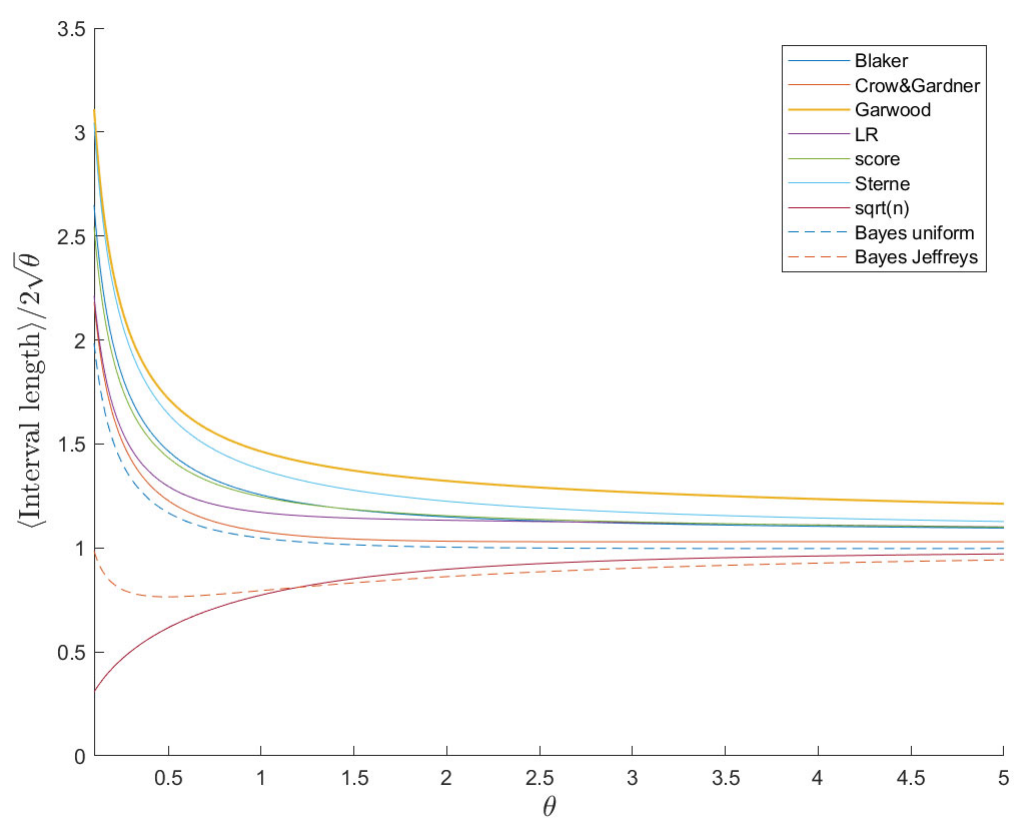}
\caption{Same as Fig.~\ref{fig:avgLength}, except divided by $2\sqrt{\theta}$, and for $\theta$ between 0.1 and 5.\label{fig:avgLengthB}}
\end{figure}

Traditionally (e.g., \cite{Sterne1954, Crow1959, Casella1989, Wendell2001, Somerville2013}), statisticians have focused on optimizing  length as well as coverage.
To compare algorithms for length, we consider the expected length of an interval, as a function of $\theta+b$. For our conventional methods, this is independent of $b$, so we may simply consider the expected length as a function of $\theta$ for $b=0$. While we discuss other measures of length in section~\ref{sec:length}, the expected length as a function of $\theta$ is sufficient for our discussion here.
For any given algorithm, giving confidence interval $(\ell(n),u(n))$, this length is
\begin{equation}
E(L;\mu=\theta+b) = \sum_{n=0}^\infty (u - \ell) \frac{\mu^n}{n!}e^{-\mu}.
\end{equation}
This is shown in Figs.~\ref{fig:avgLength}  and~\ref{fig:avgLengthB}  for several algorithms.

The detailed behavior of the lengths in Figs.~\ref{fig:avgLength}   and~\ref{fig:avgLengthB}  depend slightly on the value of $\theta$. For $\theta\sim 1$, The ordering from largest to smallest lengths is: Garwood, Sterne, Blaker, score, likelihood ratio, and Crow\&Gardner. The Kabaila\&Byrne intervals are also somewhat larger than the Crow\&Gardner intervals (Fig.~\ref{fig:KBCrow}).

\begin{figure}[h]
\centering
\includegraphics[width=0.45\columnwidth]{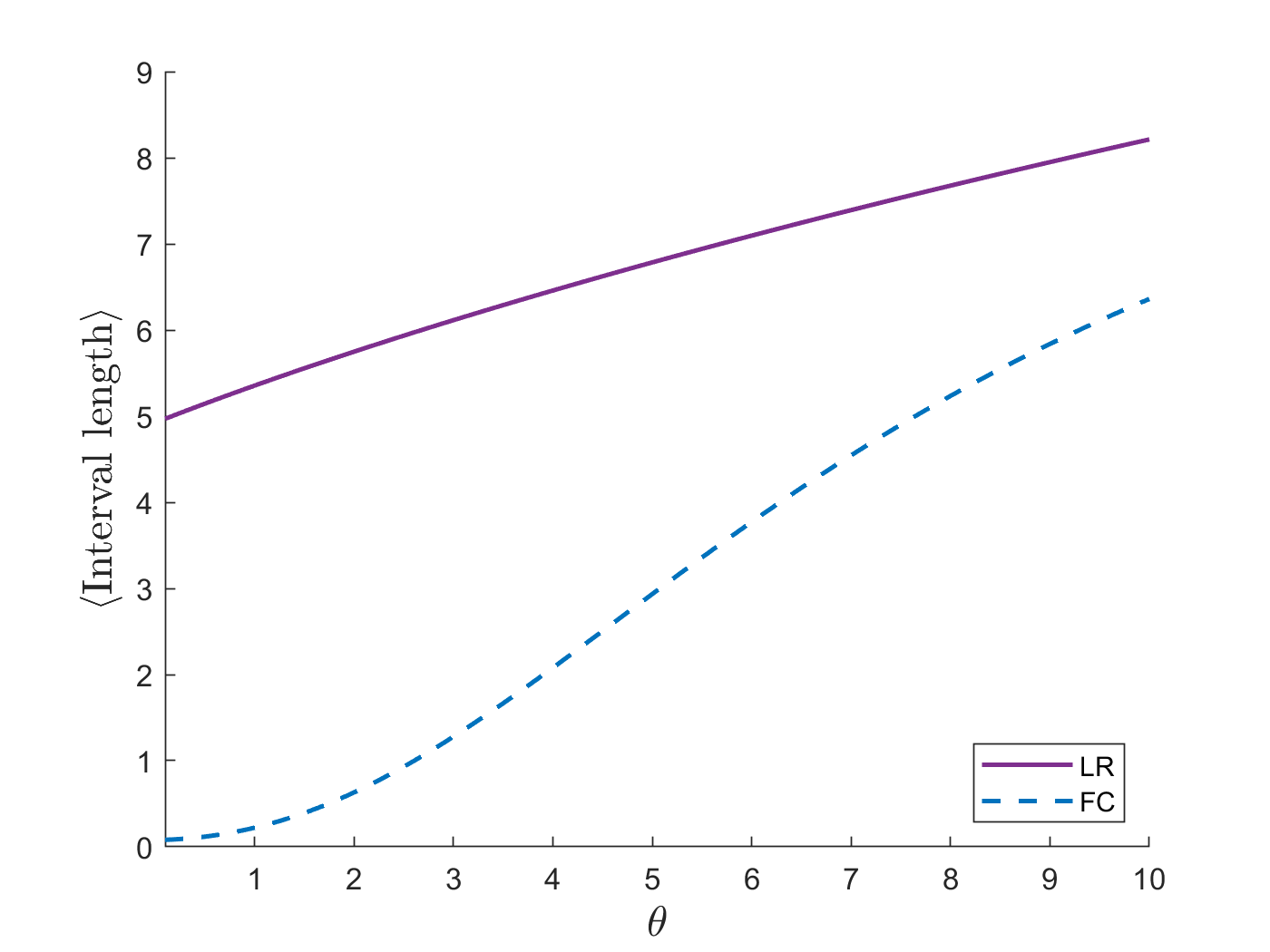}
\includegraphics[width=0.45\columnwidth]{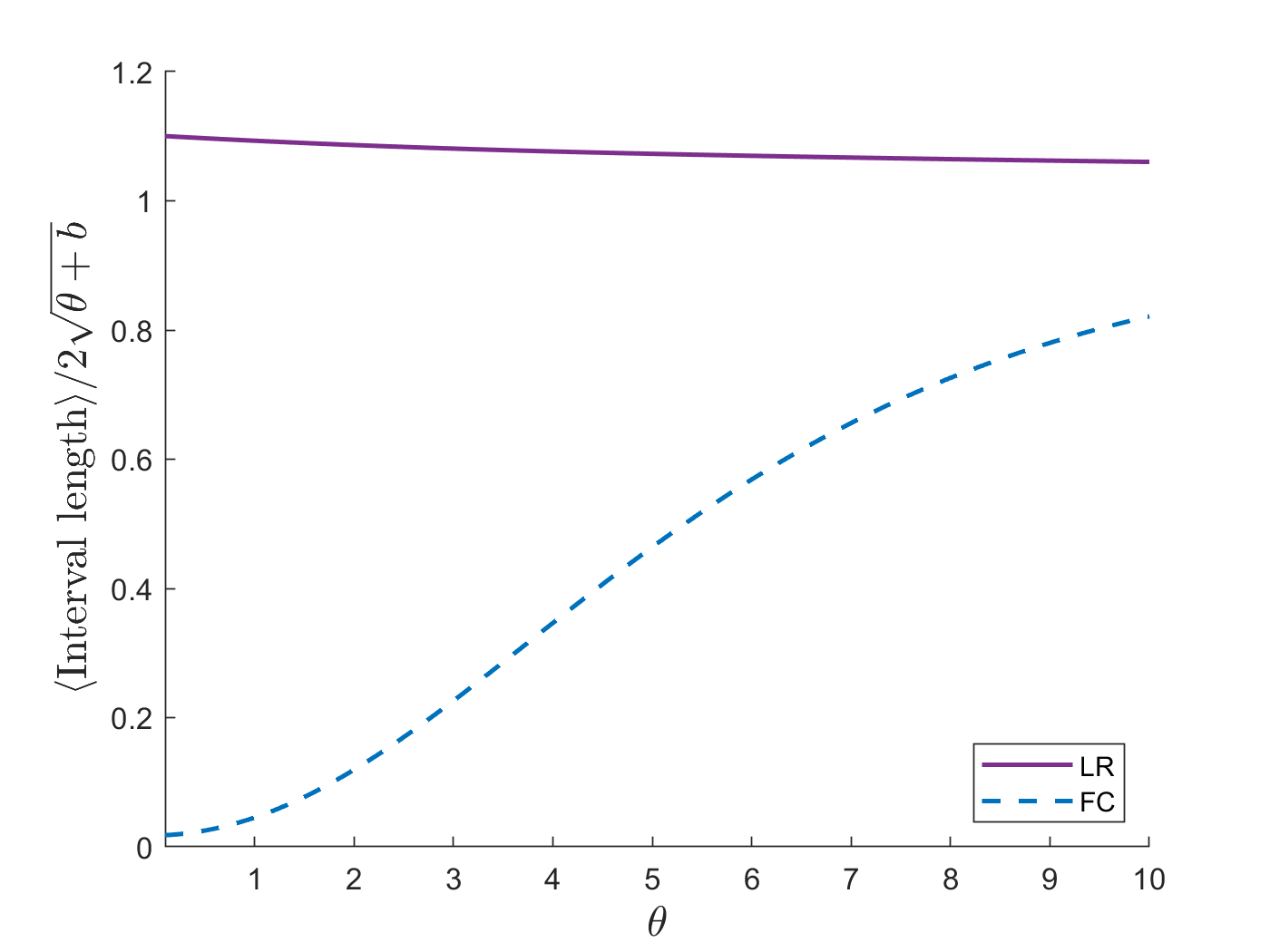}
\caption{Left: Expected length of 68\% CL intervals, as a function of $\theta$ for $b=5$, comparing FC with LR. Right: Same as left figure, except divided by $2\sqrt{\theta+b}$.\label{fig:avgLengthFCLR}}
\end{figure}

For FC intervals, with $b=0$, the expected length is identical to the likelihood ratio length. However, for $b\ne 0$, the FC interval lengths are smaller when $n$ is small, Fig.~\ref{fig:FCvsLRcut}.  This is clearly illustrated in Fig.~\ref{fig:avgLengthFCLR}, which shows a comparison of FC with LR interval expected lengths for a background of $b=5$. When $\theta$ is small (on a scale of the background),
the FC intervals are much smaller than the LR intervals, and much smaller than the anticipated scale of $\sqrt{\theta+b}$.

On the two traditional measures of performance, coverage and length, the Crow\&Gardner intervals look very attractive,\footnote{At one point I thought that the Crow\&Gardner intervals would be the recommended choice, before I realized the substantial issues with other considerations.} while the original Garwood intervals come in last. However, among strictly nested equal-tailed exact intervals, Thulin and Zwanzig~\cite{Thulin2017} show that the Garwood interval
minimizes the expected interval length for all $\theta$ and minimizes the length for all $N=n$.

\subsection{Scale and asymptopia}

 It has sometimes (e.g.,~\cite{Chernoff1951}) been suggested that other properties besides length and coverage are worthy of attention, and this has gained traction, e.g.,~\cite{Blaker2000, Kabaila2001, Vos2005, Hirji2006, Vos2008, Patil2012, Thulin2017}. We turn to these other considerations in this and following sections.
 
 It is probably not controversial that in the limit of large statistics (large $\mu$), the confidence intervals should converge on the correct coverage. Fortunately, (e.g., see Fig.~\ref{fig:avgLengthB}), all of our exact (and $\sqrt{n}$ and Bayes) intervals satisfy this requirement, though the convergence of the Garwood seems slowest, reflecting its overall length performance.
 
 More subtle is our notion that a good descriptive interval should have an intuitive scale, reflecting the measurement precision. We know that the variance of a Poisson with mean $\mu$ is $\mu$, hence our intervals should scale approximately as $\sqrt{N}$ as remarked in section~\ref{sec:scale}. From Figs.~\ref{fig:Garwood}-\ref{fig:GarwoodCrow}, \ref{fig:SternBlaker}, and \ref{fig:KBCrow}-\ref{fig:scoreLR}, we find that most of our intervals satisfy this intuition. The two Bayes intervals (Fig.~\ref{fig:Bayesb5}) also satisfy it.
 The notable exceptions are the FC intervals and the cut likelihood ratio interval in the presence of background, (Fig.~\ref{fig:FCvsLR}
 and Fig.~\ref{fig:FCvsLRcut}). In these cases, the confidence intervals are much smaller than the desired intuitive behavior when negative fluctuations occur at low statistics. That is, these intervals suggest a precision in the measurement much better than actual. FC~\cite{Feldman1997} suggest quoting the ``sensitivity'' of the experiment as a way to alert the consumer when this occurs.
 
\subsection{Ordered}
\label{sec:orderedD}

Inspection of the performance plots suggests that all of our methods are ordered, at least for the situations studied.
However, among these plots we see that the Crow\&Gardner intervals may not be strictly ordered (Fig.~\ref{fig:GarwoodCrow}). The lower bounds for $n=6$ and $n=7$, as well as for $n=9$ and $n=10$, appear to coincide. This is verified by further evaluation. This may not seem fatal, as long as the interval increases in length as the number of observations grows it may not be too concerning if one of the limits repeats.

In fact, not all methods have monotonically increasing interval sizes with observation $n$.
The Sterne and Blaker intervals violate this, as can be seen in the plot on the right in Fig.~\ref{fig:SternBlaker}, even though 
they satisfy strict ordering. Intervals derived from the  likelihood ratio test (hence also FC) also have intervals that sometimes decrease in size as $n$ increases. Score intervals are non-decreasing but sometimes have the same length for successive $n$ values. From Fig.~\ref{fig:KBCrow}, we see that the Kabaila\&Byrne intervals rather badly violate monotonicity of interval size as a function of observation. This is interesting, because Kabaila\&Byrne make a point of
demanding strict ordering. It seems that in satisfying strict ordering this may have affected the monotonicity of the
interval length. We suggest that monotonicity of interval length is more important than strict ordering, at least if ordering is 
satisfied.

\subsection{Symmetry}

The Garwood intervals are defined to be symmetric (in probability), enforcing $1-\alpha/2$ probability on each side.
We have compared the Crow\&Gardner symmetry in Fig.~\ref{fig:GarwoodCrowSymmetry} and the Blaker in Fig.~\ref{fig:GarwoodBlakerSymmetry}. In both cases, the excursions from symmetry are smaller for the Garwood intervals, as expected. We do not necessarily consider optimizing on symmetry to be crucial, but it is intuitively helpful to have a description where we know that the probability is reasonably balanced between the two tails.

\subsection{Nested}
\label{sec:DNested}

\begin{figure}[h]
\centering
\includegraphics[width=0.7\columnwidth]{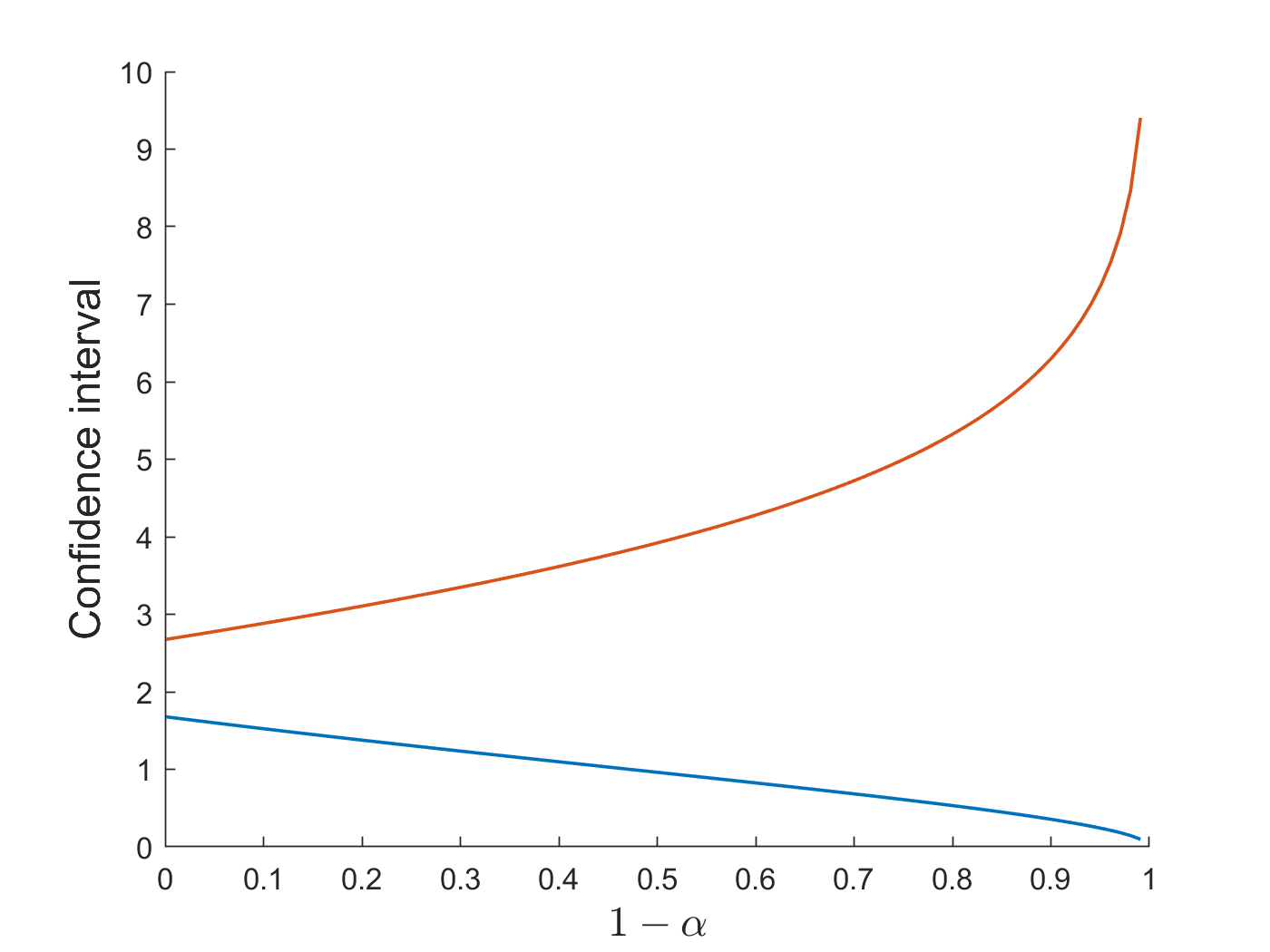}
\caption{The Garwood confidence interval $(\ell, u)$ as a function of confidence level for $n=2$. The upper curve is the upper bound $u$ and the lower curve is the lower bound $\ell$. \label{fig:GarwoodNesting}}
\end{figure}

As defined in Section~\ref{sec:nested}, a nested interval is one in which the interval for a smaller  confidence level is always a subset of the interval for a larger confidence level. Strictly nested intervals are when this is always a proper subset.
We have already seen some signs of issues with nestedness in Fig.~\ref{fig:Nesting} which shows that the Blaker intervals are nested, but not strictly nested, and  the Crow\&Gardner intervals are not even nested. 
In contrast, the Garwood interval is strictly nested,
e.g., Fig.~\ref{fig:GarwoodNesting}.

To investigate the behavior of the non-Garwood intervals, Thulin and Zwanzig~\cite{Thulin2017} define the notion of a ``strictly two-sided'' test and corresponding strictly two-sided confidence intervals. The key point in this definition is that computing a $p$-value based on the test statistic requires simultaneously comparing the test statistic to both tails under the null distribution.  The Garwood interval is not such a test, because the $p$-value for $\theta=\theta_0$ is
\begin{equation}
p_{\theta_0}(n) = 2\min\left[\sum_{k=0}^n f(n;\theta_0,b),\sum_{k=n}^\infty f(n;\theta_0,b)\right].
\end{equation}
In this case, the $p$-value is computed as twice the minimum of the two tails. This is consistent with our intuitive notion that a $p$-value is the probability (under $H_0$) of obtaining a sampling as far or farther from the null hypothesis than the observed value. On the other hand, Thulin and Zwanzig show that the likelihood ratio, Sterne, score, and Blaker intervals are all strictly two-sided intervals. They refer to the Crow\&Gardner and Kabaila\&Byrne intervals as ``strictly two-sided type'', because they are not derived from test statistics, but have similarities with the strictly two-sided intervals.
Strictly two-sided tests are found to have the following properties~\cite{Thulin2017}:
\begin{itemize}
\item The $p$-value is a discontinuous function of the null hypothesis ($\theta_0$);
\item The confidence interval bounds $(\ell_\alpha(n),u_\alpha(n))$ are not strictly monotonic in $\alpha$;
\item The confidence intervals are not strictly nested.
\end{itemize}
Hence, the lack of strict nesting or even nesting behavior we see for example in Fig.~\ref{fig:Nesting}. The likelihood
ratio, Sterne, and score intervals also are not strictly nested and the Kabaila\&Byrne intervals are not nested.

\subsection{Continuity and monotonicity}

Continuity (Section~\ref{sec:Continuity}) here refers to the continuity of the confidence intervals as a function of confidence level. We will discuss $p$-values in section~\ref{sec:Dsensiblep}. Even if the intervals are nested, if they are not continuous a small change in confidence level could correspond to a large change in interval. The Garwood intervals are continuous (e.g., Fig.~\ref{fig:GarwoodNesting}).

Similarly with the observations about nestedness, Thulin\&Zwanzig~\cite{Thulin2017} find that strictly two-sided intervals are discontinuous functions of confidence level. We see an example of this in Fig.~\ref{fig:Nesting}(left) for the Blaker interval. It seems that still worse than discontinuity is lack of monoticity.  We see that the strictly two-sided type interval, the Crow\&Gardner is not even a monotonic function of confidence level (Fig.~\ref{fig:Nesting}(right)). This is also true for the Kabaila\&Byrne strictly two-sided type interval. In such cases, increasing the confidence level can actually decrease the interval, giving seriously counterintuitive behavior.

We note that some intervals do not have lengths that are monotonically increasing with $n$ (e.g., the Kabaila\&Byrne intervals are especially notable in Fig.~\ref{fig:KBCrow} and less so for the LR and score intervals in Fig.~\ref{fig:scoreLR} and for the FC intervals in Fig.~\ref{fig:FCvsBayesU}). While this is not desirable, we consider this to be less important than some of the other considerations.

\subsection{Gives sensible $p$-values}
\label{sec:Dsensiblep}

\begin{figure}[h]
\centering
\includegraphics[width=0.49\columnwidth]{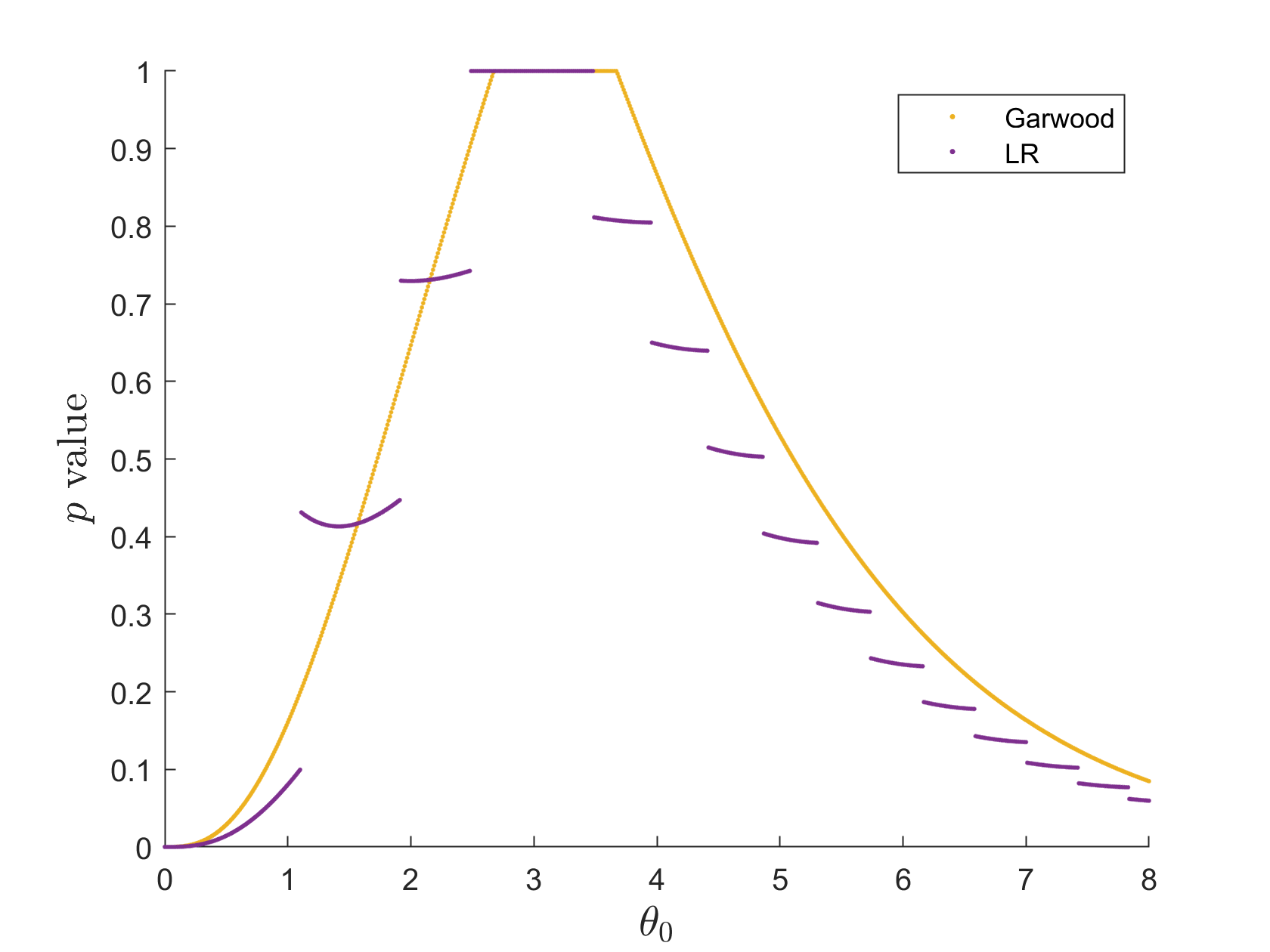}
\includegraphics[width=0.49\columnwidth]{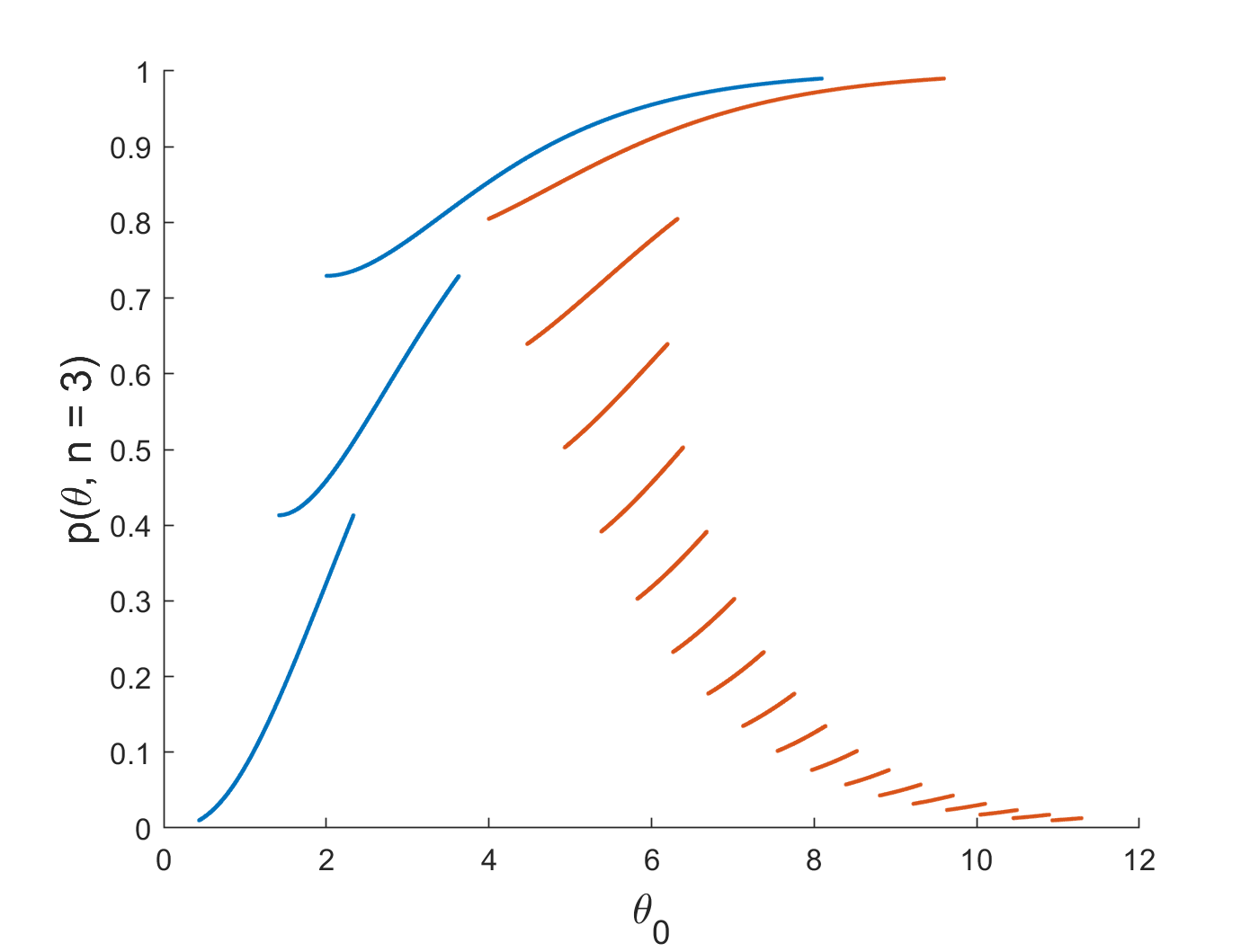}
\caption{The $p$-values as a function of null hypothesis $\theta_0$ for an observation $n=3$ (and no background). Left: The continuous orange curve is according to the Garwood interval and the discontinuous purple curve is according to the likelihood ratio interval. Right: The $p$-values according to the Kabaila\&Byrne intervals, also for $n=3$.\label{fig:Garwood-LR-pvalue3}}
\end{figure}

According to the theorem quoted in Section~\ref{sec:DNested}, $p$-values based on the strictly two-sided tests are discontinuous functions of the null hypothesis $\theta=\theta_0$. 
For example, the left side of Figure~\ref{fig:Garwood-LR-pvalue3} shows a comparison of the $p$-values derived from the Garwood and the likelihood ratio confidence intervals, as a function of null hypothesis $\theta_0$, for an observation of $n=3$.
Overall, the Garwood $p$-values tend to be larger, indicating less discrimination against the null hypothesis. This is a reflection of the fact that the Garwood intervals tend to overcover more, which is the major criticism of these intervals.

On the other hand, the Garwood $p$-values are well-behaved, being continuous and monotonically rising to a plateau of $p=1$ when $\theta_0$ is near $n=3$, and then monotonically falling when $\theta_0$ is somewhat larger than three. The Garwood $p$-values are ``bimonotonic decreasing'', in the above sense. 
In contrast, the behavior of the likelihood ratio $p$-values is much more problematic. First, they are discontinuous, so that a small change in null hypothesis ($\theta_0$) can give a large change in $p$, which is not something we would normally expect given the continuity of our probability distribution as a function of $\theta$. Second, there is a region where the $p$-value becomes higher for a more inconsistent value of $\theta_0$, contrary to anticipated reasonable behavior. The strictly two-sided type intervals are no better. For example, the right side of Fig.~\ref{fig:Garwood-LR-pvalue3} shows the $p$-values for $n=3$ computed according to the Kabaila-Byrne intervals. Now the $p$-values are not even well-defined, taking on multiple
values for many values of $\theta_0$.

We regard these issues as serious flaws in a $p$-value. A discontinuous or non-bimonotonic $p$-value is not expected behavior, and provides a non-intuitive description of the measurement. 
Hirji~\cite{Hirji2006} states that continuous bimonotonic decreasing behavior in the $p$-values implies some desirable properties: {(\it i)} A two-sided test which is significant at $\alpha_1$ is also significant at $\alpha_2$ if $\alpha_2 > \alpha_1$;
{(\it ii)} The confidence intervals are nested; {(\it iii)} The confidence intervals are connected. 
Thulin and Zwanzig~\cite{Thulin2017} show further that if the limits of a confidence interval are continuous in $\alpha$ then the derived $p$-value function is strictly bimonotone in $\theta$.

Only the Garwood intervals
provide for sensible $p$-values among the two-sided exact intervals studied.
Further, as noted earlier, the Garwood interval is optimal among nested equal-tailed exact intervals in the sense that it minimizes the expected interval length for all $\theta$ and minimizes the length for all $N=n$~\cite{Thulin2017} .  At least for the tests studied, the lack of strict nesting comes along with intervals that are not strictly momotonic in $\alpha$ and $p$-values that are discontinuous in the null hypothesis.

\subsection{\ Averages}

We also return to the matter of averaging observations, discussed in section~\ref{sec:averaging}.
Our discussion there was in the context of no backgrounds. However, if there are backgrounds
that might be different for $K$ different measurements, the sampling distribution becomes (in the notation of section~\ref{sec:averaging})
\begin{equation}
\hbox{Prob}(n) = \prod_{k=1}^K \frac{\left[(\Gamma + b_k)T_k\right]^{n_k}}{n_k!} e^{-(\Gamma+b_k)T_k}, 
\end{equation}
where $b=(b_1,\ldots, b_K)$ are background ``rate'' parameters, presumed to be known, but potentially different for each measurement. If the background parameters are not all identical, the sampling distribution is no longer in the exponential family, and in particular is not of Poisson form. In this case, we are generally forced to make numerical methods to obtain point estimators and confidence intervals.

The maximum likelihood estimator, $\hat\Gamma$, in the presence of backgrounds is obtained as the solution to:
\begin{equation}
\sum_{k=1}^K \frac{n_k}{T_k(\hat\Gamma+b_k)} = \sum_{k=1}^K T_k.
\end{equation}
The MLE may turn out to be negative. As before it is recommended to quote this as the descriptive result, rather than attempt to restrict it to a ``physical'' region. Such restriction may be adopted in a further interpretive step if desired.
It is tempting to study the variation of the likelihood function with $\Gamma$ to obtain a confidence interval.
Unless it is clear that the sample size is large enough to approximate normality, simulations are required to demonstrate the desired coverage.

\section{Conclusions}
\label{sec:conclusions}

We are ready to take our observations and come to some recommendations for descriptive intervals. We are insisting on exact confidence intervals, satisfying Eq.~\ref{eq:Calpha}, as discussed in section~\ref{sec:exact}.  We are also
requiring that our intervals be connected, as discussed in section~\ref{sec:connectedness}. Describing the observation with $\sqrt{n}$ is commonly done and we have an intuitive feeling for what it means. However, it does not provide a probabilistic interpretation that can be used in functional dependences as discussed in sections~\ref{sec:Introduction} and~\ref{sec:sqrtn}. While $\sqrt{n}$ is useful and expedient, we look here for exact intervals. The Bayes' intervals studied in section~\ref{sec:priors} do not satisfy Eq.~\ref{eq:Calpha} and in any event are not motivated as descriptive.

We have already argued that one-sided intervals (limits) are not as informative as two-sided intervals (section~\ref{sec:UL}). The motivation for quoting a one-sided interval is typically to make a statement about the true value of the parameter. In this case, a Bayesian degree-of-belief analysis is recommended.
If one insists on quoting a frequentist one-sided interval, then the ``standard'' methodology of section~\ref{sec:UL} is recommended. The $CL_s$ method (section~\ref{sec:CLs}) also produces an exact
upper limit satisfying Eq.~\ref{eq:Calpha}. However, it substantially overcovers for small $\theta$, with large intervals for small observations (Fig.~\ref{fig:CLsUL}), hence is less optimal compared with the standard limit. This is understandable because the $CL_s$ method is motivated with consideration about truth. It thus does not provide an attractive descriptive statistic.

The FC interval and the cut LR interval (or variations in section~\ref{sec:FC}), while two-sided and exact, have restrictions on the interval in the case when $b\ne0$. This restriction leads to intervals that may be very small, much smaller than the expected intuitive scale, when downward fluctuations occur. Hence the interpretation as description is clouded. While we note that FC~\cite{Feldman1997} suggest to also quote the ``sensitivity'' of the experiment when this occurs, we conclude that there are better choices for descriptive intervals that do not have this feature.

We turn to the conventional methods (section~\ref{sec:conventional}), giving two-sided exact intervals, which are of most interest as descriptive intervals. No choice stands out on all of the hoped-for features. The comparison here is more nuanced and some subjective choices must be made on which properties to emphasize over others. 

The original Garwood interval ticks the boxes on almost all of our desirable properties, including continuous monotonic and strictly nested behaviors and sensible $p$-values. The issue with it is that the interval lengths and over-coverage are larger than achievable if we relax these features. This fault is what led to the significant literature looking for ``improvements'', generally without considering some of the possible drawbacks. Unfortunately, we know of no improvement that has the other desirable properties, and have noted evidence that this may not be possible. Among intervals that do possess strict nestedness, the Garwood intervals are optimal as shortest length~\cite{Thulin2017}.

It may be debated whether the properties such as nesting, continuity, and provision of sensible $p$-values are requirements for a confidence interval. Several of our intervals provide shorter length and smaller overcoverage than the Garwood interval. If they are only used as descriptive intervals, do we care about
bizarre behavior in other respects? We contend that the interval must at least include the maximum likelihood estimator (and we insist here that for descriptive purposes the likelihood function should be evaluated without regard to knowledge of ``truth''), and we have seen that this argues against Crow\&Gardner (if low CL is considered), and possibly others. But there are intervals besides Garwood, such as the Sterne and the likelihood ratio that do contain the MLE. They also seem to have reasonable scale properties. As long as we never look too closely, maybe this is ok?

Unfortunately, as soon as we try to look at the confidence intervals for our observation at different confidence levels, as we sometimes do, we get into trouble. Intervals that change drastically for a small change in confidence level (e.g., Fig.~\ref{fig:Nesting}) lose their descriptive appeal. Of the intervals studied, the only two-sided exact interval that displays continuous dependence on confidence level is the Garwood. Another concern is that of nesting. When we look at the intervals corresponding to two different confidence levels, we naturally expect the interval with the smaller confidence level to be a proper subset of the interval with a higher confidence level. Again, our only success among two-sided exact intervals  is the Garwood interval.

As noted in section~\ref{sec:Dsensiblep}, properties of nesting and continuity are related to bimonotonicity in the $p$-values (as a function of the null hypothesis). Requiring the confidence intervals and the $p$-values to be part of a consistent framework is desirable for a
consistent description of the result of the measurement. For example, a confidence interval that is closer to the null hypothesis should correspond to a higher $p$-value than an interval which is farther. We thus consider it important that our intervals provide sensible $p$-values. In particular, the $p$-values should exhibit a continuous bimonotonic dependence on the null hypothesis. Among the studied two-sided exact intervals only the Garwood interval achieves this, and in fact, gives the two-sided $p$-values that we naively calculate (``probability of obtaining at result as different or more from the null hypothesis as observed''). 

While coverage and length are important, so are these other properties, and ensuring them seems worth erring on the 
conservative side in coverage.
It is thus our recommendation to use the Garwood interval for the Poisson distribution. It appears to be the best that we can do while satisfying the intuitively hoped-for behavior. Fortunately, it is among the easiest to compute given an observation. As we have noted, it is also the default already for
MATLAB and R, and perhaps other toolkits.

We have cautioned that relying only on stated confidence intervals in averaging result of different measurements
may be unjustified, and one should pay attention to the original sampling distribution when it is not normal.

One of the important uses of confidence intervals is in the graphical representation of a measurement, especially when 
histogramming data where each bin is a sampling from an independent Poisson distribution. If backgrounds are subtracted, do not be afraid to plot error bars that go negative, whether using Garwood or $\sqrt{n}$. Given that there are multiple choices in use, it is helpful to say how the error bars on histograms are computed.

We have examined a specific situation, sampling from a Poisson distribution with known background rate. Among several methods for computing exact confidence intervals as descriptive statistics, we recommend using the Garwood interval. Our discussion has been deliberately restricted to a ``simple case'', already with complexity and controversy.
A similar discussion, with less complexity, for the normal distribution is in~\cite{NarskyPorterNormal}.
It may be remarked that the methodologies of Garwood, Sterne, Crow\&Gardner, likelihood ratio, and score presented here, if applied to the normal case, all give the identical, familiar, results for normal two-sided confidence intervals. However, the richness of choice for the Poisson does not vanish for distributions such as the normal. Different methods have been proposed and similar considerations apply.

More generally, there may be unknown nuisance parameters that are estimated in auxillary measurements. A straightforward approach is to go ahead and use the Garwood intervals at the point estimators for the nuisance parameters. Then the errors in the nuisance parameters can be regarded as errors in the model, or ``systematic errors'', with the effect estimated by varying these parameters, i.e., by varying the model. But there are other options and the choice might depend on circumstances; such a discussion is outside the scope of this paper.

 \section*{Acknowledgements}
 
I would like to thank the anonymous referees for their impressive diligence and comments; the paper is better for it. This work is supported in part by the U.S. Department of Energy, Office of Science, Office of High Energy Physics, under Award Number DE-SC0011925.
 
\bibliographystyle{unsrt}
\bibliography{PoissonIntervals}

\end{document}